\documentclass[showkeys,superscriptaddress,10pt,twocolumn,showpacs]{revtex4-1}
\usepackage{epsfig}
\usepackage{natbib}
\usepackage{amsmath}
\usepackage{times}
\usepackage{psfrag}
\usepackage{subfigure}
\usepackage{epstopdf}
\newcommand{\hg}{ }

\begin{document}

\title{Stochastic Bifurcations induced by correlated Noise in a Birhythmic van der Pol System}

\author{R. Mbakob Yonkeu}
\address{Laboratory of Mechanics and Materials, Department of Physics, Faculty of
Science,
University of Yaound\'e I, Box 812, Yaound\'e, Cameroon.}
\author{R. Yamapi}
\address{Fundamental Physics Laboratory, Physics of Complex System group, Department of Physics, Faculty of
 Science, University of Douala, Box 24 157 Douala, Cameroon.}

\author{G. Filatrella}
 \address{Department  of Sciences and Technologies
\small and Salerno unit of CNISM, University of Sannio, Via Port'Arsa 11,
I-82100 Benevento, Italy.}
\author{C. Tchawoua}
\address{Laboratory of Mechanics and Materials, Department of Physics, Faculty of Science,
 University of Yaound\'e I, Box 812, Yaound\'e, Cameroon.}
\date{5 August 2015}


\begin{abstract}
We investigate the effects of exponentially correlated noise on  birhythmic van der Pol type oscillators.
The analytical results are obtained applying the quasi-harmonic assumption to the Langevin equation to derive an approximated Fokker-Planck equation.
This approach allows to analytically derive the probability distributions as well as the activation energies associated to switching between coexisting attractors.
The stationary probability density function of the van der Pol oscillator reveals the influence of the correlation time on the dynamics.
Stochastic bifurcations are discussed through a qualitative change of the stationary probability distribution,
which indicates that noise intensity and correlation time can be treated as bifurcation parameters.
Comparing the analytical and numerical results, we find good agreement both when the frequencies of the attractors are about equal or when they are markedly different.

\noindent
\textbf{\emph{Keywords:}}:\emph{Stochastic bifurcation; colored noise; birhythmic system; Fokker-Planck equation\\
\textbf{Corresponding author:} {\it E-mail address:} filatrella@unisannio.it (G. Filatrella).\\
\textbf{ To appear in:} {\it  Commun. Nonlinear Sci. Numer. Simulat. }}
\end{abstract}

\maketitle

\section{Introduction }

When studying a phenomenon, one  is generally interested in overriding effects.
Nonlinear science has been used to account for some phenomena in physics, biology and
 modeling of social facts that could not be otherwise explained, even at a qualitative level.
Among these phenomena, we can number self-sustained oscillations and noise activated transitions.
A paradigmatic model for oscillating systems is the van der Pol circuit, named after the work of  Balthazar van der Pol, who introduced relaxation oscillations to describe the cycles produced by self-sustained oscillating systems such as the triode circuit.\cite{a.blondel,vanderpol,Ginoux12}.
With its many variants, van der Pol oscillators  have served as a paradigm in various physical, chemical, and biological processes \cite{Jewett99,Anishchenko07,Phillips10}.
A special merit of van der Pol has been to propose a dimensionless (or reduced) equation in the terminology of Curie \cite{a.blondel,Curie891}, for it can be used to explain various system regardless their origins.
The van der Pol system was not the first one where self oscillating behaviors were observed, but this is the first reduced equation in Curie's sense.
With its graphical representation \cite{c.hayashi new york} van der Pol also contributed to change the way in which these nonlinear phenomena had to be investigated.

The oscillatory motion can be  influenced by internal or external noise \cite{Moss89,Zakharova10,Berglund14} and the resulting interplay between nonlinearities and noise can induce transitions similar to the standard transition between two stable points  \cite{Kramers40,N.G.Van Kampen}.
In the case of self-sustained oscillations, one observes a transition between two stable orbits  rather than two points. However, it has been shown that the escape is indeed of the Kramers' type for uncorrelated noise \cite{Yamapi10,Yamapi12} even in presence of time delay \cite{Ghosh11,Chamgoue12,Chamgoue13}, and it can be ascribed to the existence of a quasipotential or pseudopotential \cite{Graham85,Kautz88}.
The quasipotential allows to estimate the energy barriers, and indicates that for most parameters, a single attractor is dominantly stable \cite{Yamapi10,Yamapi12}, while the other is characterized by a small energy barrier.
The effect of a finite correlation time has also been investigated, and it has been found that, as expected for static potential, also the periodic orbits are stabilized by the finite correlation \cite{Mbakob14}.

 Noise, however, can produce also effects that are more radical, for it can induce a structural (or topological) change of the solutions \cite{Arnold03}.
 This is the case when noise induces a stochastic bifurcation, that may be characterized with a qualitative change of the stationary probability distribution.
 Moreover, noise can stabilize unstable equilibria and shift bifurcations, or it can even induce new stable states that have no deterministic counterpart (for instance noise can excite internal modes of oscillation and it can even enhance the response of a nonlinear system to external signals \cite{Wiesenfeld89,Gammaitoni98,Addesso12}).
When the noise driven stochastic differential equation is colored, the nature of the stochastic  process becomes non-Markovian and cannot be treated with the standard Fokker-Planck techniques.
The study of stochastic nonlinear systems driven by colored noise has been undertaken by a number of investigation \cite{Risken90,Hanggi95,Fiasconaro09}.
{\hg On a general ground correlated noise appears as the result of the gross approximation of a system over some finite scale that masks the effect of slow variables \cite{Hanggi95}, as it has been argued in ecology \cite{Spagnolo04} or electronic circuits \cite{Gammaitoni98}.
Birhythmic oscillators are  no exception, as they emerge in biochemical reactions \cite{Kaiser81} or sleeping patterns \cite{Jewett99,Phillips10} where such coarsening could occur.
}
In many treatments, one is essentially concerned with an equilibrium thermal bath at a finite temperature, which stimulates the reaction coordinate to cross the activation energy barrier.
In this paper, we wish to investigate the transition between orbits \cite{Roberts96,Bag06,Xu11,Valenti14}, characterized by different frequencies (and hence birhythmic systems). To do so we analytically treat the effects of Gaussian correlated noise on a special van der Pol oscillator  system that displays a birhythmic behavior, for the  two coexisting attractors characterized by different frequencies.
We search the relation between stochastic bifurcations and noise correlation time on the dynamical properties, taking the system parameters and the statistical characteristics of the noise (e.g., noise intensity and noise correlation time) as bifurcation parameters.
Our approach of the bifurcations is based on the stochastic averaging method \cite{Anishchenko07}, that is separating fast and slow variables of the oscillator.

The work is organized as follows. In Sect. \ref{sec:model} we discuss the model and the numerical approach to the discretized equations. In  Sect. \ref{analysis} we derive an analytic approximation that includes the effects of  a finite correlation time. Section \ref{comparison} compares the analytical and numerical results. Section \ref{conclusion} concludes.

\section{The correlated noisy model and birhythmic properties}
\label{sec:model}

In this Section we present the van der Pol model that we use as a prototype of birhythmic model. We do so first presenting the electrical equivalent circuit and then deriving the normalized equations.
\subsection{The correlated noisy model}


\begin{figure*}
\begin{center}
\includegraphics[height=7cm,width=14cm]{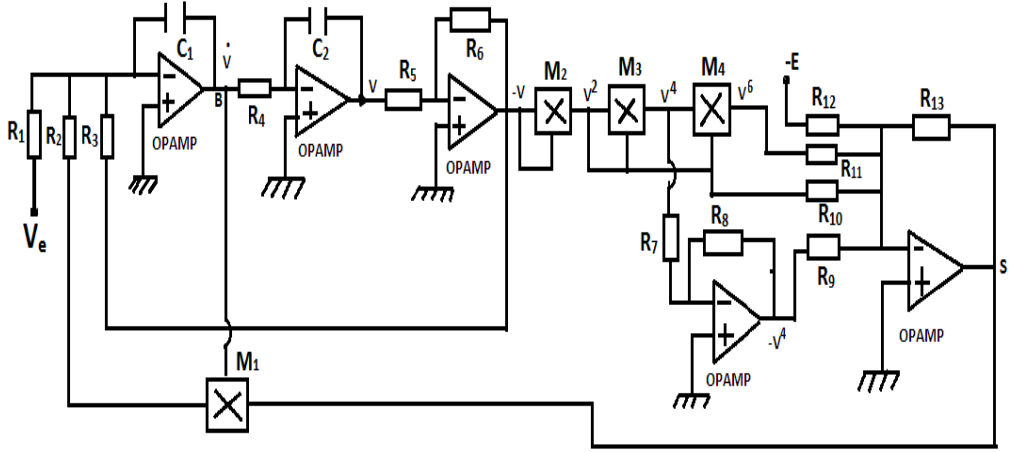}\\
\caption{\it Circuit diagram for an electronic circuit approximately described by a birhythmic van der Pol type equation. Here the $M_i$'s are analog multipliers circuits while the operation amplifiers with feedback capacitors perform analogue integrations.
}
\label{fig1}
\end{center}
\end{figure*}

The model used in our analysis can be reproduced with electronic circuits. An example of equivalent circuit is shown in Fig. \ref{fig1}.
It consists of electronic multipliers $M_i$ $(i=1,...,4)$, integrators that are operational amplifiers with a feedback capacitor, and sommators realized by operational amplifiers with multiple input resistors.
Using Millman law, the characteristic of each component and the contributions of the electrical voltage $V_e$  supplied by an external source (that in this study is assumed to be a random term) we find for the potential at the left of the point $S$:
\begin{equation}\label{eqvdp1a}
    V_s=E\frac{R_{13}}{R_{12}}
    -\frac{R_{13}}{R_{10}}\frac{V^{2}}{K}+\frac{R_{13}}{R_9}\frac{V^{4}}{K^3}-\frac{R_{13}}{R_{11}}\frac{V^{6}}{K^5}.
\end{equation}
where $K$ is a scaling factor, which has the dimension of voltage.
The multiplicator $M_1$ gives the end voltage  (the dots denote the time derivative):
\begin{widetext}
\begin{eqnarray}
&   &  \int   \left[ \frac{V_e}{R_1C_1}-\frac{V}{R_3C_1}+
    \left( \frac{R_{13}E}{R_{12}}
    -\frac{R_{13}}{R_{10}}\frac{V^{2}}{K}+\frac{R_{13}}{R_9}\frac{V^{4}}{K^3}-\frac{R_{13}}{R_{11}}\frac{V^{6}}{K^5}\right)
  \frac{  \dot{V}}{K}     \right] \frac{dt'}{R_2C_1} \nonumber \\
&=&  \dot{V}.
\label{eqvdp1b}
\end{eqnarray}
If we take the time derivative of Eq.(\ref{eqvdp1b}), we get
\begin{equation}\label{eqvdp1c}
R_2C_1\ddot{V} -\frac{E}{K}\frac{R_{13}}{R_{12}}(1-\frac{R_{12}}{R_{10}E}\frac{V^{2}}{K}+
   \frac{R_{12}}{R_9E}\frac{V^{4}}{K^3}-    \frac{R_{12}}{R_{11}E}\frac{V^{6}}{K^5})\dot{V}+\frac{1}{R_3C_1}V \nonumber \\
=  \frac{1}{R_1C_1}V_e.
\end{equation}
\end{widetext}
It is convenient to introduce the variables $t'=\omega_0 t$ and $ V=V_0x$,  where $V_0$  is the reference voltage.
With the constraint  $R_{12}V_0^2=R_{10}EK$, Eq. (\ref{eqvdp1c}) becomes the following non-dimensional differential equation:
\begin{equation}
\label{eq4}
    \ddot{x}-\mu(1-x^{2}+\alpha x^{4}-\beta x^{6})\dot{x}+x=\eta(t),
\end{equation}
where
$$\mu=\frac{R_{13} E}{KR_{12}R_2C_1\omega_0};
\quad \alpha=\frac{R_{12}V_0^4}{R_9EK^3};
\quad \beta=\frac{R_{12}V_0^6}{R_{11}EK^5}; $$
$$ \omega_0^{2}=\frac{1}{R_2 R_3 C_1^2};
\quad    \eta(t)=\frac{R_3}{R_1 V_0}V_e.
$$
A similar equation was originally obtained by van der Pol in the analysis of a triode, when the triode  was approximated  with a cubic function  truncating the Taylor-Mclaurin series expansion \cite{vanderpol,Ginoux12}.
Analogously, one can consider a further expansion to the $5^{th}$ order \cite{Kadji07a,Kadji07b} to retrieve an equation as the above (\ref{eq4}).

Equation (\ref{eq4}), even when $\eta(t)\neq 0$, is of the van der Pol type equation, inasmuch most of the practical applications are driven by weak noise intensities.
It is also worth mentioning that van der Pol provided a stability criterion for periodic solutions when the coefficients are sufficiently small.
This equation is also used to  model coherent oscillations in biological system \cite{Phillips10,Kaiser81,Kadji07a}.
In the context of enzymatic reactions \cite{Kaiser81} the parameters $\alpha$ and $ \beta$ are positive  parameters which measure the degree of tendency of the system to a ferroelectric instability  compared to its electric resistance, while $\mu$ is the parameter  that tunes non linearity.

We assume that both the noise  $\eta(t)$ is stationary and  independent of the memory Kernel, thus one cannot apply the fluctuation-dissipation theorem.
We also assume that the random excitation $\eta(t)$ is a zero average:
\begin{equation}
\label{eq8a} \langle \eta (t) \rangle  = 0
\end{equation}
correlated noise, so we have
  \begin{equation}
  \label{eq5}
  \langle \eta(t), \eta(\tilde{t}) \rangle =D\lambda \exp{\left[ -\lambda(t-\tilde{t}) \right]}.
 \end{equation}
 (the parameter $D$ is the intensity of the noise and $\tau=\lambda^{-1}$ is the correlation time for the colored noise.)

The correlated noise $\eta(t)$ can be generated as the solution of the Langevin equation \cite{l.borlandphys}:
\begin{equation}
\dot{\eta}(t) = -\lambda \eta(t) + \lambda g_w(t)
\label{eq:correlated}
\end{equation}
where $g_{w}(t)$ is a standard Gaussian distribution of unit variance.

To complete the description, one further specifies  the distribution of initial $\eta_0$ values in Eq.(\ref{eq8b}), denoted by the parenthesis $\{...\}$:, which is essential for the stationary correlation
{\hg  ( one should consider different initial conditions, with different initializations of the correlated noise, a caution not necessary for uncorrelated noise):
}
\begin{subequations}
\begin{eqnarray}
\label{eq8b} \{\langle \eta (t) \eta (\tilde{t})\rangle \} &=& D \lambda \exp(-\lambda|t-\tilde{t}|).
\end{eqnarray}
\end{subequations}
The distribution of the initial values is given by:
\begin{equation}
\label{eq9}
P(\eta_0)=\frac{1}{(2\pi D\lambda)}\exp\left[ -\frac{\eta_0^2}{2D\lambda}\right].
\end{equation}
The overall noise intensity $D$ is the zero-frequency part of the power spectrum of the (stationary) noise source $\eta(t)$
  \begin{equation}
  \label{eq10}
    2D=\int_{-\infty}^{\infty}|\langle \eta(t)\eta(0)\rangle |dt.
 \end{equation}
The parameter $\tau$ can be also referred to  the intrinsic correlation time:
  \begin{equation}
  \label{eq11}
  \tau=\frac{2D}{\langle \eta^{2}\rangle }
\end{equation}


\begin{figure}[h]
\begin{center}
\includegraphics[height=8cm,width=11cm]{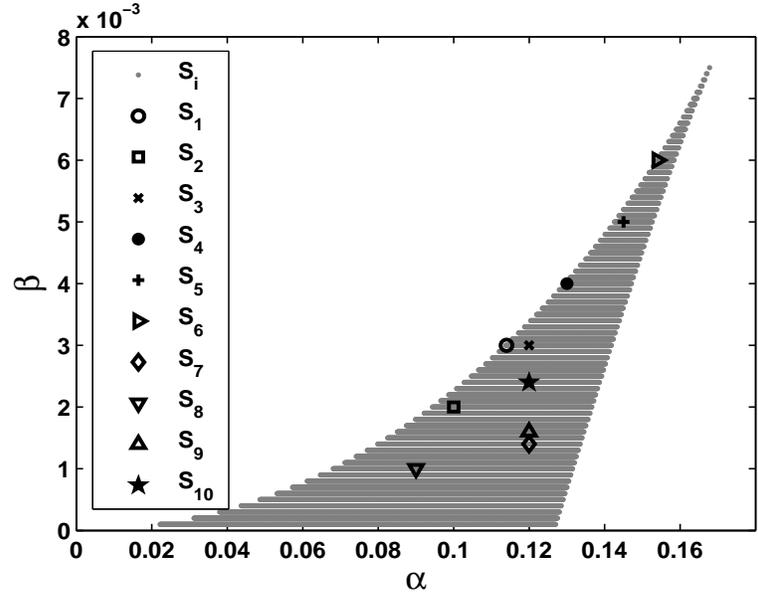}
\caption{\it
Parameters domain for the existence of a single limit cycle (white area) and three limit cycles (colored area) with $\mu=0.01$ and the parameter $S_i (\alpha,\beta) (i=1,2,3,4,5,6)$
where $\Omega_1=\Omega_3=1$ (Table \ref{identical}) and $S_i (i=7,8,9,10)$ where $\Omega_1\neq\Omega_3$ (Table \ref{different}).}
\label{fig3}
\end{center}
\end{figure}

{\footnotesize
\begin{table*}
\begin{tabular}{|l|c|c|c|}
\hline
$S_i(\alpha,\beta)$& Amplitudes $A_i$ of the orbits& Frequencies $\Omega_i$ of the orbits&Periods  $P_i$ of the orbits \\
\hline
                   &$A_1$=2.3772                    & $\Omega_1$=1.0021      & $P_1=6.270$\\

$S_1(0.114;0.003)$& $A_2$=5.0264                     & $\Omega_2$=1.0011           &$P_2=6.275$ \\

                 & $A_3$=5.4667                     & $\Omega_3$=1.0231      &$P_3=6.141$ \\
\hline
                        & $A_1$=2.3069           & $\Omega_1$=0.9870      &$P_1=6.366$ \\

$S_2(0.1;0.002)$     & $A_2$=4.8472                    & $\Omega_2$=1.0001             &$P_2=6.275$ \\

                        & $A_3$=7.1541                 & $\Omega_3$=0.97123     &$P_3=6.468$ \\
\hline
                        & $A_1$=2.4269                   & $\Omega_1$=0.9850       &$P_1=6.379$ \\

$S_3(0.12;0.003)$     & $A_2$=4.2556                  & $\Omega_2$=0.9990            & $P_2=6.289$ \\

                        & $A_3$=6.3245             & $\Omega_3$=0.9865       &$P_3=6.369$ \\
\hline
                        & $A_1$=2.4903               & $\Omega_1$=1.0002   &$P_1=6.282$ \\

$S_4(0.13;0.004)$     & $A_2$=4.4721                    & $\Omega_2$=1.0001        & $P_2=6.282$\\

                        & $A_3$=5.0791               & $\Omega_3$=0.9991       & $P_3=6.289$\\
\hline
                         & $A_1$=2.6605              & $\Omega_1$=1.0002      & $P_1=6.282$\\

$S_5(0.145;0.005)$    & $A_2$=3.8305                    & $\Omega_2$=1.0001          & $P_2=6.282$\\

                        & $A_3$=4.9643                 & $\Omega_3$=1.0005    &$P_3=6.280$ \\
\hline
                        & $A_1$=2.7864                & $\Omega_1$=0.9992     &$P_1=6.288$ \\

$S_6(0.154;0.006)$     & $A_2$=3.8821                   & $\Omega_2$=1.0001          & $P_2=6.282$\\
                        & $A_3$=4.2698                & $\Omega_3$=1.0002      & $P_3=6.282$\\
\hline
\end{tabular}
{\it \caption{ \small
 Characteristics of the approximated \cite{Kadji07a} limit cycles, as per Eq.(\ref{eq:amplitude}), of the birhythmic van der Pol system, Eq.(\ref{eq6}) for the noiseless case ($D=0$) when the two stable frequencies are about equal (\emph{i.e.} $\Omega_1 \simeq \Omega_3$). All data refer to the case $\mu=0.01$.
\label{identical} }  }
\end{table*}
}

\begin{table*}
\begin{tabular}{|l|c|c|c|}
\hline
$S_i(\alpha,\beta)$& Amplitudes $A_i$ of the orbits& Frequencies $\Omega_i$ of the orbits&Periods $P_i$ of the orbits\\
\hline
                   &$A_1$= 2.491378                 & $\Omega_1$=1.00210& $P_1=6.2698$\\

$S_7(0.12;0.0014)$& $A_2$=3.52558                    & $\Omega_2$=0.99994     &$P_2=6.2834$ \\

                 & $A_3$=10.88605                     & $\Omega_3$=0.57300 &$P_3=10.9651$ \\
\hline
                        & $A_1$= 2.26969              & $\Omega_1$=0.99986 &$P_1=6.2839$ \\

$S_8(0.09;0.001)$     & $A_2$= 4.59373                & $\Omega_2$=0.99974         &$P_2=6.2847$ \\

                        & $A_3$= 10.85109            & $\Omega_3$=0.68107&$P_3=9.2252$ \\
\hline
                        & $A_1$=  2.48185           & $\Omega_1$=1.00034 &$P_1=6.2809$ \\

$S_9(0.12;0.0016)$     & $A_2$=  3.58637             & $\Omega_2$=0.99993        & $P_2=6.2845$ \\

                        & $A_3$=   10.04878           & $\Omega_3$= 0.72935&$P_3=9.2352$ \\
\hline
                        & $A_1$=  2.44836             & $\Omega_1$=0.98930 &$P_1=6.3510$ \\

$S_{10}(0.12;0.0024)$     & $A_2$=  3.88657         & $\Omega_2$=0.99989       & $P_2=6.2837$\\

                        & $A_3$=  7.67464         & $\Omega_3$=0.86703 & $P_3=7.2466$\\
\hline
\end{tabular}

{\it \caption{ \small
 Characteristics of the approximated \cite{Kadji07a} limit cycles, as per Eq.(\ref{eq:amplitude}), of the birhythmic van der Pol system, Eq.(\ref{eq6}) for the noiseless case ($D=0$) when the two stable frequencies are different, (\emph{i.e.} $\Omega_1 \ne \Omega_3$). All data refer to the case $\mu=0.01$.
\label{different}}  }
\end{table*}

The analysis of the noisy dynamical flow involves a study in  terms of a two-parameter space ($D$,$\tau$).
Accurate approximation schemes for colored noise are only valid in the asymptotic limits of one the both  parameters $D$ and/or $\tau$, as discussed in Ref.  \cite{Hanggi95}.

\subsection{Birhythmic properties}

The properties of the dynamical attractors of the modified  van der Pol model (\ref{eq4}) can be analyzed as follows. One first assumes that the periodic solutions of Eq.~(\ref{eq4}) are represented by  \cite{Kadji07a}
\begin{eqnarray}
\label{eq:amplitude}
x(t)=A_i\cos \Omega_i t.
 \end{eqnarray}
The  amplitudes $A_i$ and the frequencies  $\Omega_i $
can be analytically approximated \cite{Kadji07a}; in this scheme the amplitude $A_i$ is independent of the
coefficient $\mu$, that only enters in the frequency $\Omega_i$ of the orbits.
The number of cycles of the modified van der Pol
system Eq. (\ref{eq4}) is determined by the parameters $\alpha$ and $\beta$.
It is found that in some region of the parameter space
one obtains three limit cycles (two of them are stable, and one is unstable).
For each of the two stable limit cycles it corresponds to a different frequency, and hence the system is birhythmic.
The parameter plane  ($\beta$-$\alpha$)  is represented in Fig.\ref{fig3}, where a shadowed area denotes the portion of  the  region where birhythmicity occurs \cite{Kadji07a,Kadji07b}.
At the border of the shadowed area one observes the passage from a single limit cycle to three limit cycles.
The limit cycles exhibit very similar frequencies in the region of  low $\alpha$  (Table \ref{identical}) and clearly different frequencies when $\alpha$ increases (Table \ref{different}).
By way of conclusion of this part, we have described the equation of motion of the system of Fig. \ref{fig1} when the current is truncated to the fifth order. The system is known to be birhythmic, and the main properties, approximating the solution as in Eq.(\ref{eq:amplitude}) are summarized in Tables \ref{identical} and \ref{different}.

\section{Analytic estimates}
\label{analysis}
In this Section we derive the properties of the attractors described in Sect. \ref{sec:model} when subject to correlated noise.

\subsection{Residence times of the attractors}

The evolution of the parameter $x(t)$ is subjected to the influence of random forces, which permanently tend to destabilize it.
Random fluctuations, in the case of three attractors (two stable and one unstable) can induce a transition between the two stable attractors \cite{Yamapi10}.
The ubiquitous problem of noise induced escape from a metastable state is characterized by a rate; when the phenomenon occurs  on a long time scale compared to dynamic time scales of the localized solution it assumes the features of an Arrhenius like behavior for a dynamical attractor \cite{Graham85,Kautz88}, and in particular in van der Pol birhythmic systems \cite{Yamapi10,Yamapi12}.

\begin{figure}[htbp]
\begin{center}
(a)\\
\includegraphics[scale=0.5]{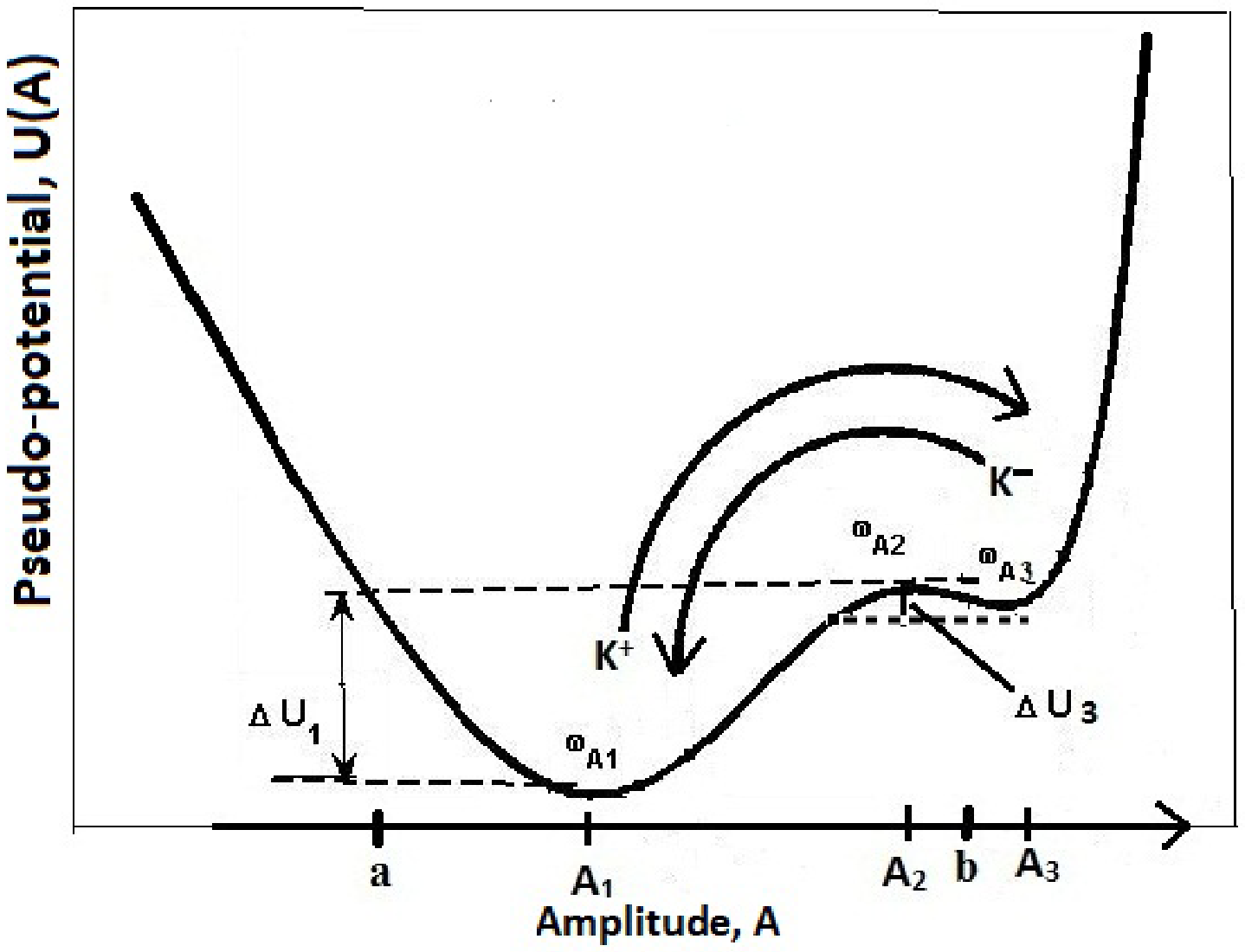}\\
(b) \\
\includegraphics[scale=0.5]{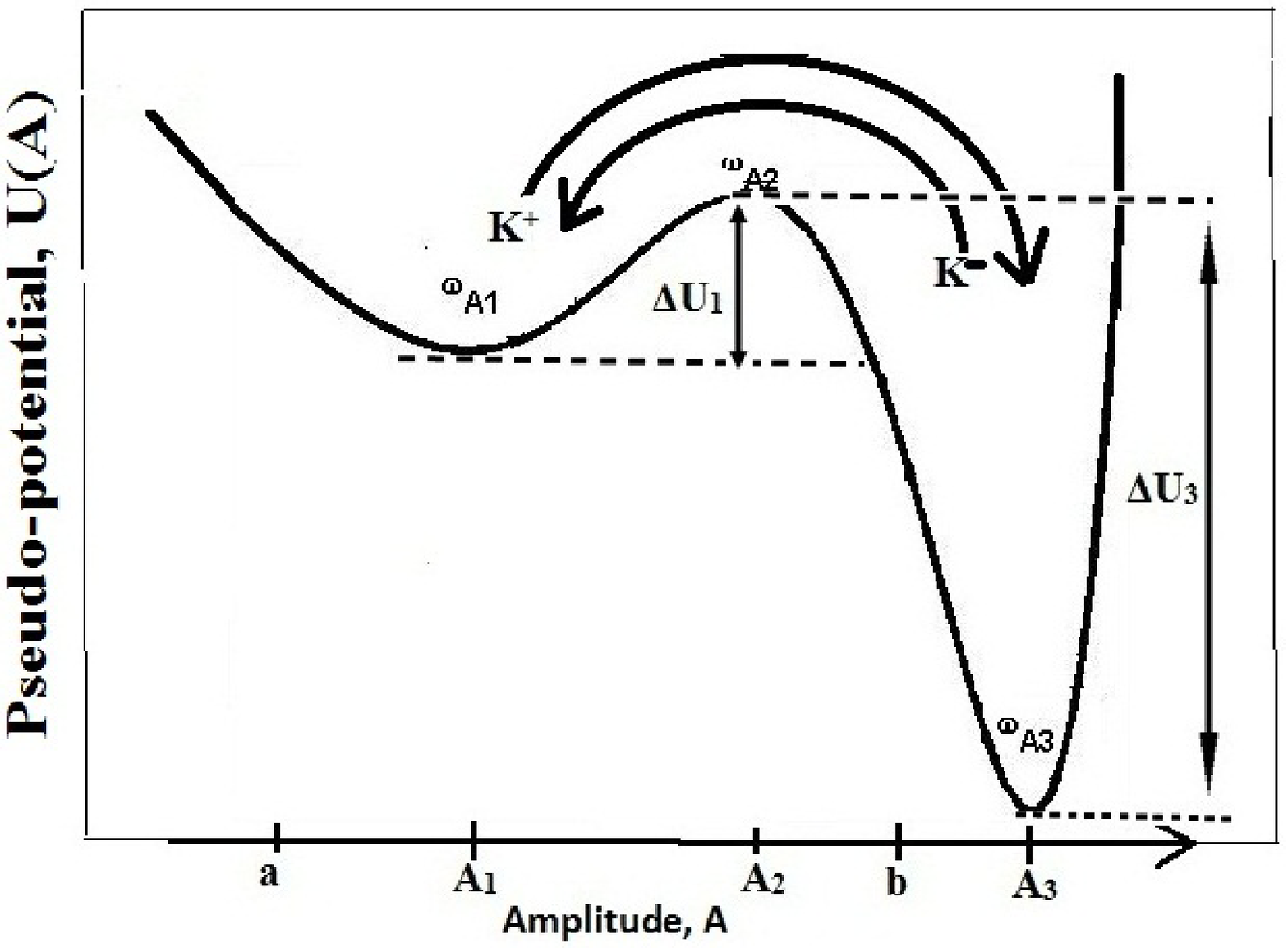}
\caption{\it
Schematic illustration of the escape process in an asymmetric  metastable potential.
Escapes occur via the forward rate $K^{+}$ and the backward rate $K^{-}$,
respectively, while $\Delta U_{1,3}$ are the corresponding activation energies.}
\label{fig4}
\end{center}
\end{figure}

Fig. \ref{fig4} is an illustration of the escape process from a state of local stability via noise-assisted hopping.
The stable states are characterized by an asymmetric double well pseudo-potential $U(A)$  with angular frequency of the metastable state $\omega_{A_{1,3}}$ and positive-valued angular
frequency of the unstable state at the barrier $\omega_{A_{2}}$ defined through the relation
 $\omega_{A_{1,3}}^{2}=M_0^{-1}U''(A_{1,3})$ and $\omega_{A_{2}}^{2}=M_0^{-1}\mid U''(A_{2})\mid$.
Thus, considering the particle of mass $M_0$ moving in a double well quasi-potential, the rate coefficients $K^{+}$ and $K^{-}$ are the forward and backward rates, respectively:
\begin{equation}
\label{eq21}
    K^{\pm}=\nu_{\pm} \exp \left(\frac{- \Delta U_{\pm}}{D}\right)
\end{equation}
where $\Delta U_\pm $ denote the threshold energies for activation and $\nu_\pm $ are the corresponding pre-factors.

The inverse of the Kramers escape rate is the escape time from one well to another.
The two times might be very different \cite{Stambaugh06,Chamgoue12}.
To measure the different properties, we compute the average persistence or residence time $T_{1,3}$ on the attractor with limit cycle amplitude $A_{1,3}$ as:
\begin{subequations}
\label{eq27}
\begin{eqnarray}
T_{1}&=&T(A_{1}\rightarrow A_{3})=\frac{1/K^{+}}{1/K^{+}+1/K^{-}}, \\
T_{3}&=&T(A_{3}\rightarrow A_{1})=\frac{1/K^{-}}{1/K^{+}+1/K^{-}}.
\end{eqnarray}
\end{subequations}
We wish to emphasize here the special character of both the coordinate $A$ and of the quasipotential $U${\hg , see Fig. \ref{fig4}}.
In fact $A$ is the amplitude of the oscillations approximated by Eq. (\ref{eq:amplitude}), and therefore it  represents the amplitude of an approximated orbit,  rather that a single point. Also, the quasipotential  $U$ is a Lyapunov function that characterizes the asymptotic behavior of the nonequilibrium state  \cite{Graham85}{\hg, not a bona fide potential. Therefore Fig. \ref{fig4} is, at this stage, a pictorial description of the process. However, we will show in the next subsection that a function playing the role of an activation energy (i.e., determining the rate of rare escapes in the low noise intensity limit), can be analytically retrieved.}

\subsection{Stationary probability distribution}

In the quasi-harmonic regime, on the assumption that the noise intensity is small, it is convenient to use a change of variables suggested by the approximation Eq.(\ref{eq:amplitude}), i.e. $x=A\cos \theta$ and $\theta = \varphi + t$ (we assume $\Omega = 1$).
The instantaneous amplitude $A(t)$ and phase $\varphi(t)$ are given by the following effective  Langevin equation \cite{Mbakob14,Xu11,Chamgoue12,Chamgoue13}:
\begin{widetext}

\begin{figure}
\begin{center}
\includegraphics[scale=0.5]{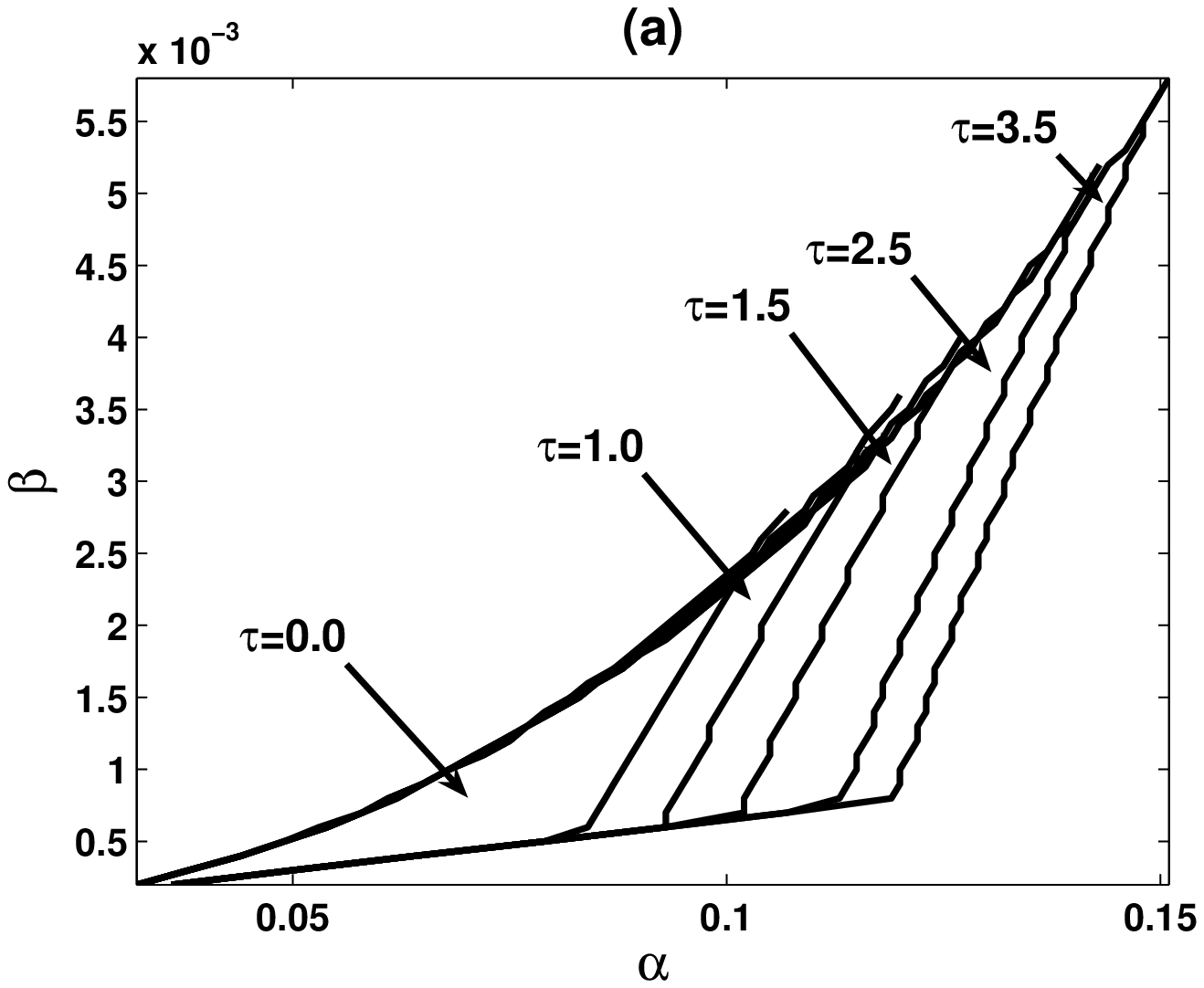}
\includegraphics[scale=0.5]{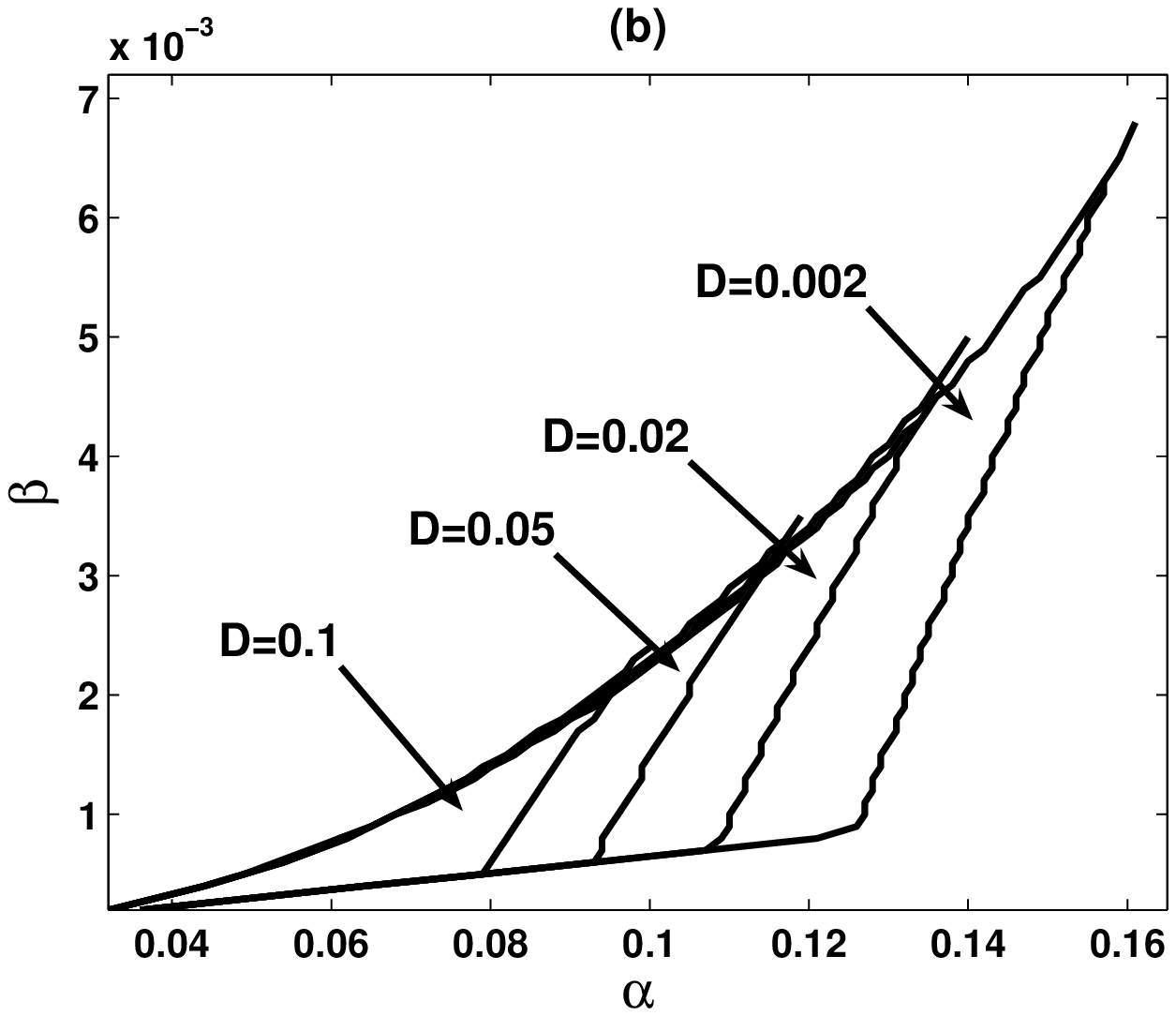}
\caption{\it
Effect of the noise and of the correlation time $\tau$ on the boundary between the region
of one and three limit-cycle solutions in the parametric $(\alpha,\beta)$-plane for the stationary probability current governed by Eq.(\ref{eq27}) as per the analytic solution Eq. (\ref{eq32}).
(a) The effect of correlation time $\tau$ for $D=0.1$;
(b) The effect of noise intensity $D$ for correlation time $\tau=0.1$.}
\label{fig5}
\end{center}
\end{figure}

\begin{subequations}
\begin{eqnarray}
 \label{eq29}
 \frac{dA}{dt}&=&
\frac{\mu A}{128}\left[ 64-16A^{2}+8\alpha A^{4}-5\beta A^{6}\right] +
 \frac{D}{2A(1+\tau^{2})} + \xi_A (t) =
   -\frac{dU}{dA}+ \xi_A (t) \\
\label{A_Langevin}
\frac{d\varphi}{dt}&=&
\xi_\varphi (t),
\label{phi_Langevin}
   \end{eqnarray}
   \end{subequations}

\noindent here $\xi_A (t)$ and $\xi_\varphi (t)$ represent  the noise sources, and $U$ is an effective potential or quasipotential:

\begin{equation}
U= -\frac{\mu A}{128}\left[ 32A^2-4A^{4}+
\frac{4}{3}\alpha A^{6}-\frac{5}{8}\beta A^{8}\right] - \frac{D}{2(1+\tau^{2})} \ln A.
\label{potential}
\end{equation}
\begin{figure}[htbp]
\begin{center}
\includegraphics[scale=0.4]{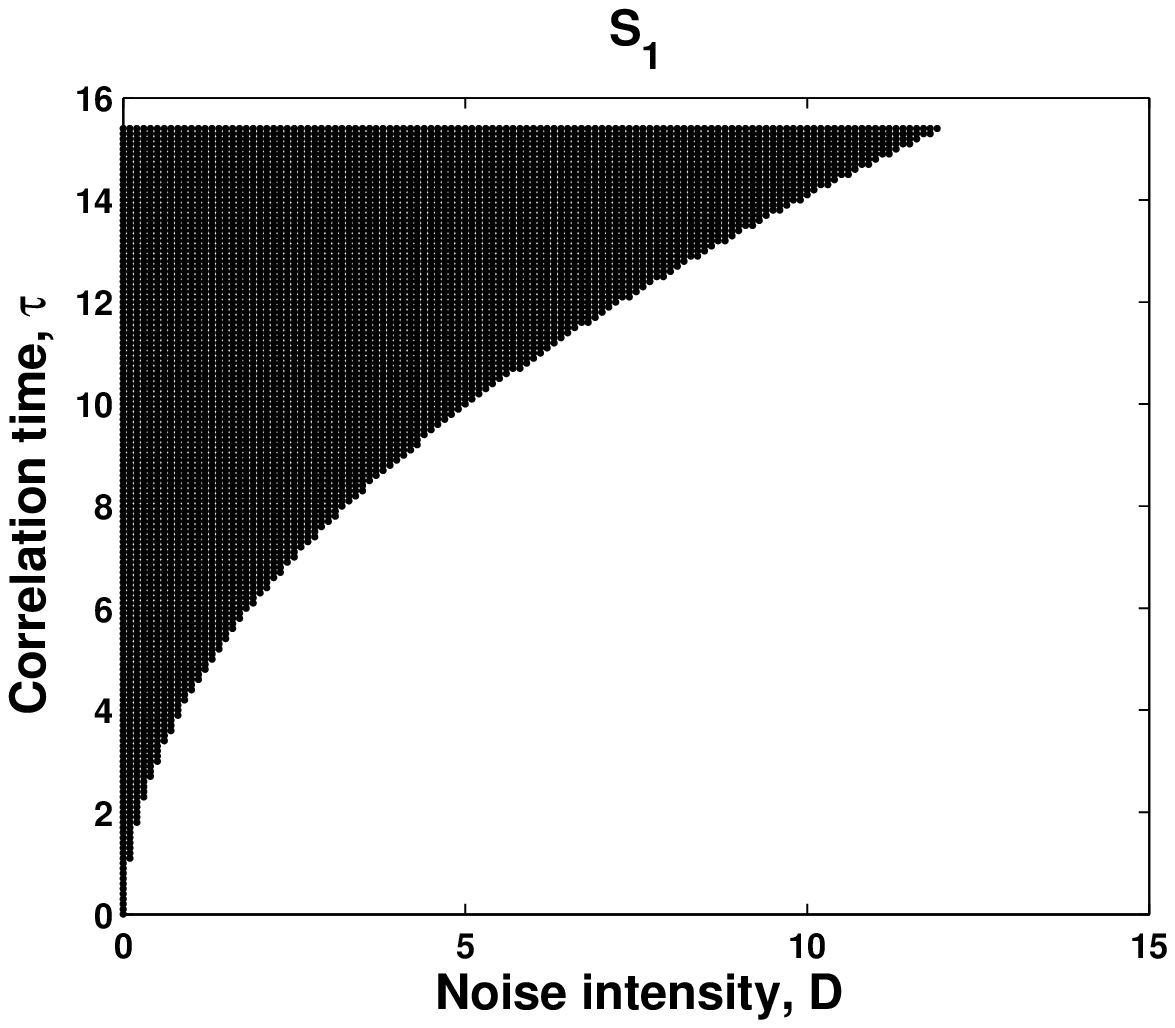}
\includegraphics[scale=0.4]{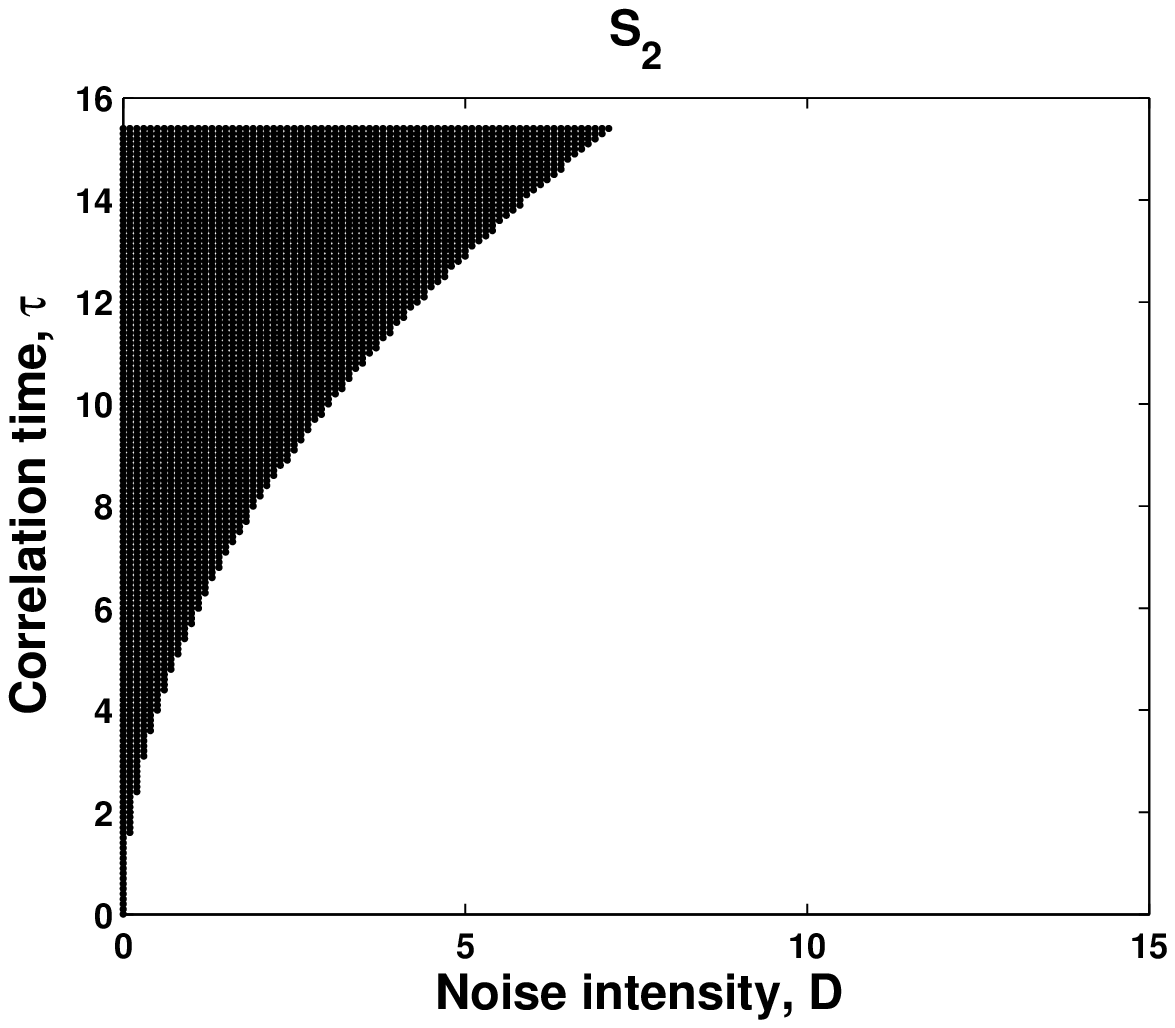}\\
\includegraphics[scale=0.4]{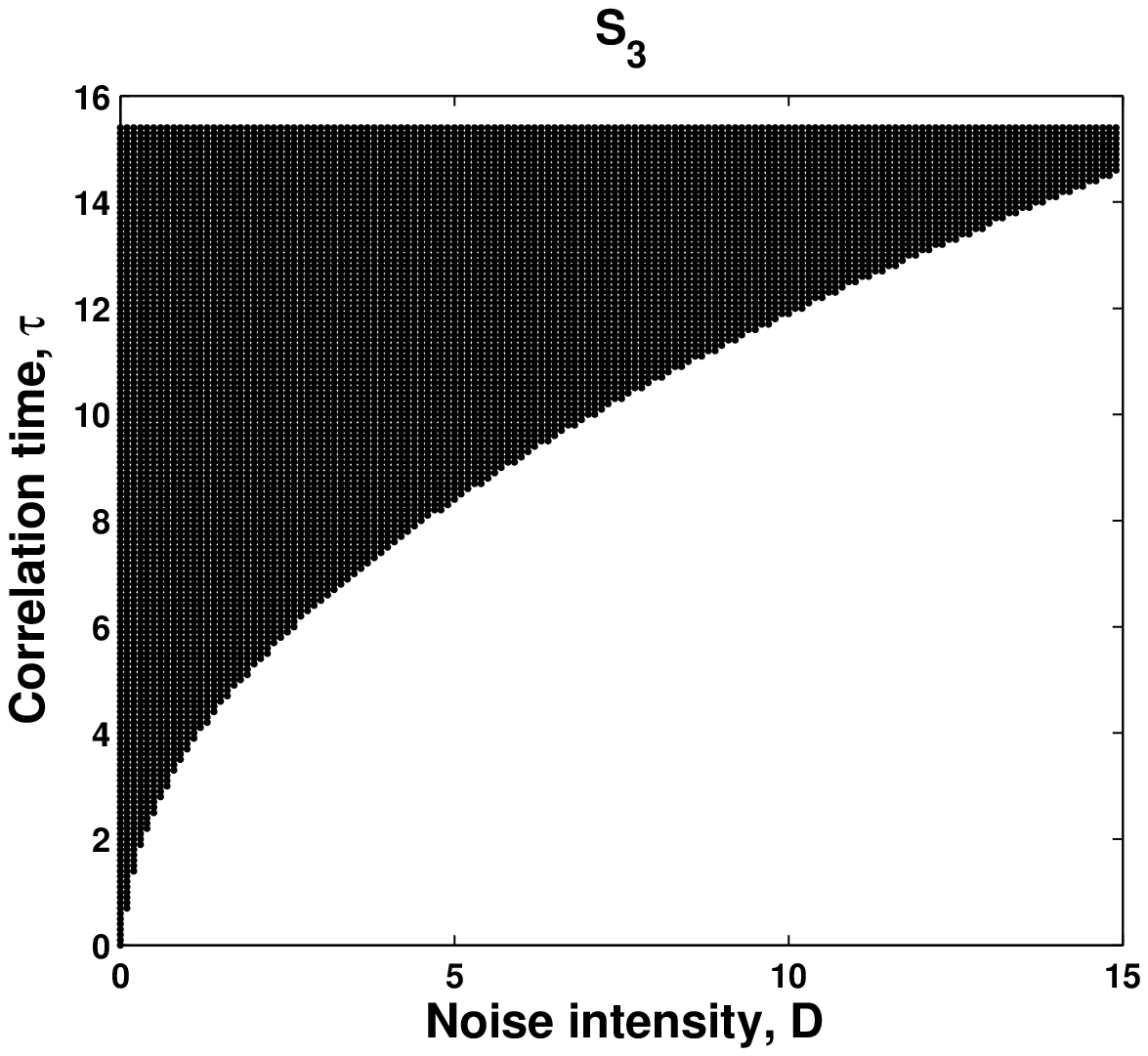}
\includegraphics[scale=0.4]{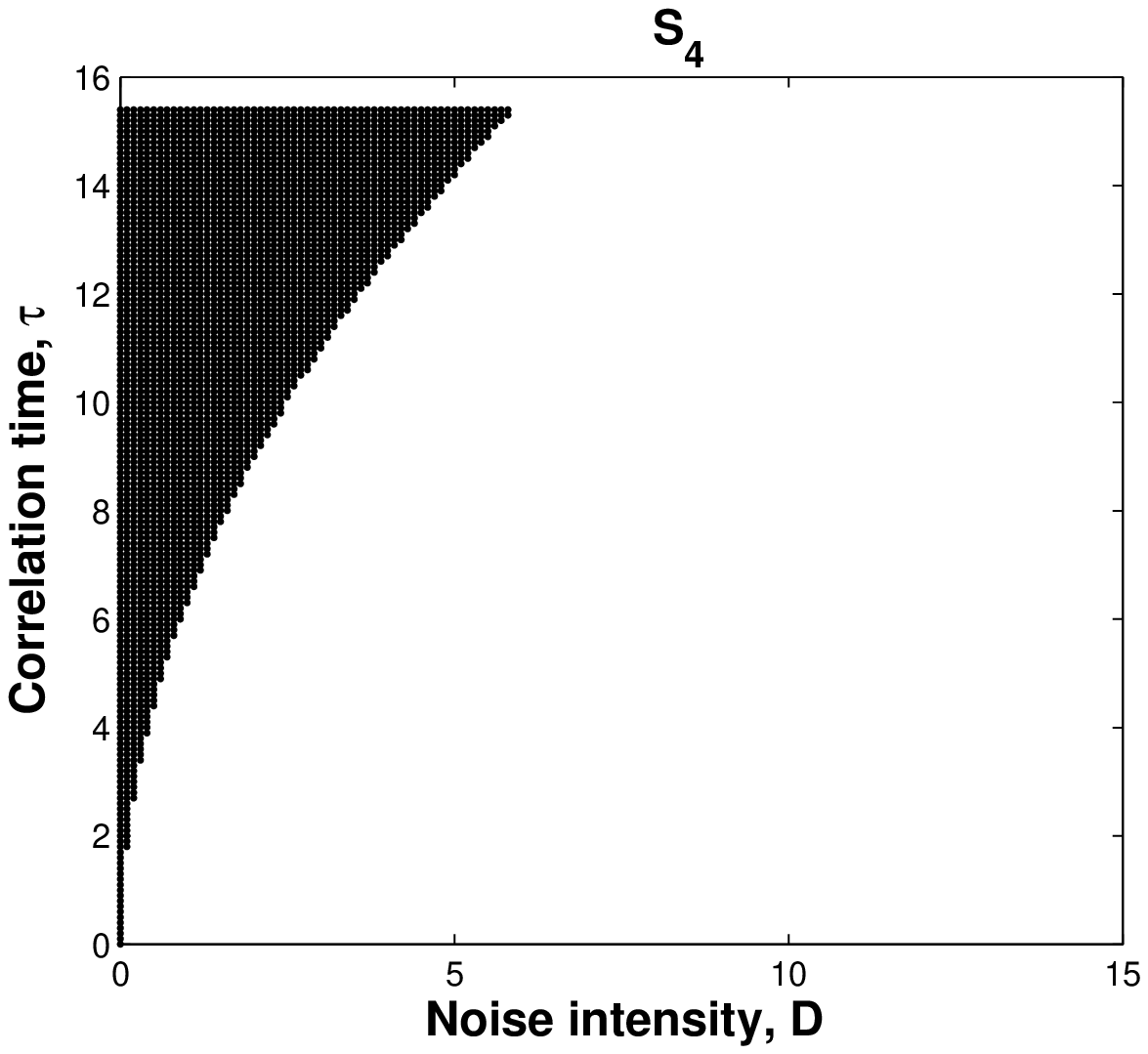}\\
\includegraphics[scale=0.4]{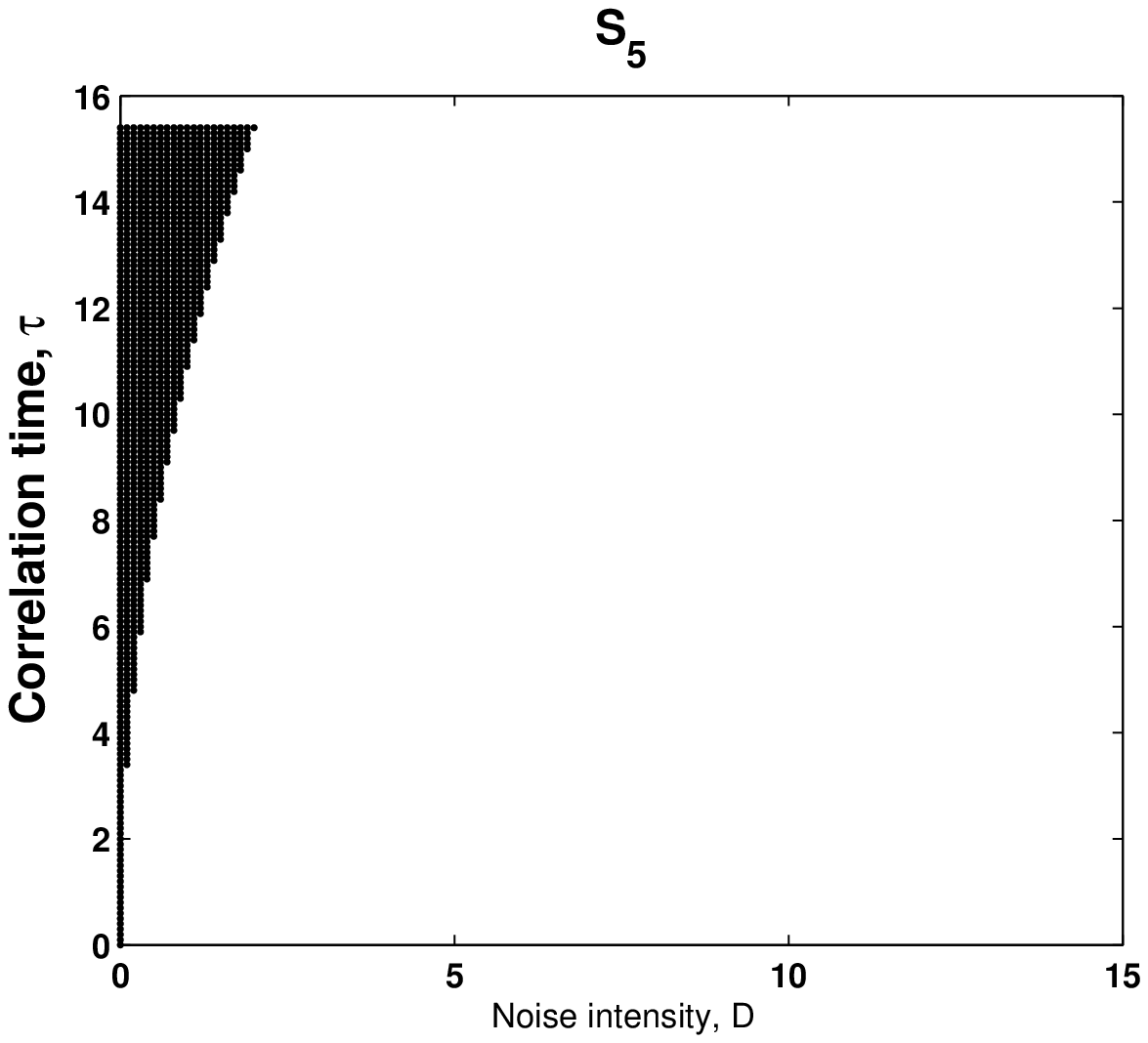}
\includegraphics[scale=0.4]{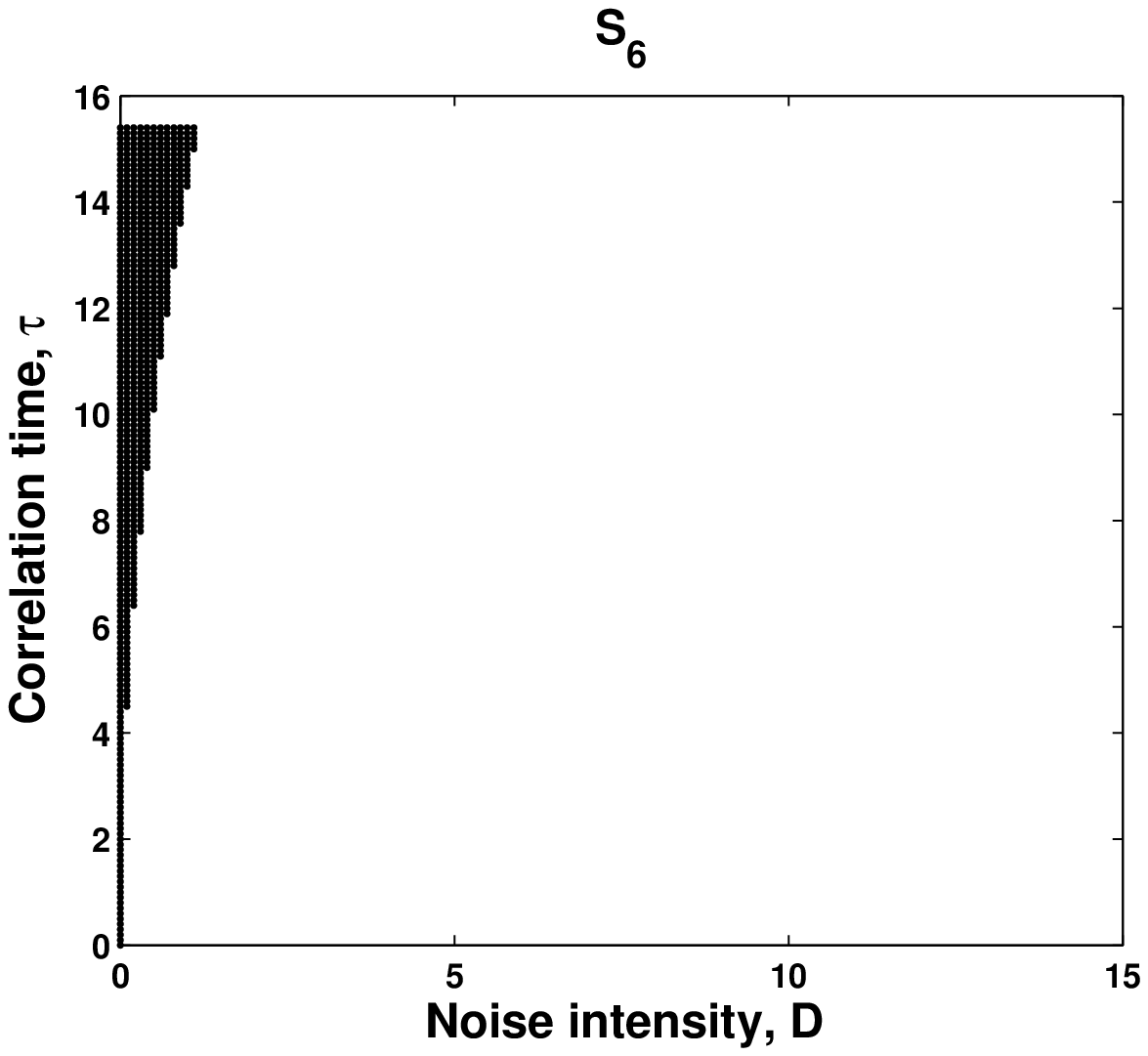}\\
\caption{\it
Stochastic P-Bifurcation in the parameter plane $(D,\tau)$ in the case where
the two stable frequencies are equal, $\Omega_1=\Omega_3$. The dotted area represents the parameter region where two sable orbits coexist. Parameters of the system are  $\mu=0.01$  and $S_i(i=1,2,3,4,5,6)$}
\label{fig6}
\end{center}
\end{figure}
\end{widetext}

Contrary to the case of uncorrelated sources, a rigorous derivation of the noise characteristics
 cannot be obtained, for the additional variable $\eta$ introduces an extra dimension.
However, $dA$ does not depend on  $ \varphi$, thus we assume that we can develop a probability
 density for $A$, rather than a joint density for A and $ \varphi$.
For $D \neq 0$, we also assume that Eq.(\ref{A_Langevin})  amounts to an equation for the coordinate $A$ driven by  correlated noise and we ignore the fact that it represents  an orbit rather than a single point.

Consequently, we treat the effect of noise as in the case of a standard double well potential associated to a driven oscillator \cite{Jung93}.
In conclusion, the probability density function $p(A,t |A_{0},t_{0})$ of the averaged amplitude equation is assumed to be governed by the condition that the probability flux $S_A$ vanishes, i.e.:

\begin{widetext}
\begin{equation}
p(A) : S_A = 0 \rightarrow D_1^A p(A) - \frac{\partial}{\partial A} \left( D_2^{A,A}  p(A) \right) = 0,
\label{prob_flux}
\end{equation}
 where
 \begin{subequations}
\begin{eqnarray}
 \label{eq30a}
D_1^A &=& \frac{\mu A}{128}\left[ 64-16A^{2}+8\alpha A^{4}-5\beta A^{6}\right] +\frac{D}{2A(1+\tau^{2})}
\\
\label{eq30b}
D_2^{A,A} &=&  \frac{D}{2(1+\tau^{2})}
\end{eqnarray}
\end{subequations}

Hence, the stationary solution is:

\begin{equation}
\label{eq32}
p(A)=N A \exp\left[\mu \frac{1+\tau^{2}}{512D}\left( 256A^{2}-32A^{4}+\frac{32}{3}\alpha A^{6}-5 \beta A^{8}\right) \right],
\end{equation}
where

\begin{figure}[htbp]
\begin{center}
\includegraphics[height=4.3cm,width=5.5cm]{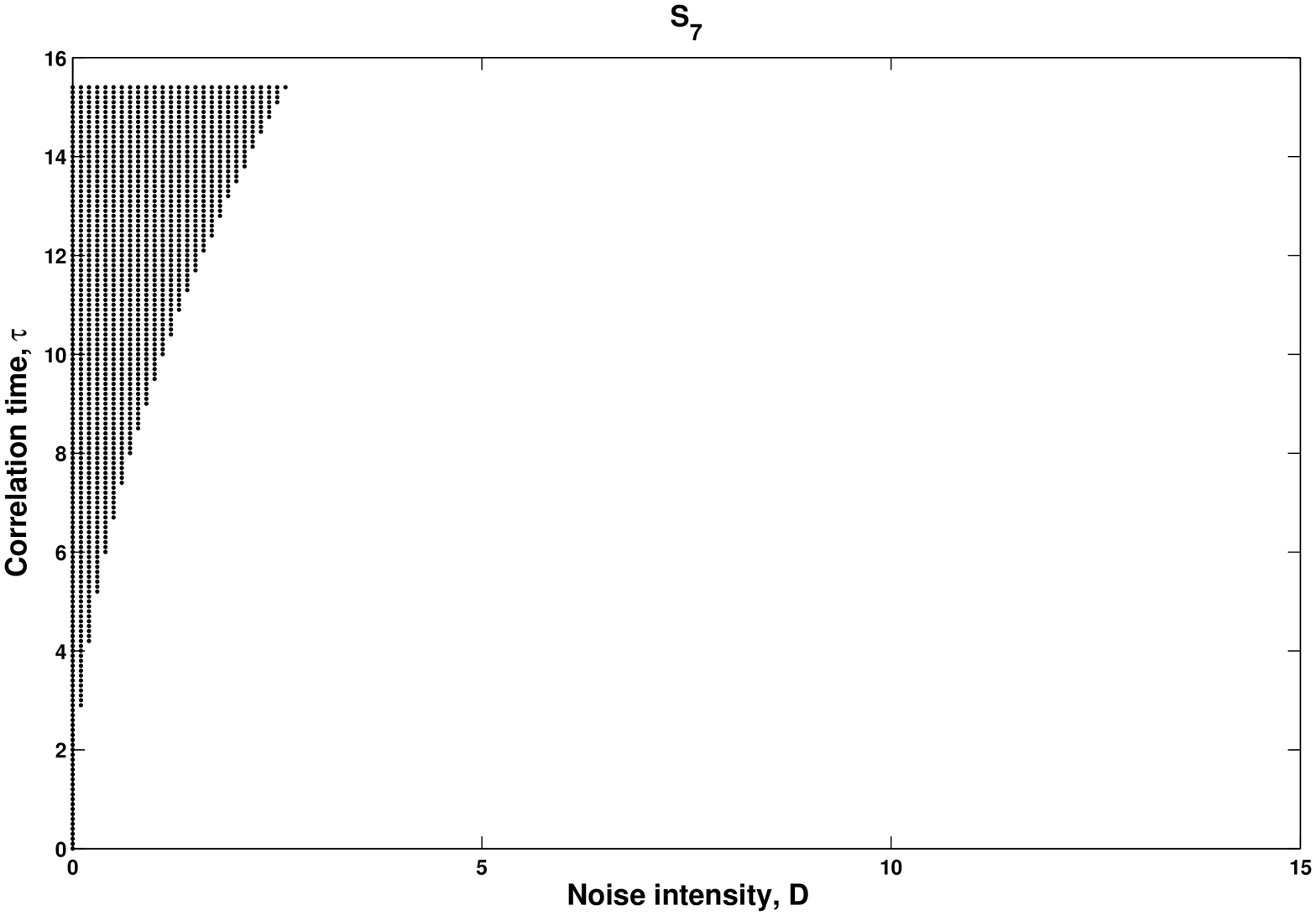}
\includegraphics[scale=0.4]{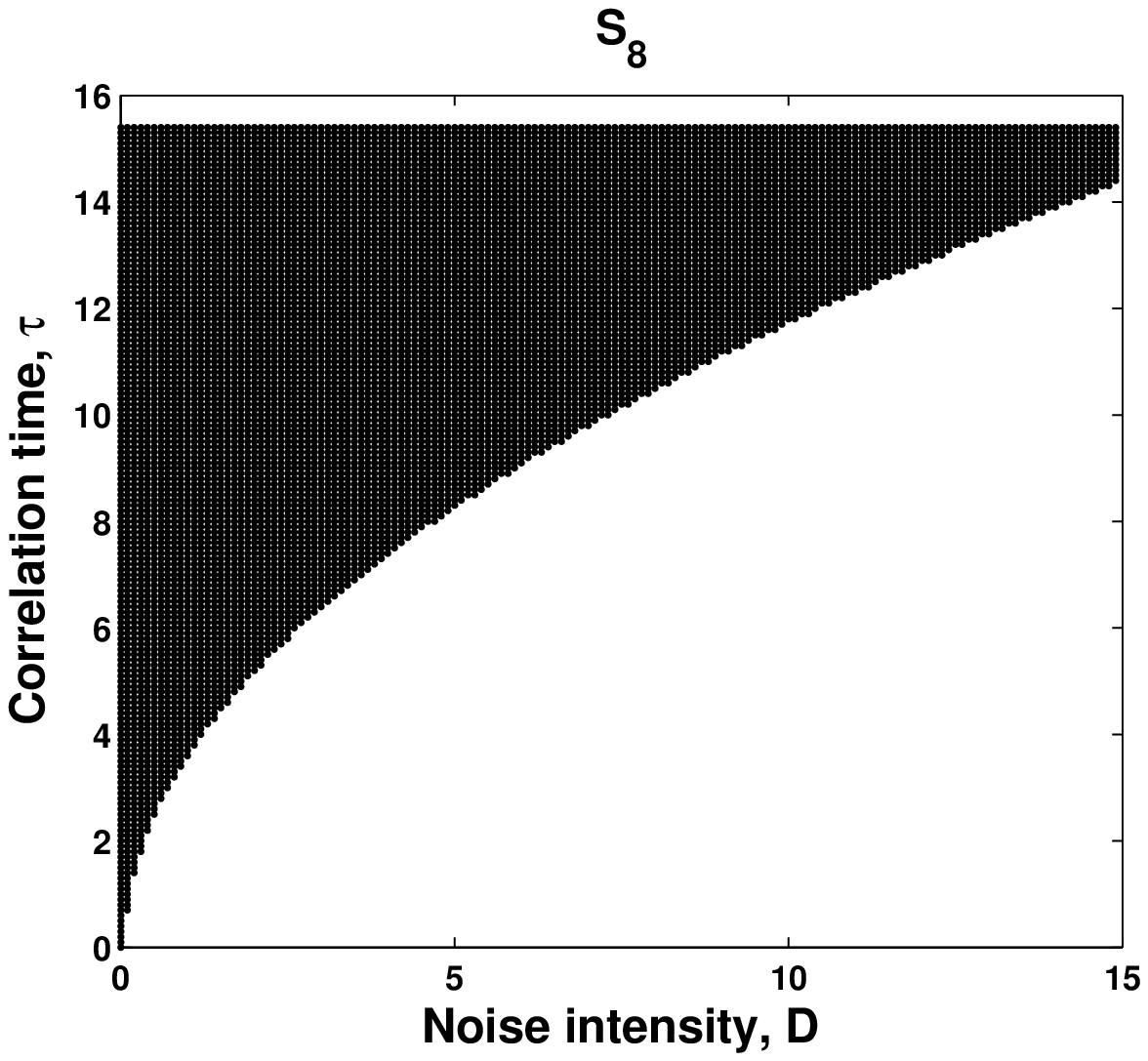}\\
\includegraphics[scale=0.4]{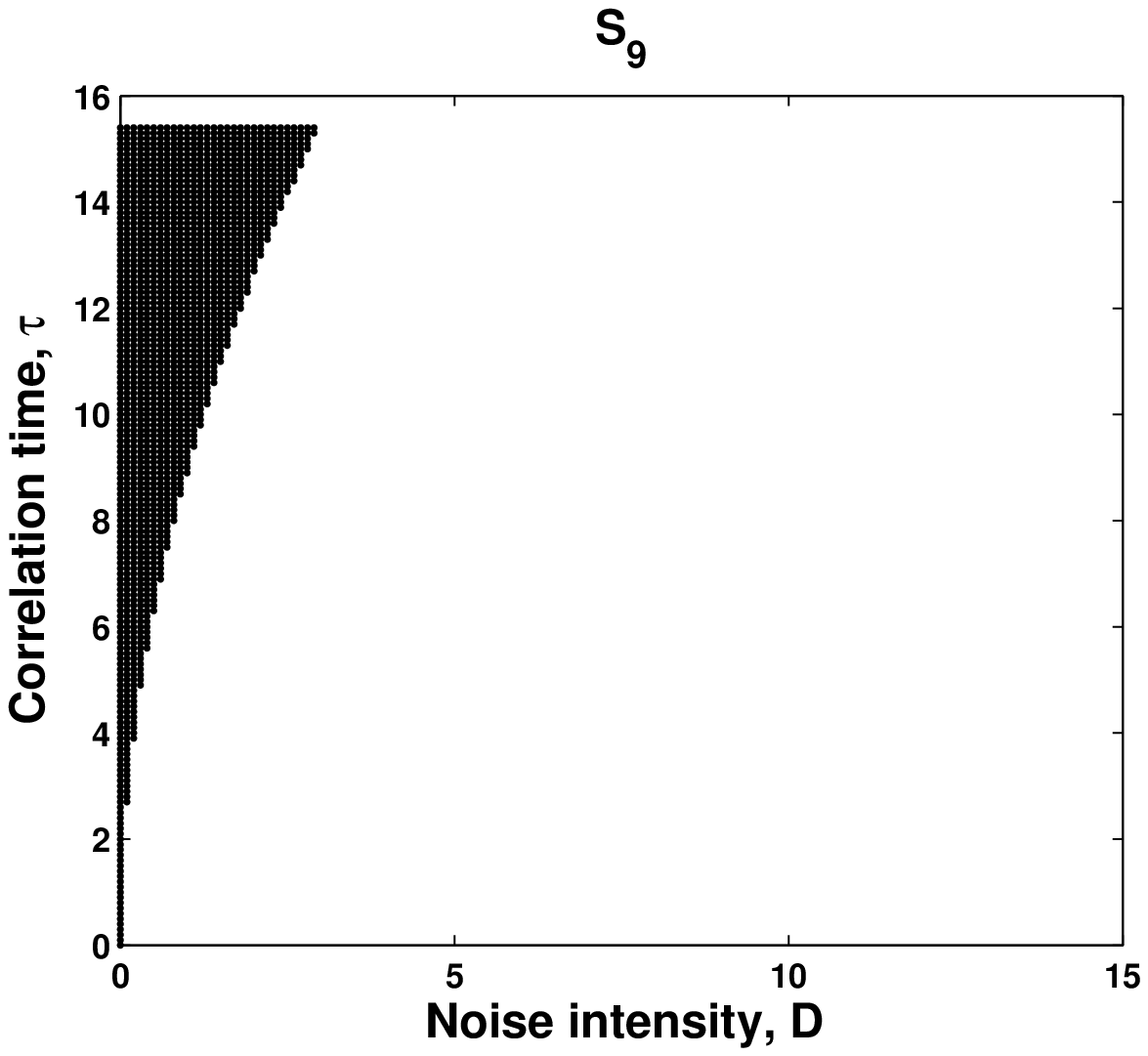}
\includegraphics[scale=0.4]{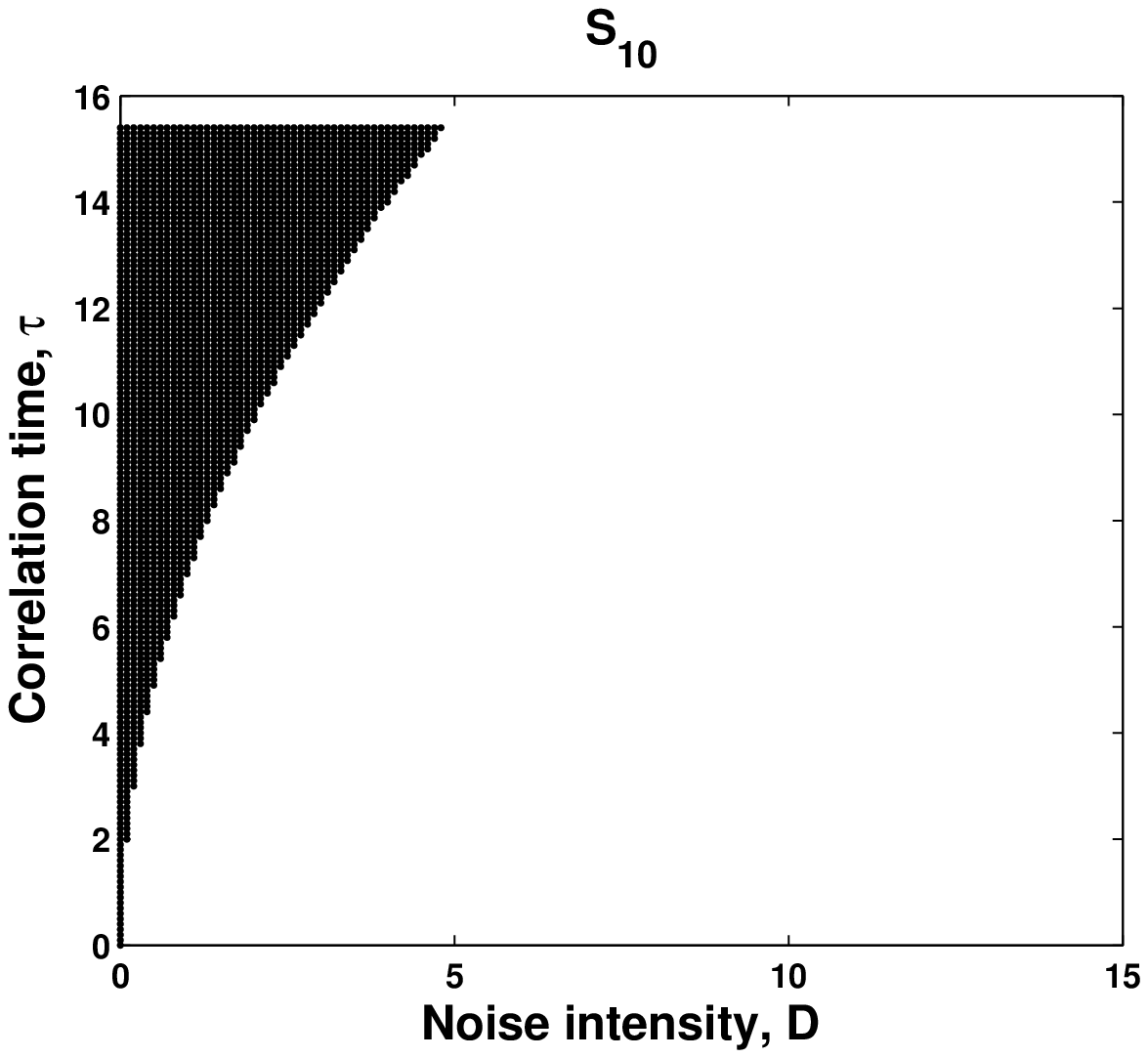}
\caption{\it
Stochastic P-Bifurcation in the parameter plane $(D,\tau)$ in the case where
the two stable frequencies are different, $\Omega_1\neq \Omega_3$. The dotted area represents the parameter region where two sable orbits coexist. Parameters of the system are  $\mu=0.01$  and $S_i(i=7,8,9,10)$}
\label{fig6a}
\end{center}
\end{figure}

\begin{equation}
\label{eq33}
N= \left\{  \int^{+\infty}_{0}  A \exp\left[\mu \frac{1+\tau^{2}}{512D}\left( 256A^{2}-32A^{4}+\frac{32}{3}
\alpha A^{6}-5 \beta A^{8}\right) \right]  \right\}^{-1}
\end{equation}
is the normalization constant.
\end{widetext}

Equations (\ref{eq32},\ref{eq33})  show that in the
limit $\tau \rightarrow 0$ the colored noise $\eta(t)$ tends to a white noise \cite{Hanggi95}. It is important to note that the transitions between the unimodal and the bimodal stationary probability densities are also referred to as the noise induced transitions, and stochastic bifurcation discussed here is closely connected to the noise induced transition \cite{Chamgoue13}.
The probability distribution is in general very asymmetric, therefore for most the parameters  $\alpha$ or $\beta$ one can localize the probability function around a  single orbit.
The peaks of the probability distribution can be located deriving the logarithm of the probability distribution, $
\partial \left[ \ln p(A) \right] / \partial A=0.$
Thus, to determine the extrema of the distribution Eq.~(\ref{eq32}) amounts to determine the roots of the equation:
\begin{equation}
\label{eq35}
1-\frac{A^{2}_{k}}{4}+\alpha\frac{A^{4}_{k}}{8}-\frac{5\beta A^{6}_{k}}{64}+\frac{D}{\mu A^{2}_{k}(1+\tau^{2})}=0
\end{equation}
For $D=0$, the amplitude equation
(\ref{eq35}) coincides with the deterministic amplitude equation \cite{Chamgoue12}.
Generally this equation can admit three  positives roots \cite{Chamgoue12,Chamgoue13}
varying the parameters $\alpha$, $\beta$, $\mu$, and $D$ the number of the  real roots of Eq.(\ref{eq35}) change.
The effects of the parameters in the system Eq.~(\ref{eq32},\ref{eq33})  can be seen as a type of a stochastic P-bifurcation  \cite{Arnold03}, \emph{i.e} a sudden change of the probability density function.

In Fig. \ref{fig5}, we report the influence of the correlation time $\tau $ and the noise intensity $D$ on the birhythmic behavior; in particular we show  in Fig. \ref{fig5}(a) the effect of the correlation  time for fixed $D=0.1$, and in Fig. \ref{fig5}(b)  the effect of noise intensity for fixed correlation   time ($\tau=0.1$) as predicted by the analysis of   Eq.(\ref{eq35}).
From the Figure, it is clear the  opposite effect of the noise intensity $D$ and of    the correlation time $\tau$, as the region of  birhythmicity increases when $\tau $ increases,   and decreases when $D$ increases.

 In the same vein to determine the effects of such noise, we represent in Figs. \ref{fig6},\ref{fig6a}   the bifurcation diagram  for the system parameter plane $(D, \tau)$ respectively in the case where    the two stable frequencies are equal  ($\Omega_1=\Omega_3$) and different $(\Omega_1\neq \Omega_3)$.
The domain of the existence of three limit cycles decreases  when the correlated time $\tau$ and the noise intensity $D$ increase,  and disappears when these parameters become large enough.
One finds the same phenomena for all the set of parameters $S_i$.

\begin{figure}[htbp]
\begin{center}
\includegraphics[height=4cm,width=5.5cm]{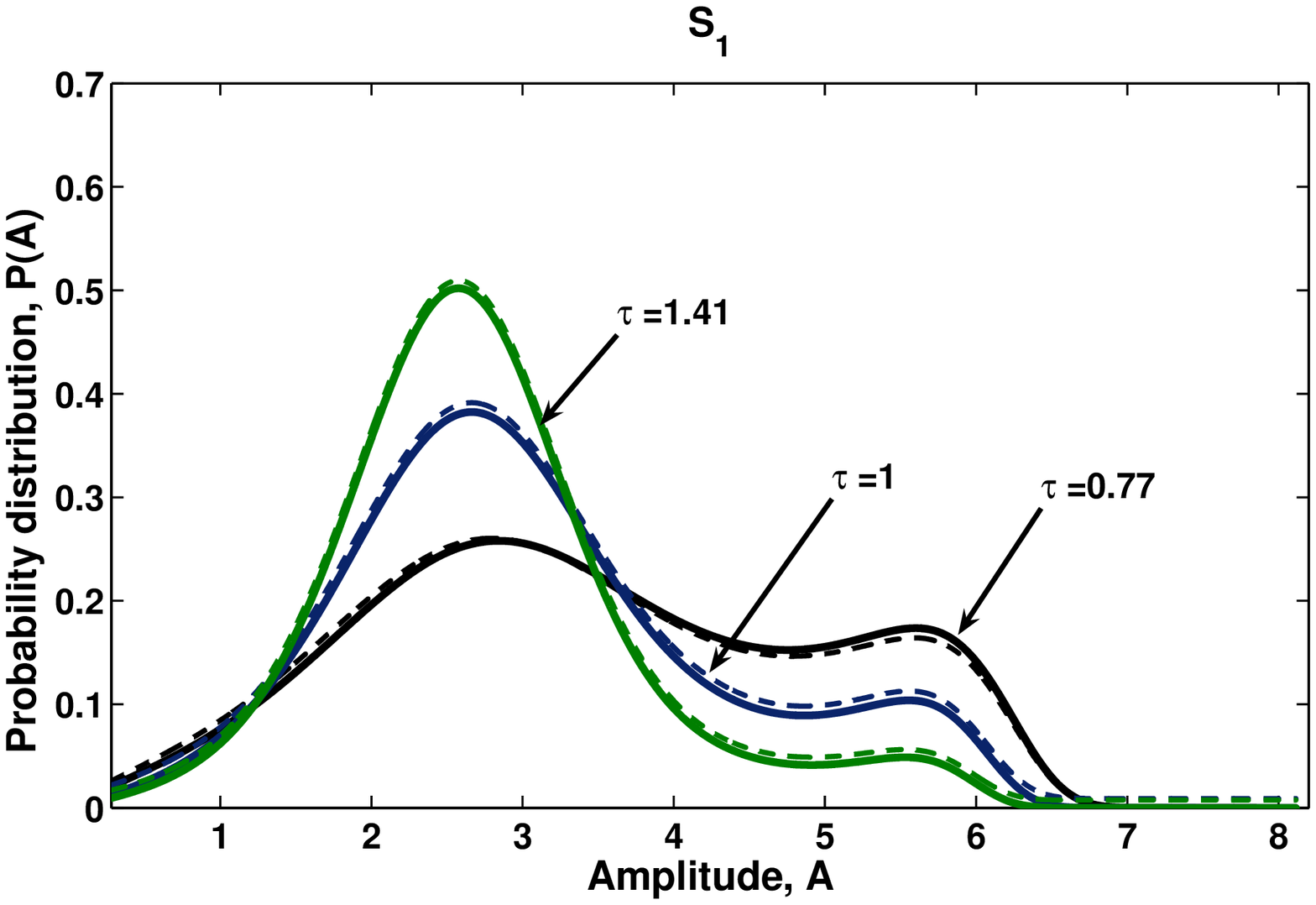}
\includegraphics[height=4cm,width=6.5cm]{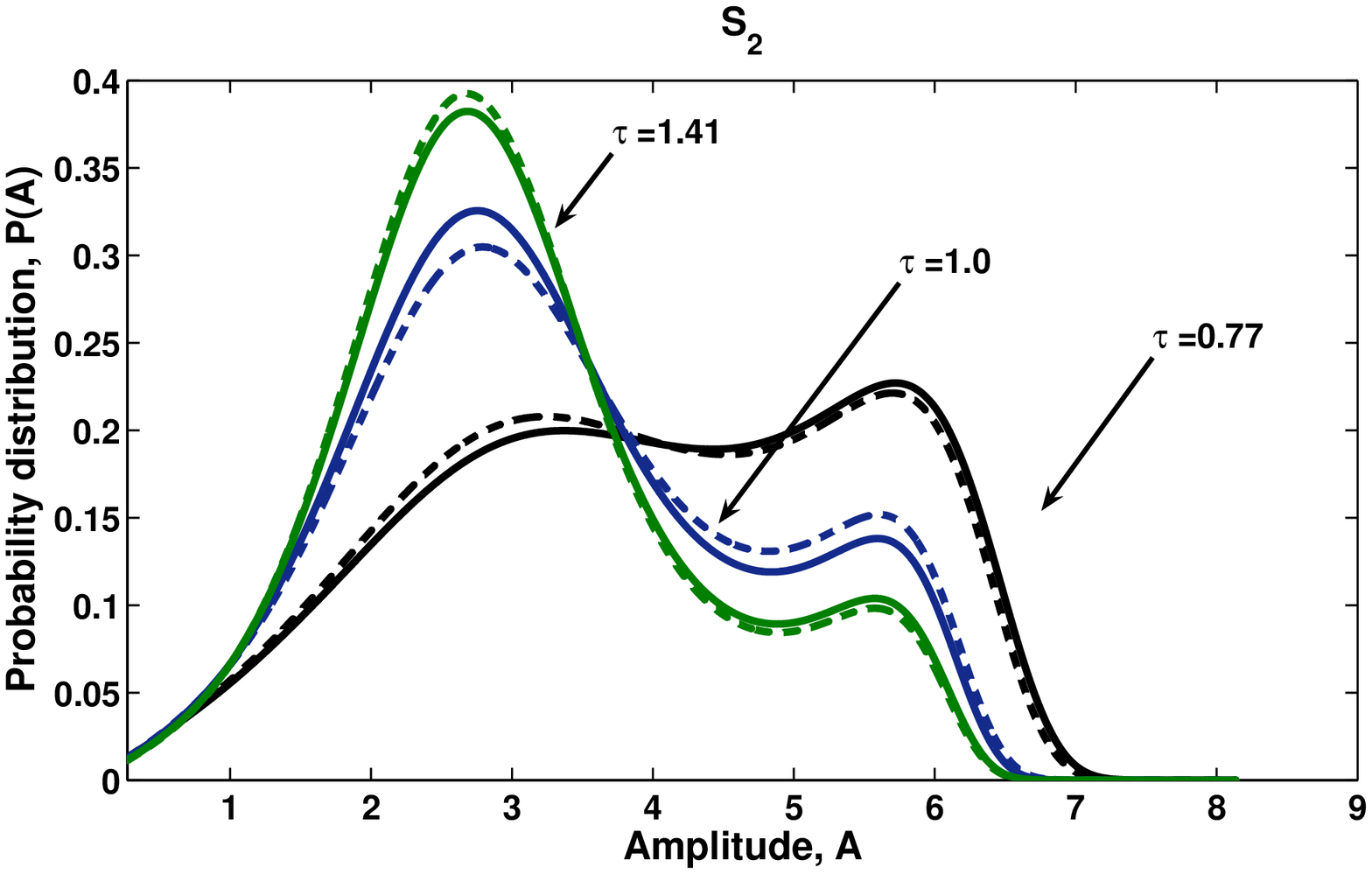}\\
\hspace{-0.5cm}\includegraphics[height=3.6cm,width=6cm]{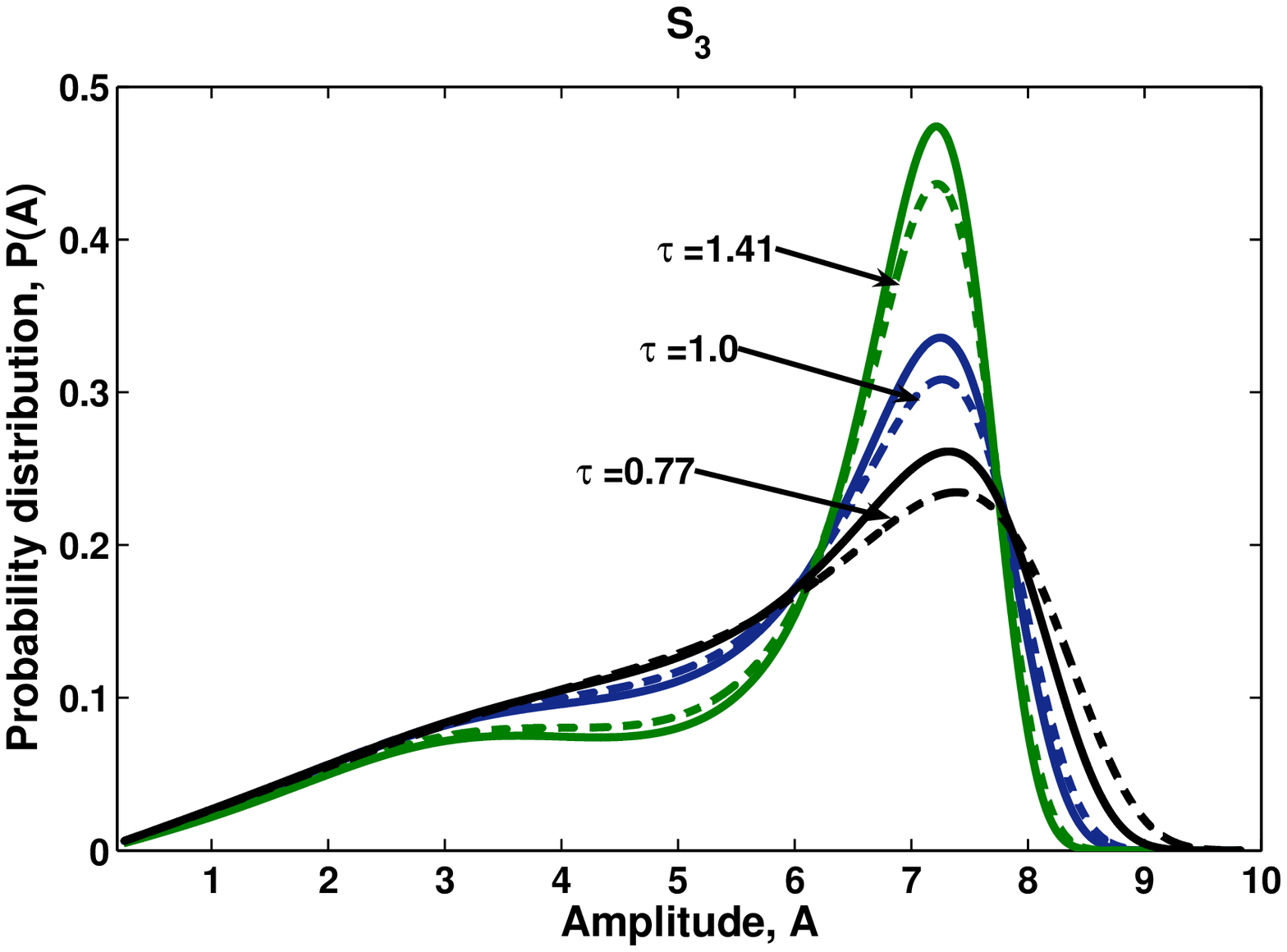}\hspace{0.3cm}
\includegraphics[height=3.7cm,width=5.5cm]{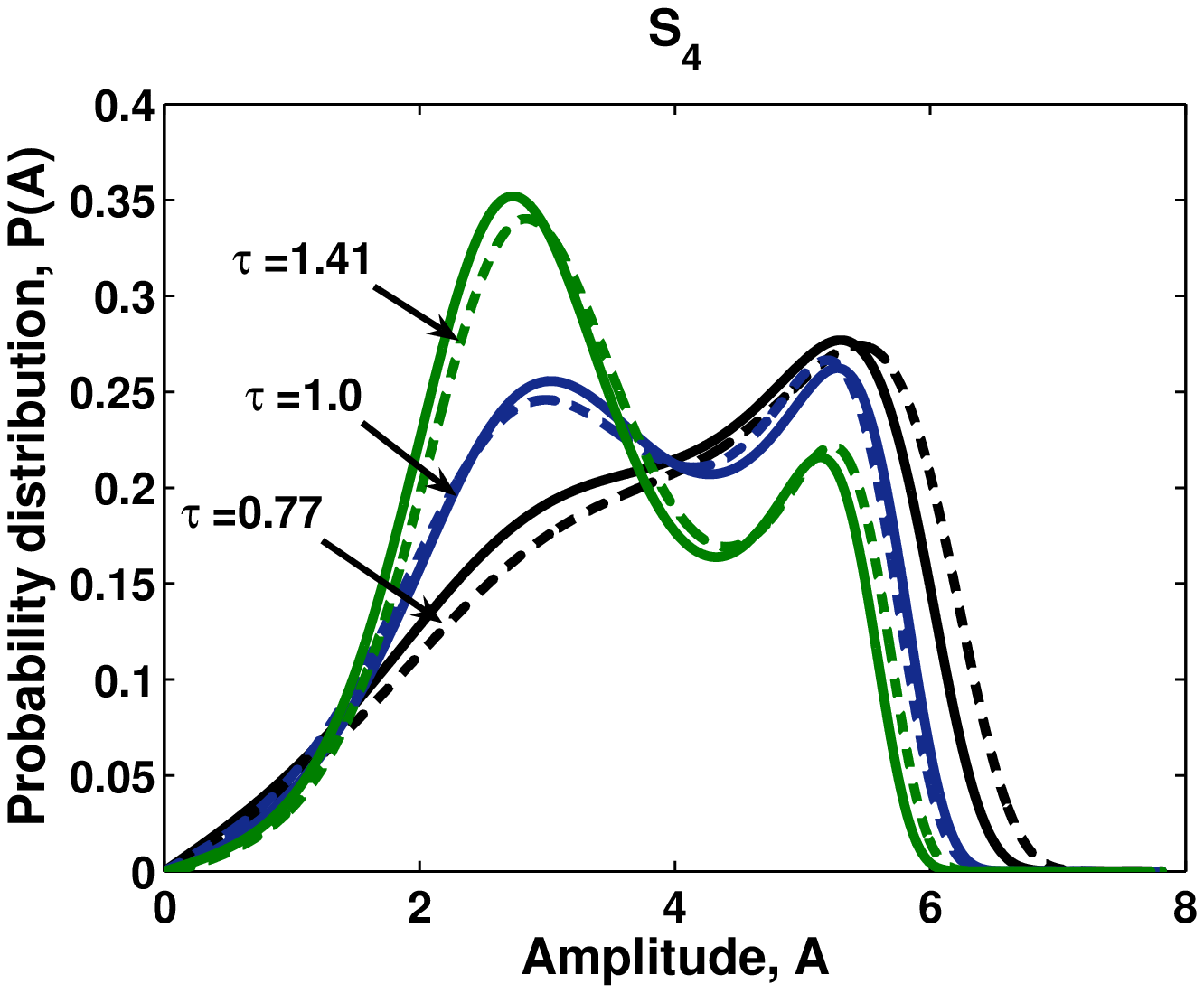}\\
\hspace{-0.5cm}\includegraphics[height=3.7cm,width=5.5cm]{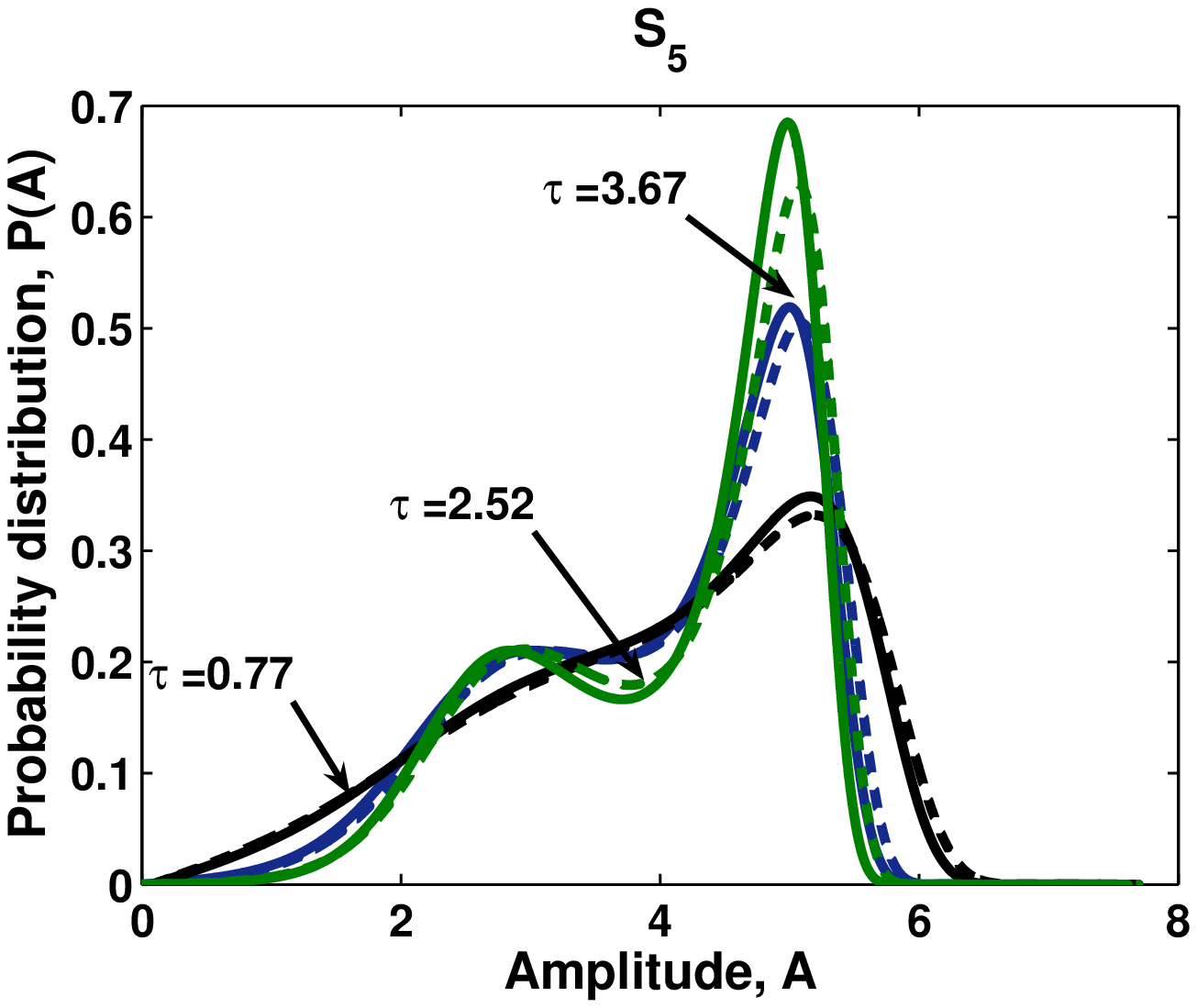}\hspace{0.3cm}
\includegraphics[height=3.7cm,width=5.5cm]{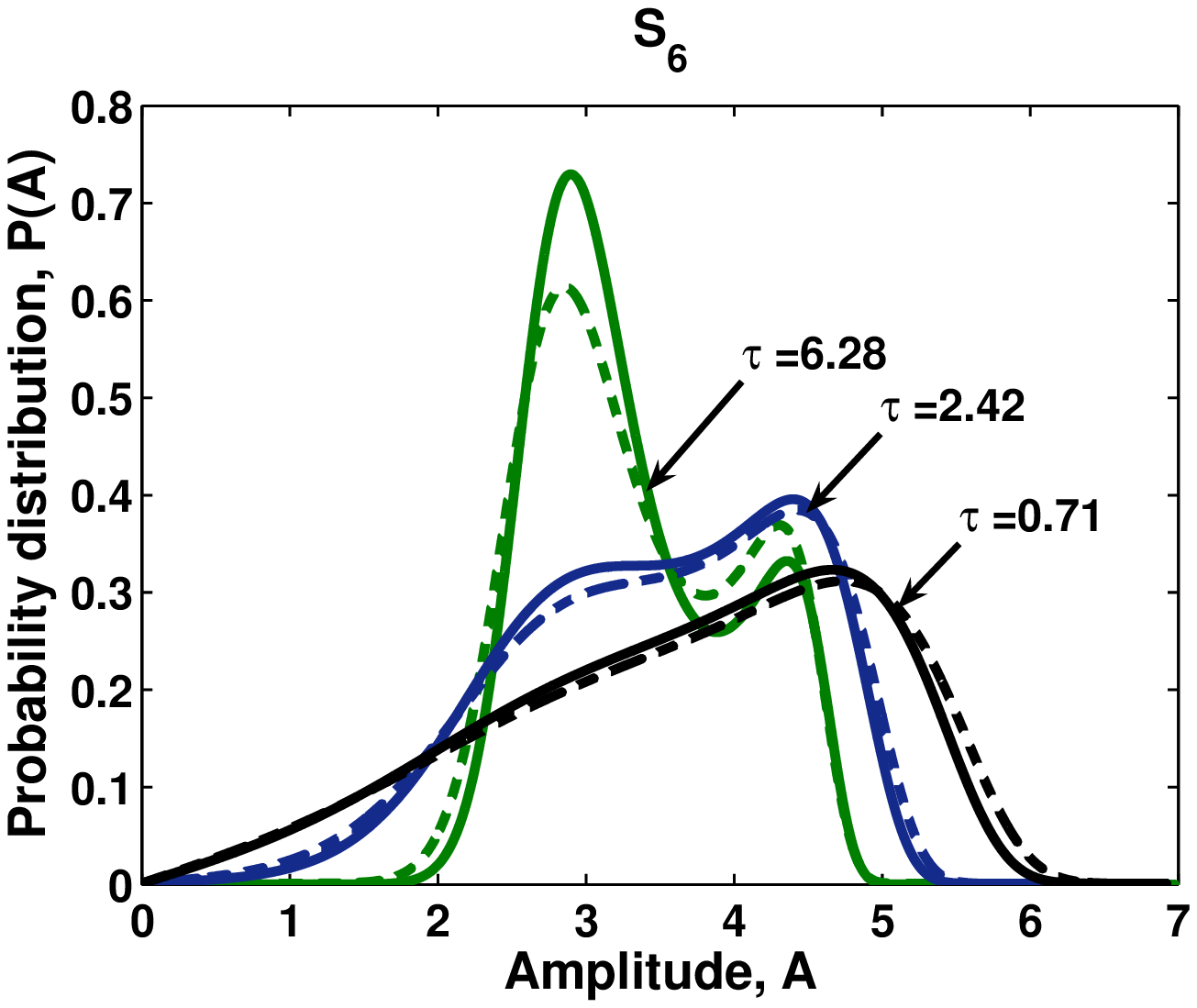}
\caption{\it
Probability distributions of the amplitude $A$ for different values of the correlation times $\tau$ when the frequencies of both attractors are identical, \emph{i.e }$\Omega_{1}\cong \Omega_{3}\cong 1 $.
Solid lines denote the analytical results, while dashed lines  denote numerical results.
Parameters of the system refer to  $ S_{i}$ of Table \ref{identical} with  $\mu = 0.01$, $D=0.01$}
\label{fig9}
\end{center}
\end{figure}

\begin{figure}
\begin{center}
\includegraphics[height=4cm,width=6cm]{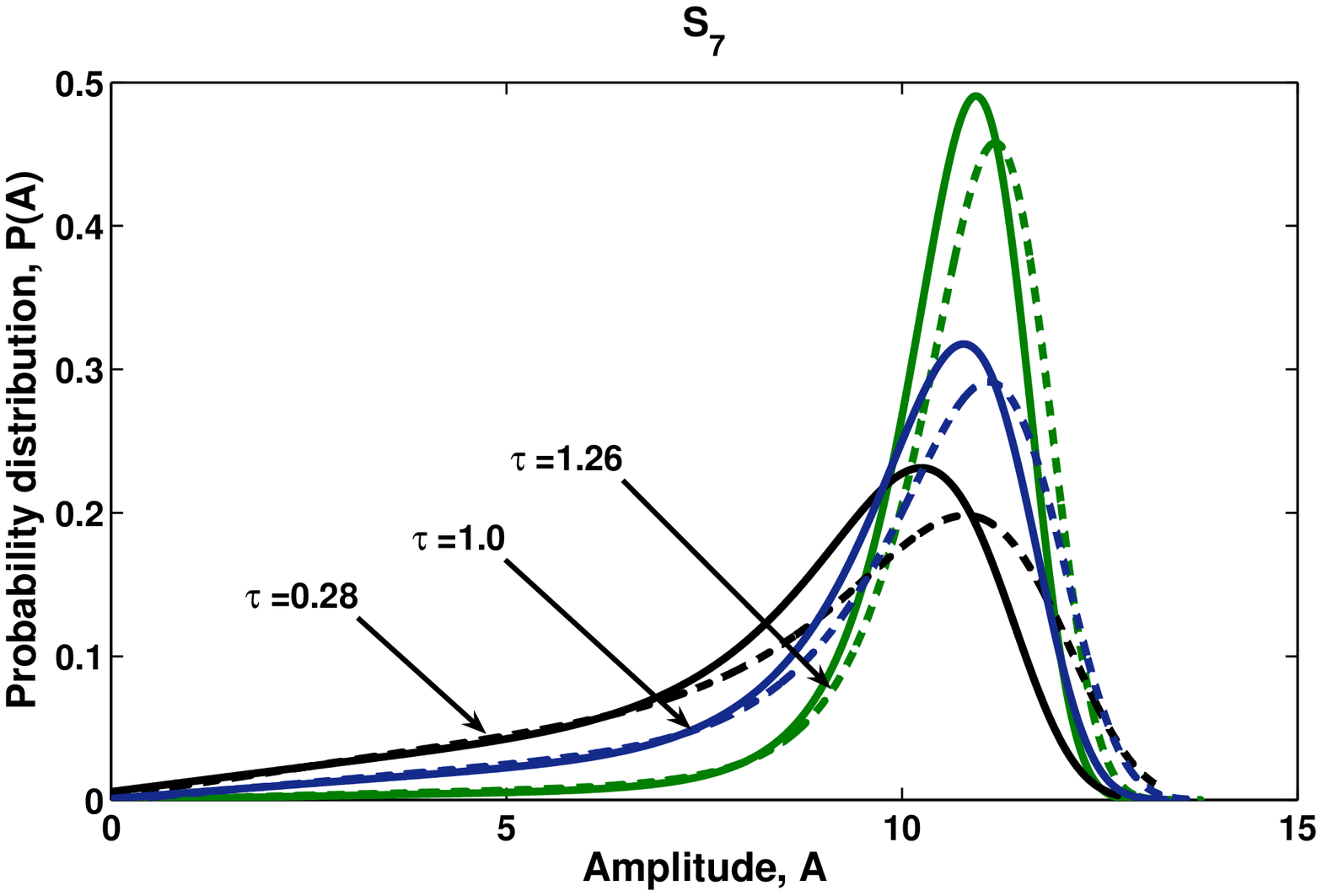}
\includegraphics[height=4cm,width=6cm]{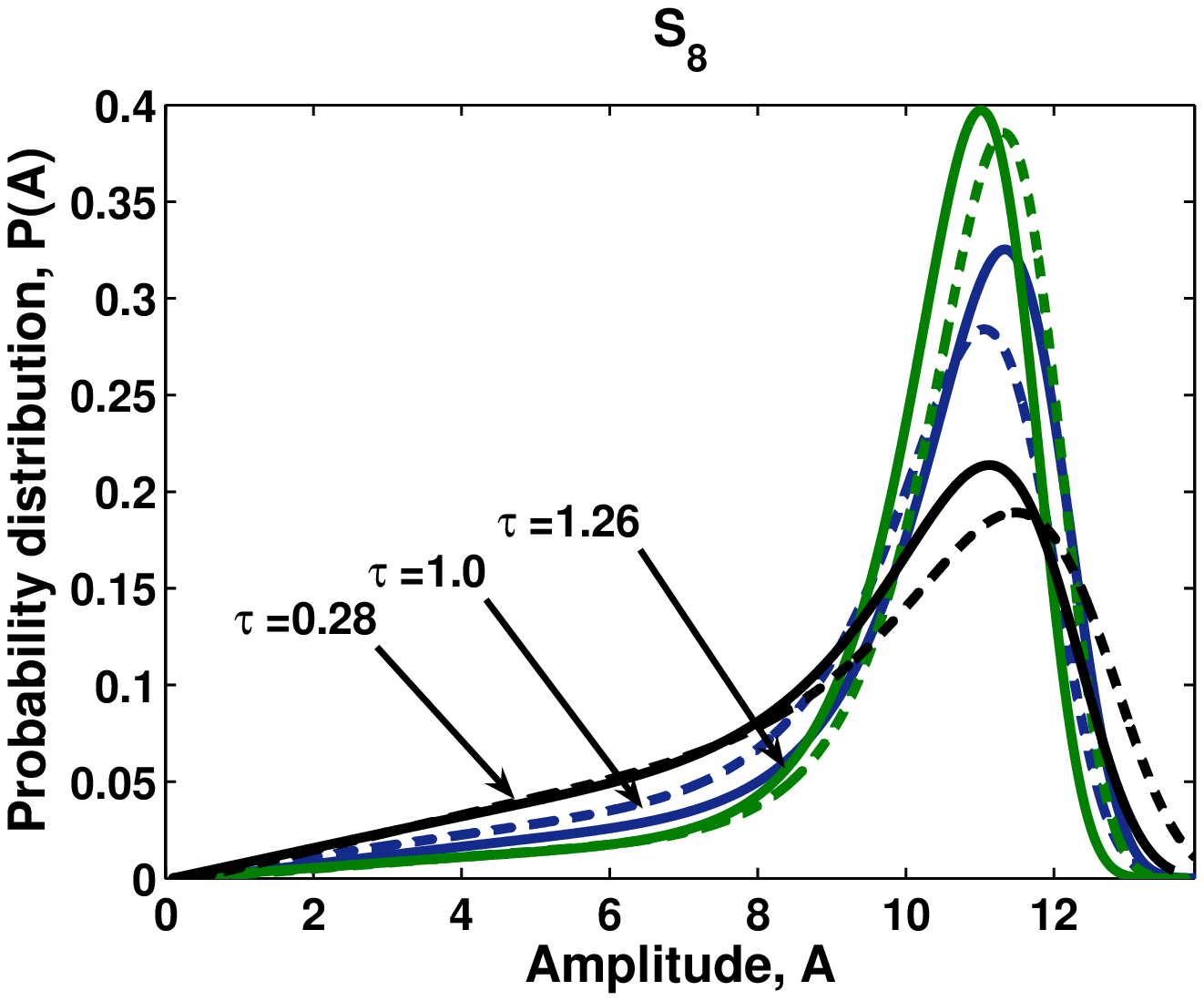}\\
\includegraphics[height=4cm,width=6cm]{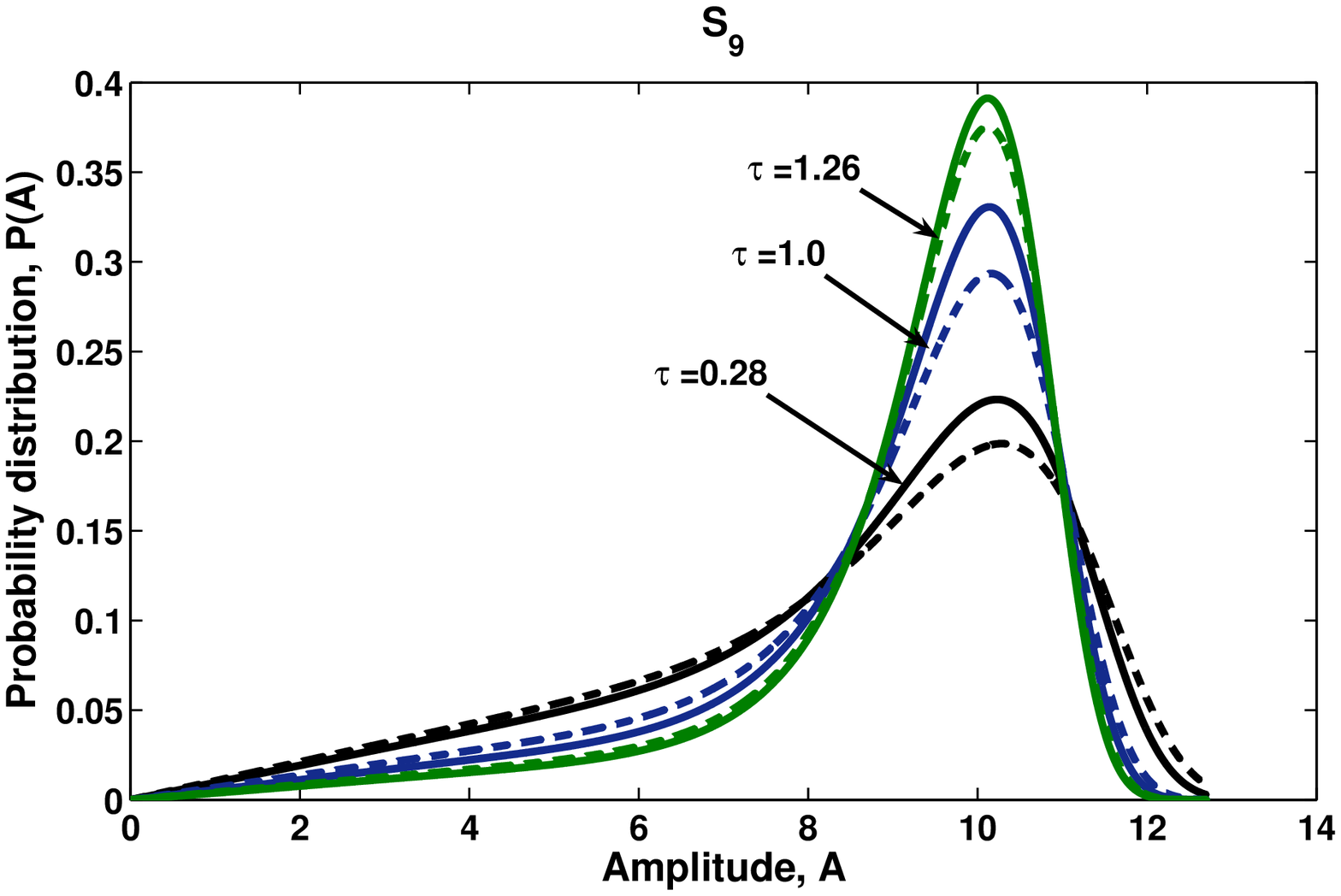}
\includegraphics[height=4cm,width=6cm]{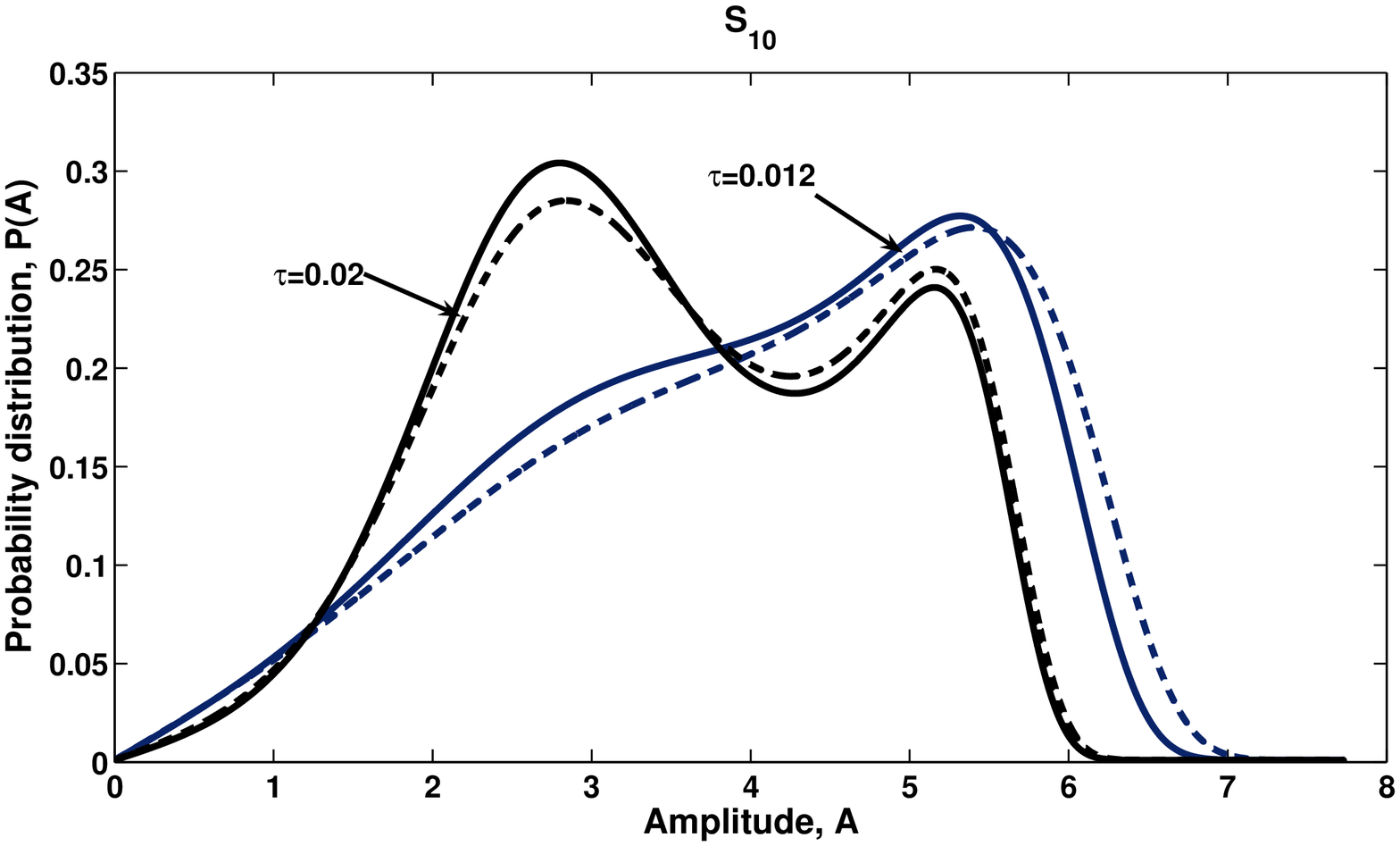}
\caption{\it
Probability distributions  of the amplitude $A$ for different values of the correlation times $\tau$ when the frequencies of the attractors are different, $\Omega_{1}\neq \Omega_{3}$.
Solid lines denote the analytical results, while dashed lines denote numerical results.
Parameters of the system refer to  $ S_{i}$ of Table \ref{different} with $\mu = 0.01$, $D=0.4$}
\label{fig10}
\end{center}
\end{figure}

\begin{figure}[htbp]
\begin{center}
\includegraphics[height=4cm,width=6cm]{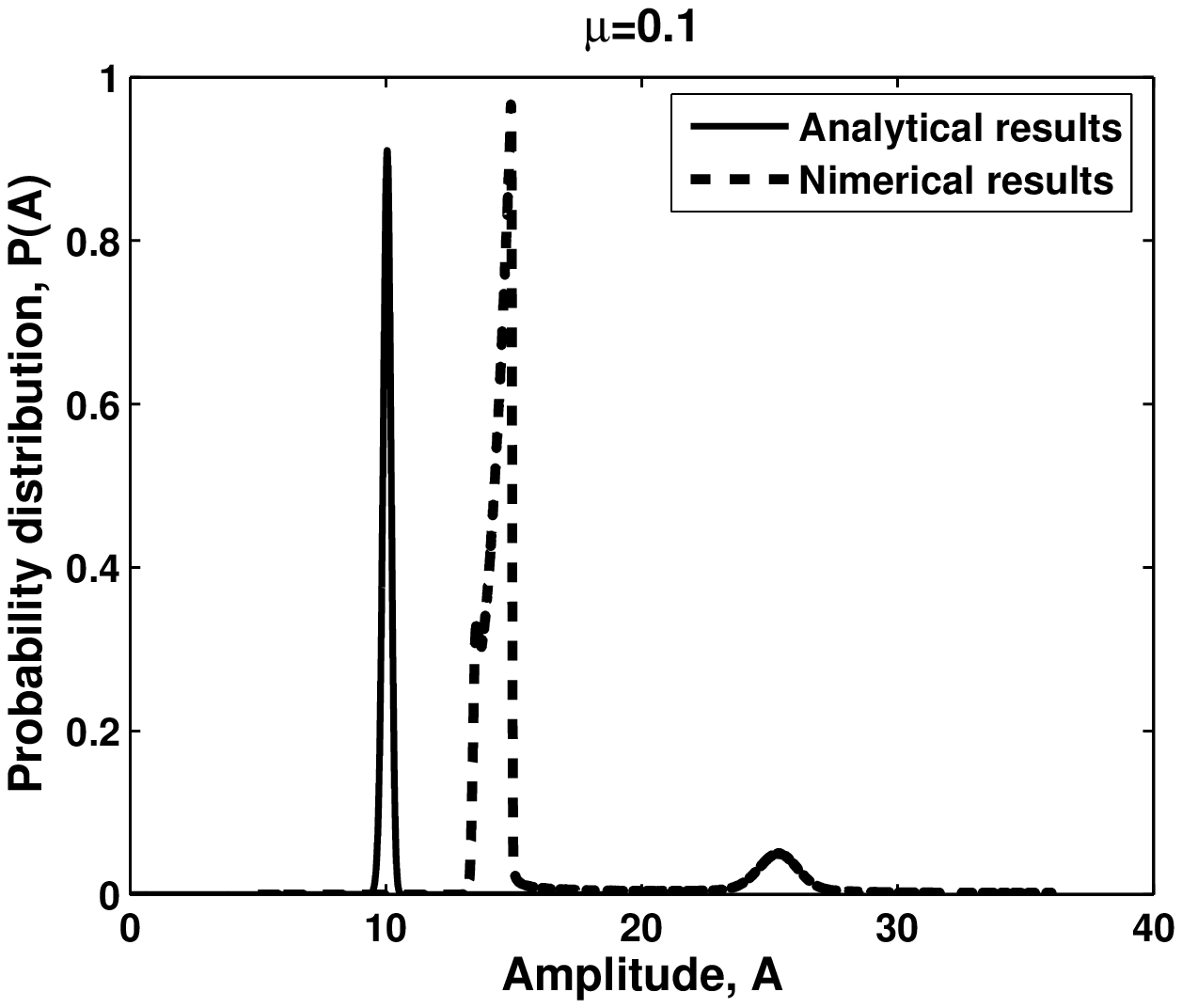}
\includegraphics[height=4cm,width=6cm]{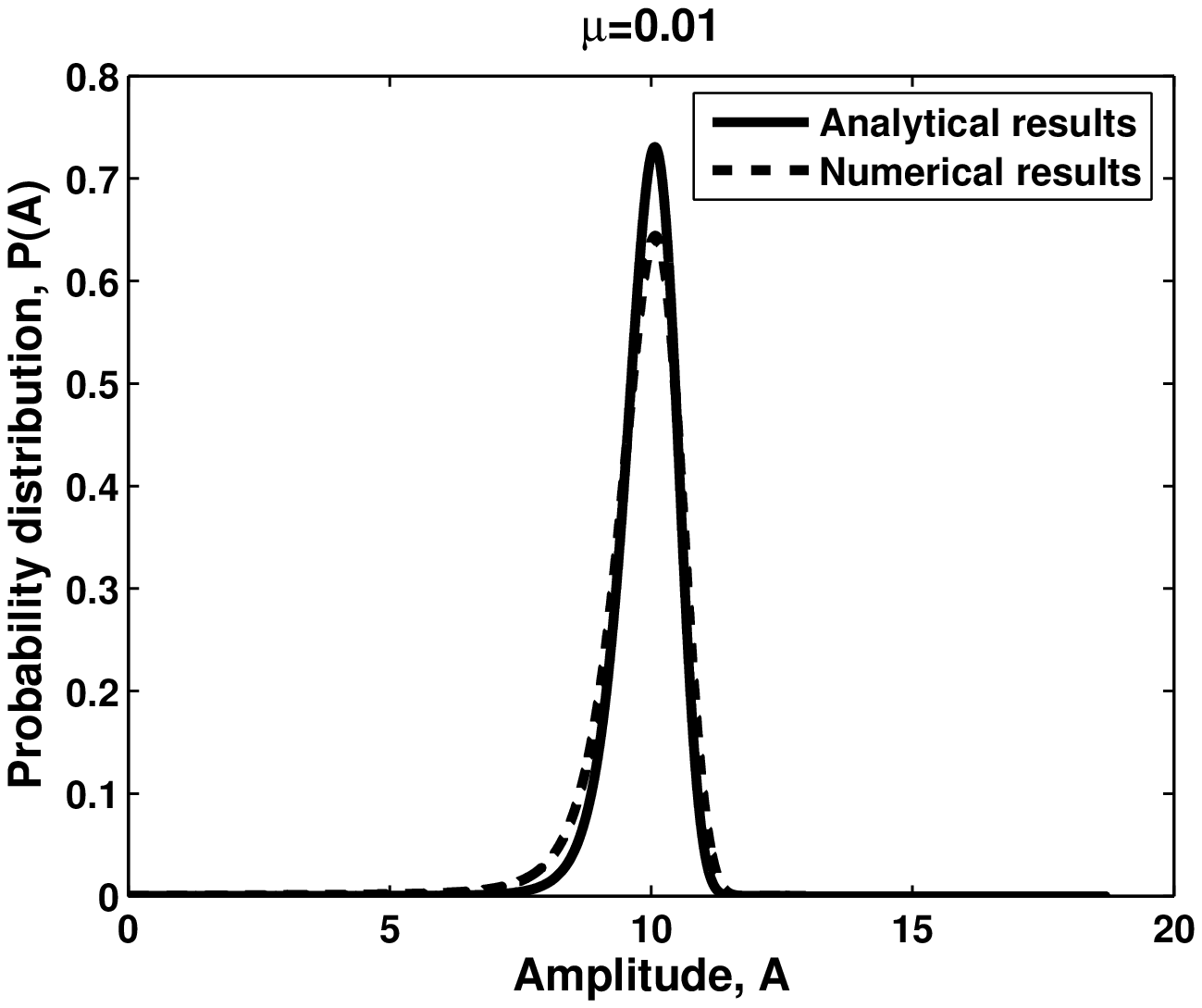}\\
\includegraphics[height=4cm,width=6cm]{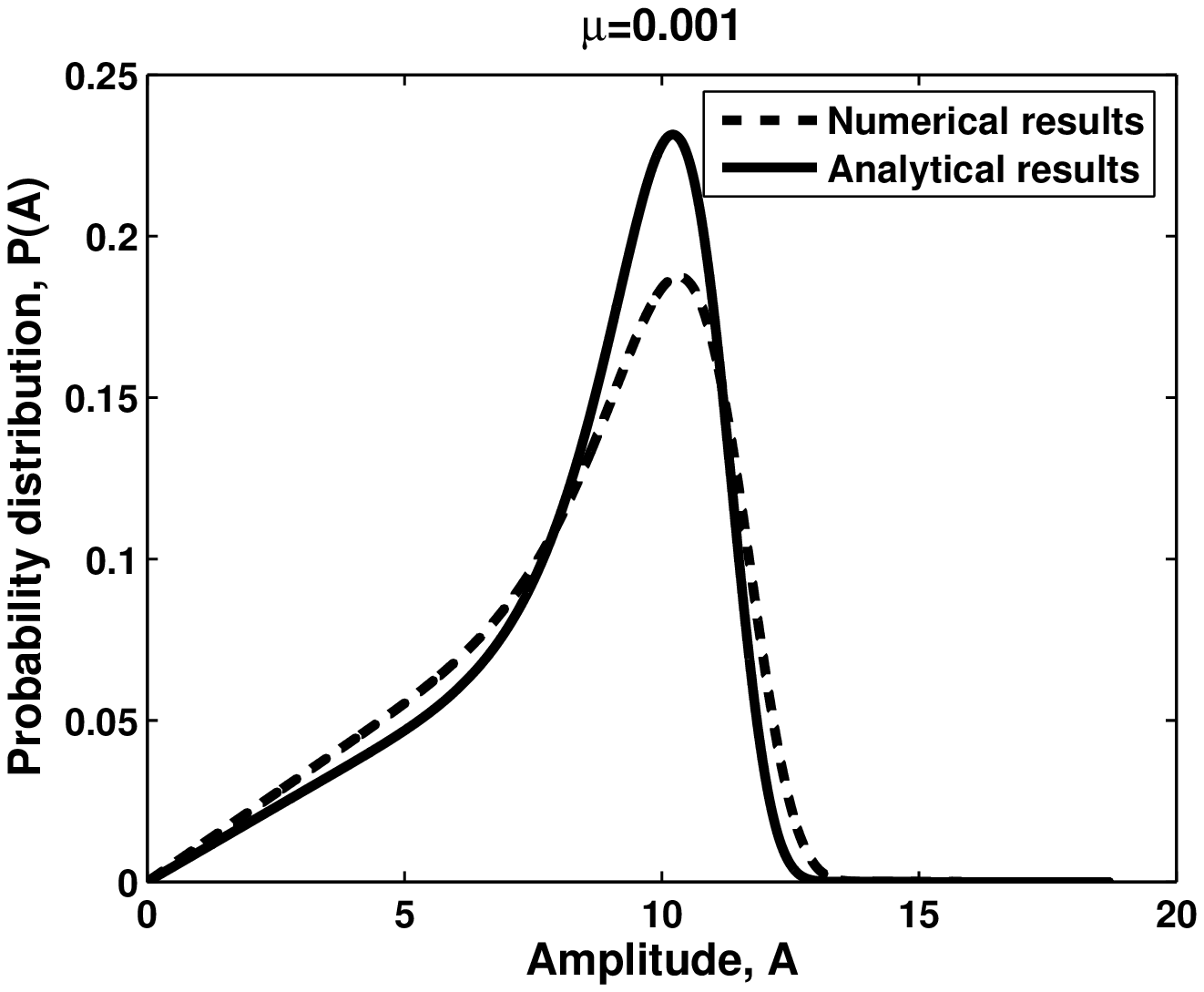}
\includegraphics[height=4cm,width=6cm]{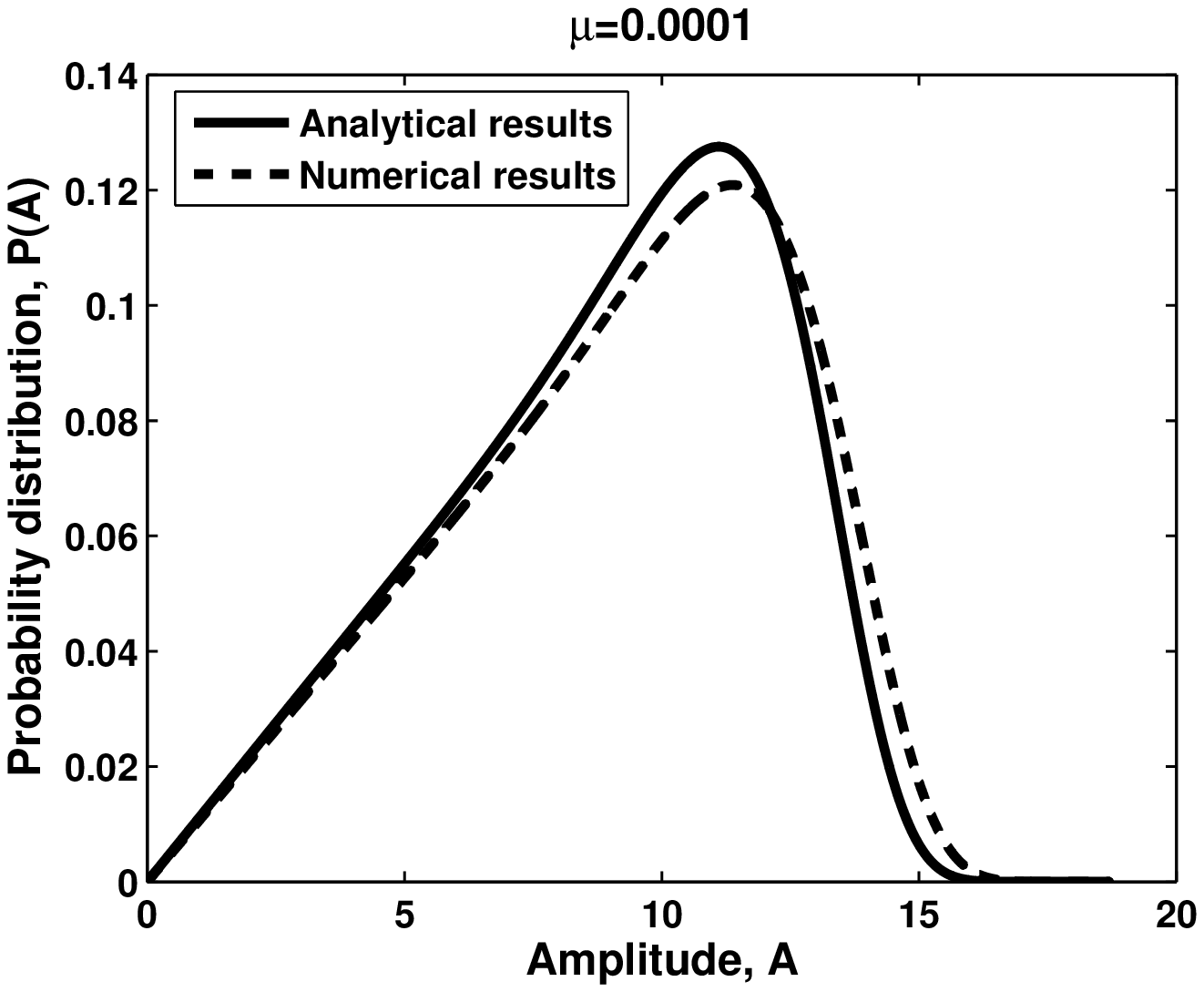}
\caption{\it
Probability distributions versus the amplitude $A$ for different values of the coefficient $\mu$ when
the frequencies of both attractors are different, $\Omega_{1}\neq \Omega_{3}$.
Solid lines denote the analytical results, while dashed lines denote numerical results.
Parameters of the system refer to corresponding to $S_8$ of Table \ref{different} with $D=0.05$ and $\tau=0.0001$}
\label{fig11}
\end{center}
\end{figure}

\section{Comparison with numerical simulations}
\label{comparison}

To check the validity of the approximations behind the analytic treatment that has led to the probability density function  (\ref{eq32}), we have performed numerical simulations of the Langevin dynamics (\ref{eq4}) with the numerical scheme to be described in Subsect. \ref{numerical}.

\subsection{Numerical algorithm for colored noise}
\label{numerical}
There are several methods and algorithms to solve stochastic differential equations \cite{mannella2}, as the implicit midpoint rule with \emph{Heun} and \emph{Leapfrog} methods or faster numerical
algorithms such as the stochastic version of the \emph{Runge-Kutta} methods and a quasisymplectic algorithm \cite{mannella1,mannella2,mannella3}.
The starting point is the \emph{Box-Muller} algorithm \cite{Knuth69} to generate
exponentially correlated colored noise distributed random variable $\eta_{\Delta t}$ from the Gaussian white noise and two random numbers $a$ and $b$ which are uniformly distributed on the unit interval $[0, 1]$.

To simulate the exponentially correlated colored noise, Eqs.(\ref{eq4}) are replaced with the following equations \cite{fox}:
\begin{subequations}
 \label{eq6}
 \begin{eqnarray}
\label{eq6a}  \dot x&=&u,  \\
\label{eq6b}  \dot u&=&\mu(1-x^2+\alpha x^4-\beta x^6)u-x+\eta, \\
\label{eq6c}  \dot \eta&=&-\lambda \eta+\lambda g_w,
 \end{eqnarray}
\end{subequations}
where $g_w$ is a Gaussian white noise:
\begin{eqnarray}
\label{eq7} &&\langle g_w (t)\rangle=0\nonumber \\
&&\langle g_w (t)g_w (t') \rangle=2D \delta (t-t').
\end{eqnarray}
Combining Eq.(\ref{eq6c}) and Eqs.(\ref{eq7}) one gets an exponentially correlated colored noise $\eta$.
To numerically solve Eqs.(\ref{eq6}), we use the Box-Mueller algorithm \cite{Knuth69} to generate a Gaussian white noise $g_w$  from two random  numbers $a$ and $b$, which are uniformly distributed on the unit interval, as in Refs. \cite{Chamgoue12,Chamgoue13}.

The equations have been integrated halving the step size until consistent results are obtained (the problem is particularly delicate in the proximity of an absorbing barrier \cite{mannella3}). Furthermore, the procedure has been calibrated with a standard activation barrier to retrieve the Kramers escape rate \cite{Kramers40}, as modified by the correlation time $\tau$ \cite{Hanggi95}.
In the end, we have find that a step size $\Delta t=0.0001$ is generally appropriated, but in few cases it has been necessary an even smaller step.
Moreover, we have averaged the results over $200$ realizations, that ensures converge within few percents.
This scheme has been employed to check the validity of the approximations behind the analytic treatment that has led to the probability density function  (\ref{eq32}).

In Fig. \ref{fig9}, we plot the behavior of the probability distribution $p$ as a function of the amplitude $A$ for low noise intensity $D=0.01$ and several values of the correlation time $\tau$, when the frequencies of both attractors are similar, \emph{i.e.} $\Omega_1\simeq\Omega_3\simeq 1$. It appears that the system is more likely found at two distinct distances from the origin, the essential feature of birhythmicity.
In fact, for the parameters here considered, there are two attractors in the deterministic system, see Table \ref{identical}. However, as underlined in the previous Section, we expect that noise can destroy birhythmicity (see Fig. \ref{fig5}). In fact, for short correlation times, the amplitude has sometimes a single peak (parameters $S_4,_5,_6$). We also notice that the correlation noise tends to restore the birhythmic behavior, and in fact for the largest values of the noise correlation time $\tau$ we always find a two-peak solution.
The probability distribution $p$ is asymmetric for the set of parameters $S_{1,2,3,4,5}$.
Although the correlation time  varies from the values $\tau=0.77$ to $\tau=1.47$, this asymmetric property of the density probability distribution $p$  persists.
In all these configurations, the comparison between the analytic and numerical results is acceptable.

Fig. \ref{fig10} shows the variation of the probability distribution density $p$ versus the amplitude $A$ when the frequencies of both attractors are different,  (\emph{i.e.} $\Omega_1\neq \Omega_3$, see Table \ref{different}) for $D=0.4$ and different values of the correlation  times $\tau$.
Also in this case it is evident that the correlation time $\tau$ tends to stabilize a second orbit, and thus gives rise to a birhythmic behavior.

Finally, in  Fig. \ref{fig11} we show the probability distribution function when the frequencies that characterize the attractors are different varying the nonlinear parameter $\mu$.
The Figure demonstrates that the analytic and numerical estimates appear to be quite close, except at high levels of the parameter $\mu$.

To measure the effect of the correlation time on the stability of the two solutions, let us return to Eq. (\ref{eq27}) that shows the connection between the average persistence (or residence) times $T_{1,3}$ on the attractor with limit cycle amplitude $A_{1,3}$ and the corresponding escape rates determined by the quasipotential (\ref{eq21}).
The analytical and numerical behaviors of the residence times versus $\tau$ are reported in Figs. \ref{fig7} and \ref{fig8},  for $\Omega_1\simeq\Omega_3$ and $\Omega_1\neq \Omega_3$, respectively.
In Fig. \ref{fig7}, three relevant cases can be found for the first occurrence ($\Omega_1 \simeq \Omega_3$, Table \ref{identical}):

\begin{widetext}

\begin{figure}[htbp]
\begin{center}
\includegraphics[scale=0.35]{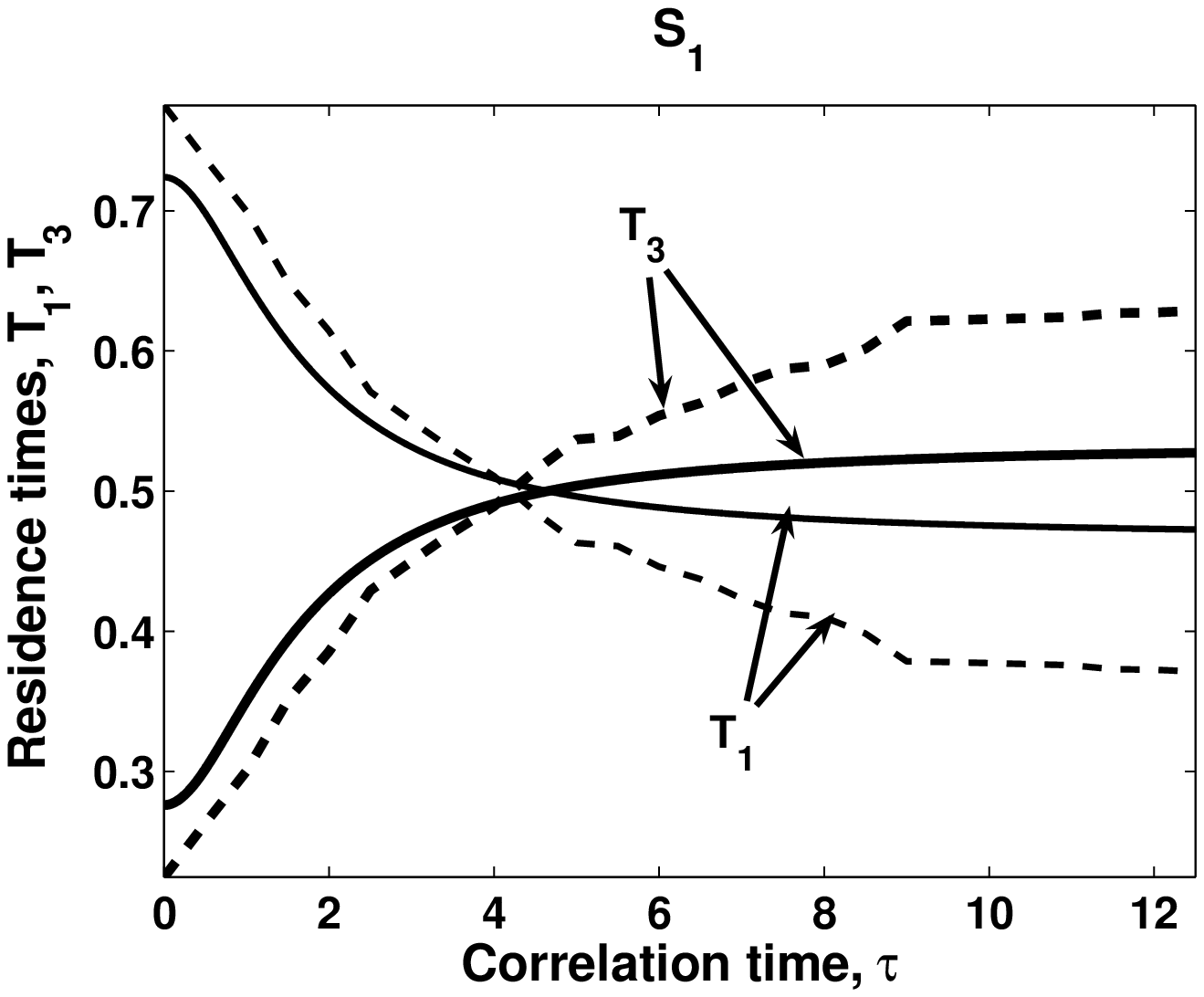}
\includegraphics[scale=0.35]{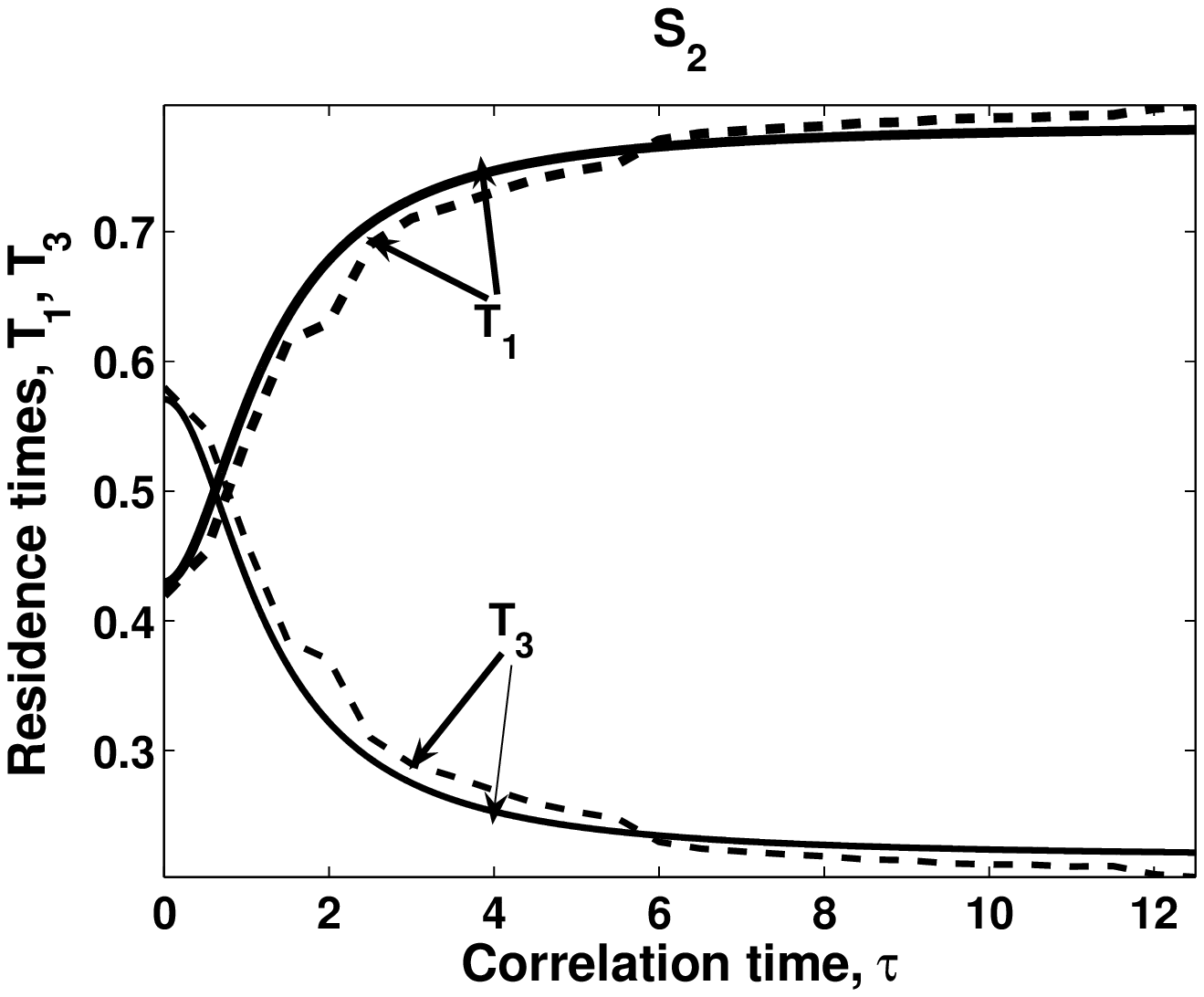}\\
\includegraphics[scale=0.35]{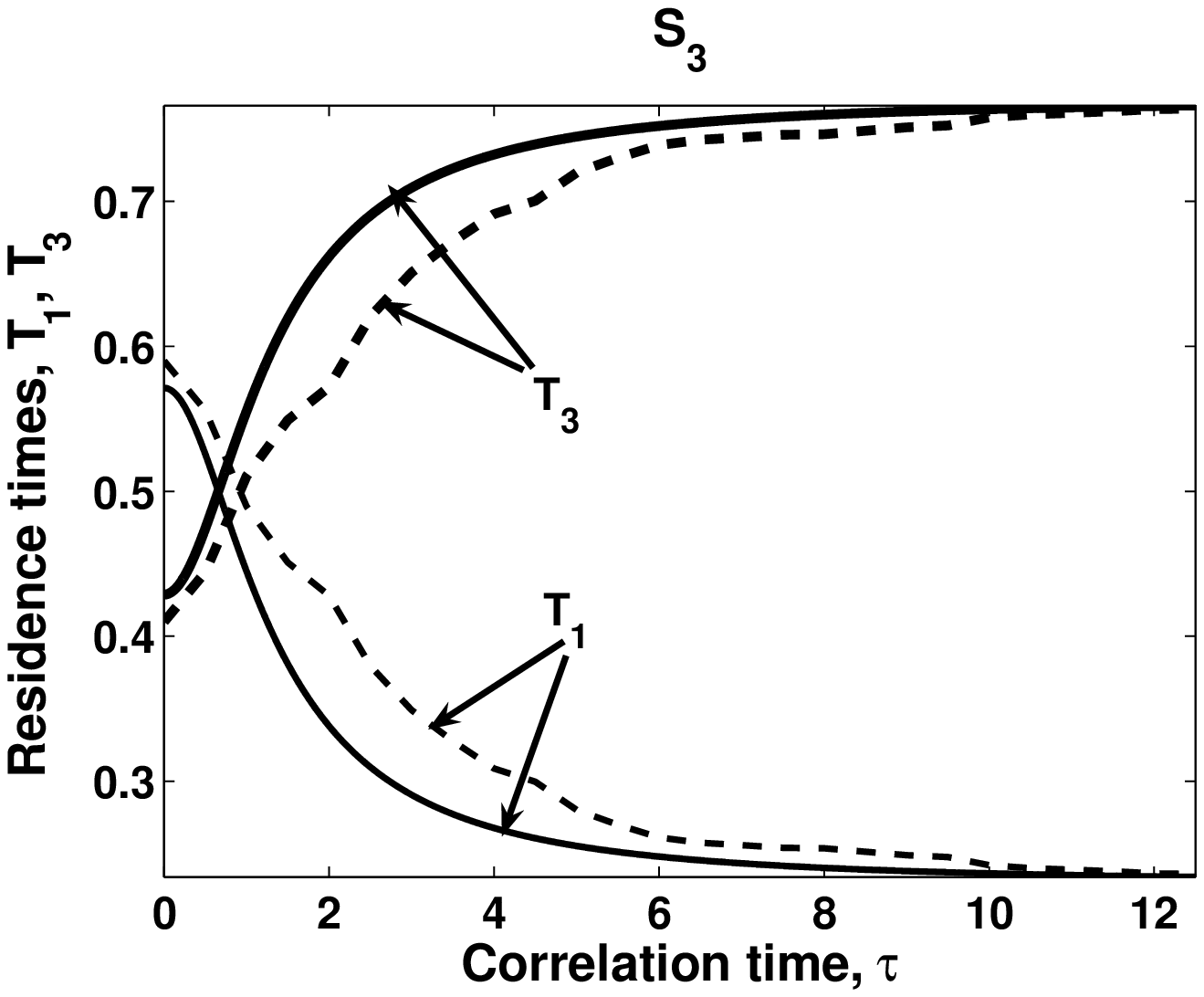}
\includegraphics[scale=0.35]{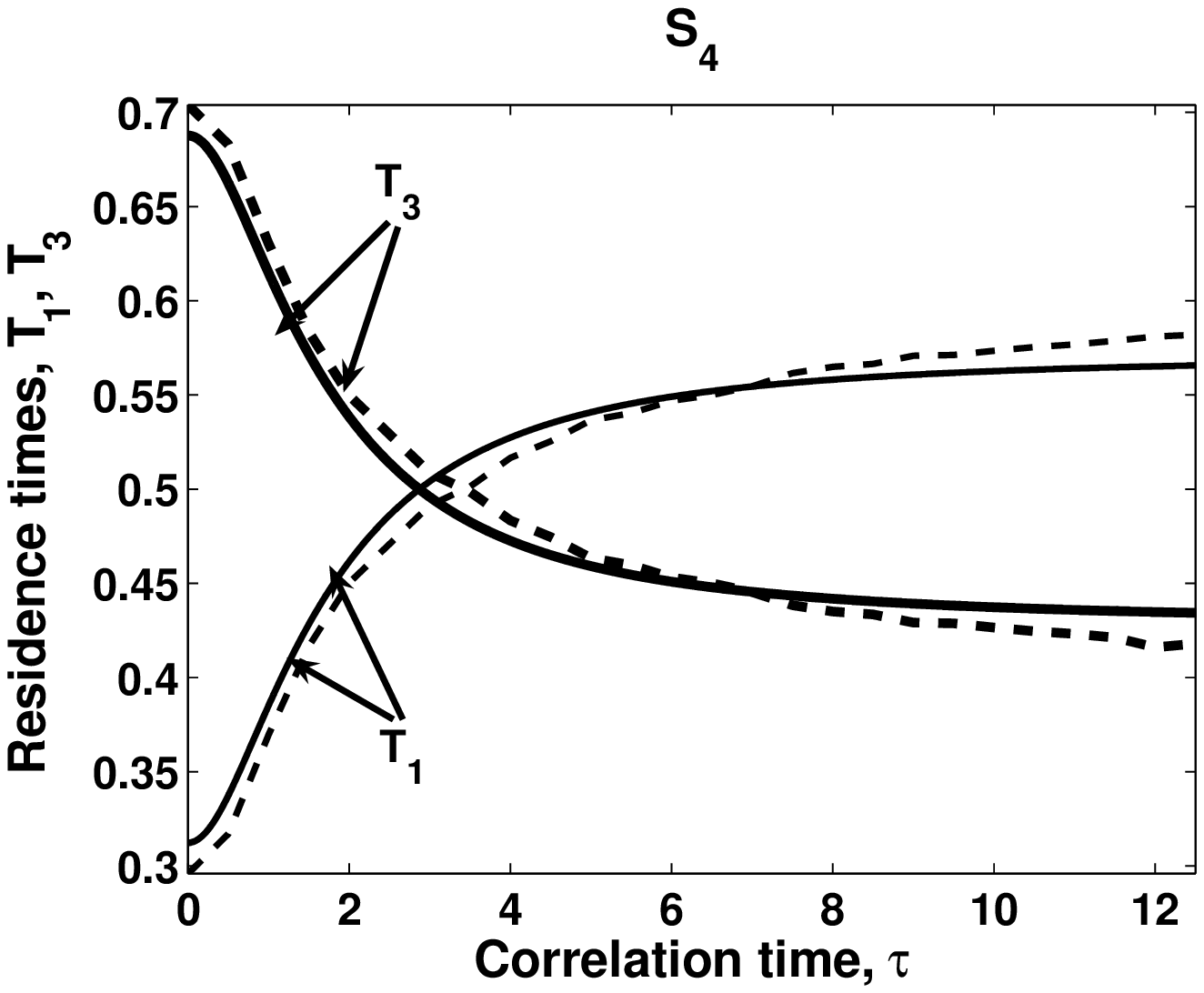}\\
\includegraphics[scale=0.35]{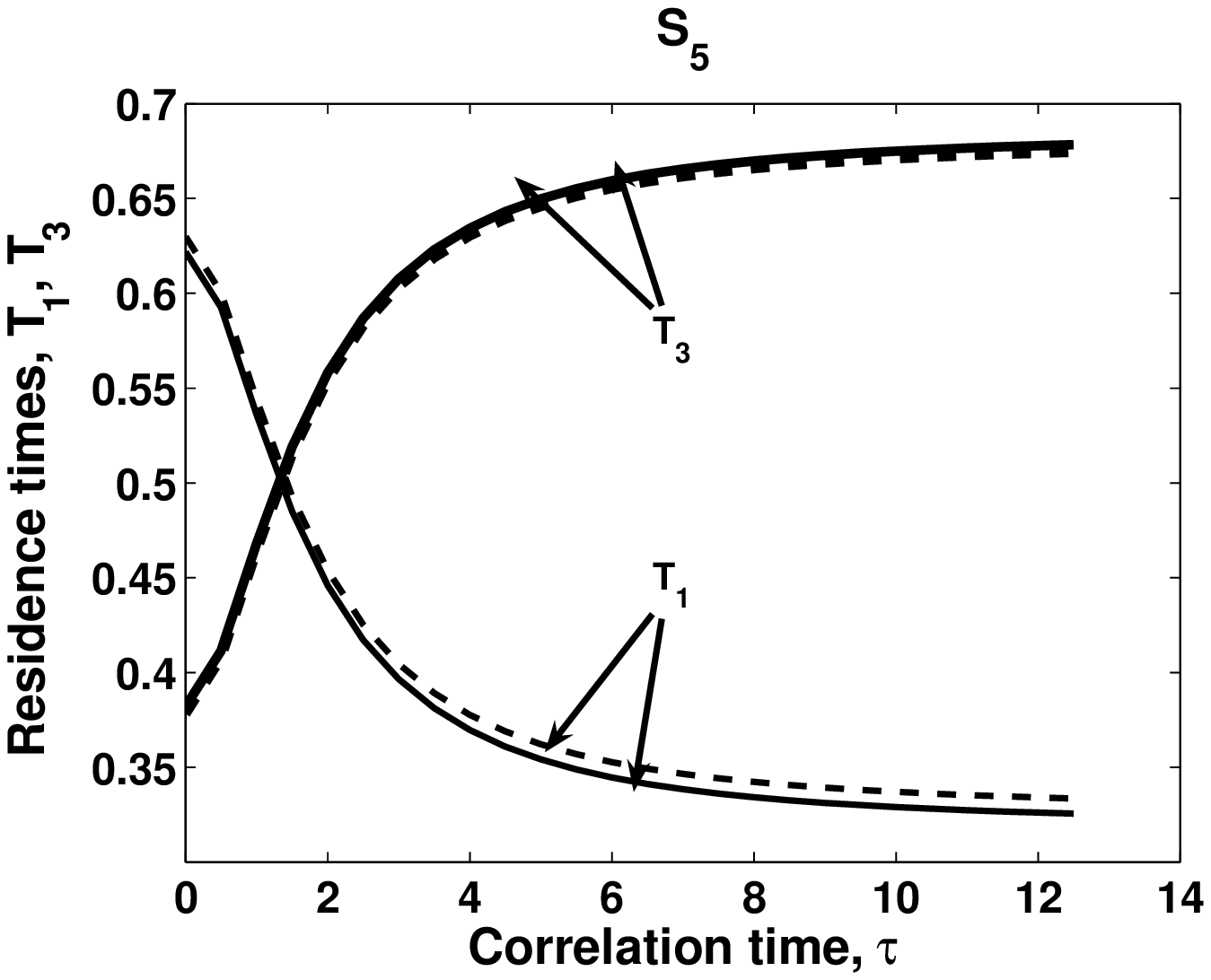}
\includegraphics[scale=0.35]{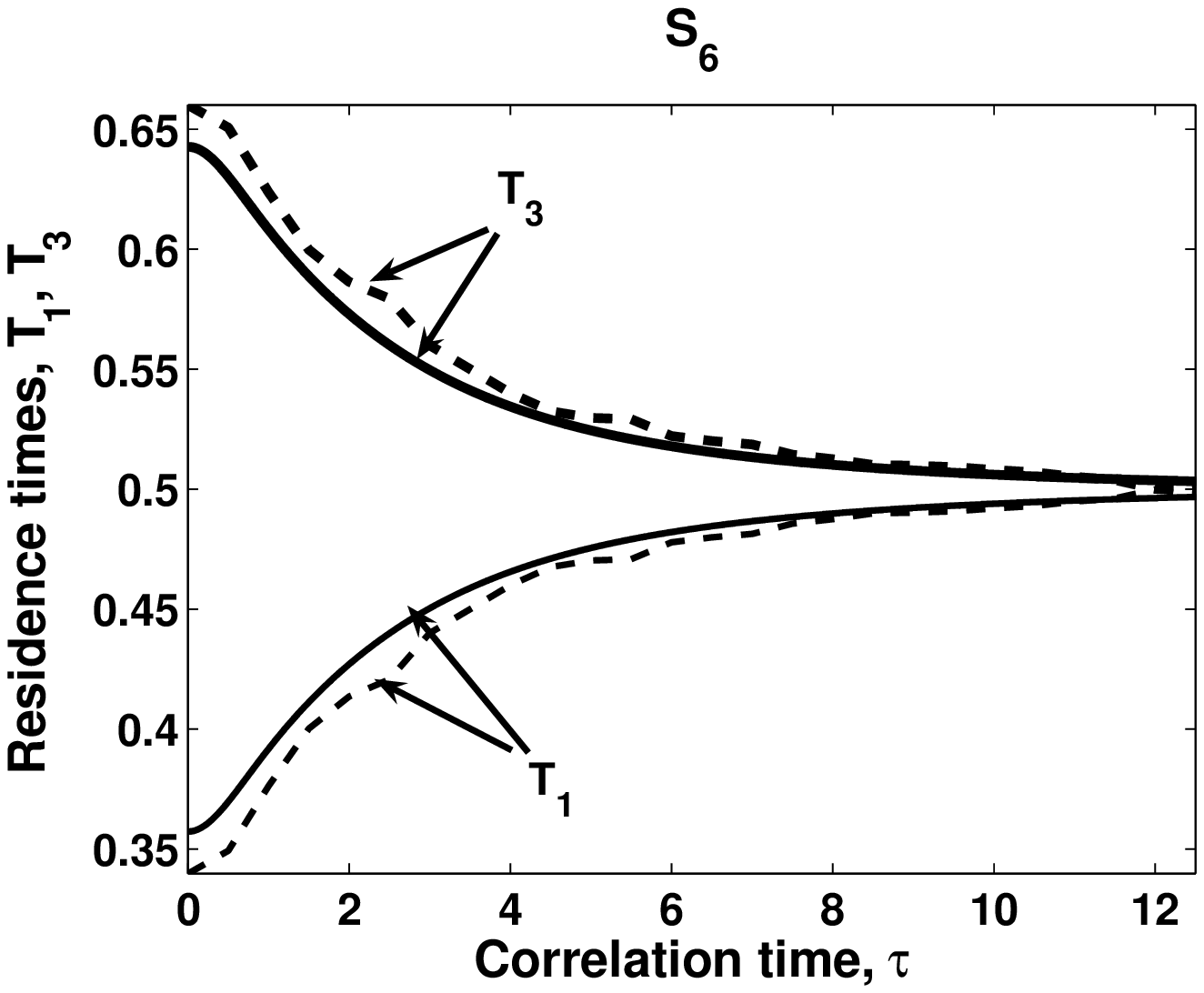}
\caption{\it
Residence times versus the correlation time $\tau$ for different values of the parameter $S_i=(\alpha; \beta)$ when the frequency are almost identical ( $\Omega_{1}\cong \Omega_{3}\cong 1 $).
Solid lines denote the analytical results, while dashed lines denote numerical results.
 Parameters of the system refer to  $ S_{i}$ of Table \ref{identical} with $\mu=0.01$, $D=0.01$.}
\label{fig7}
\end{center}
\end{figure}

\begin{figure}[htbp]
\begin{center}
\includegraphics[scale=0.4]{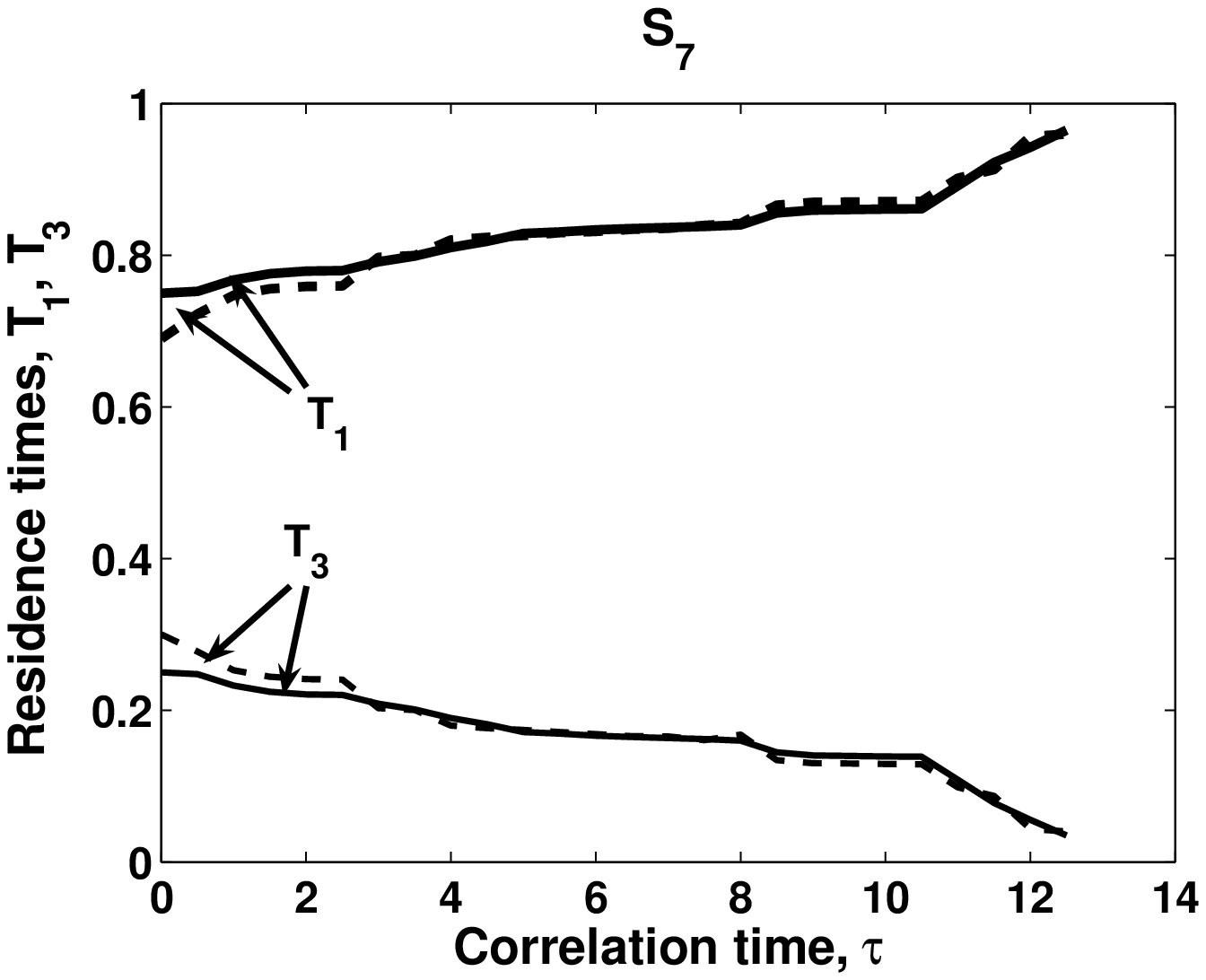}
\includegraphics[scale=0.4]{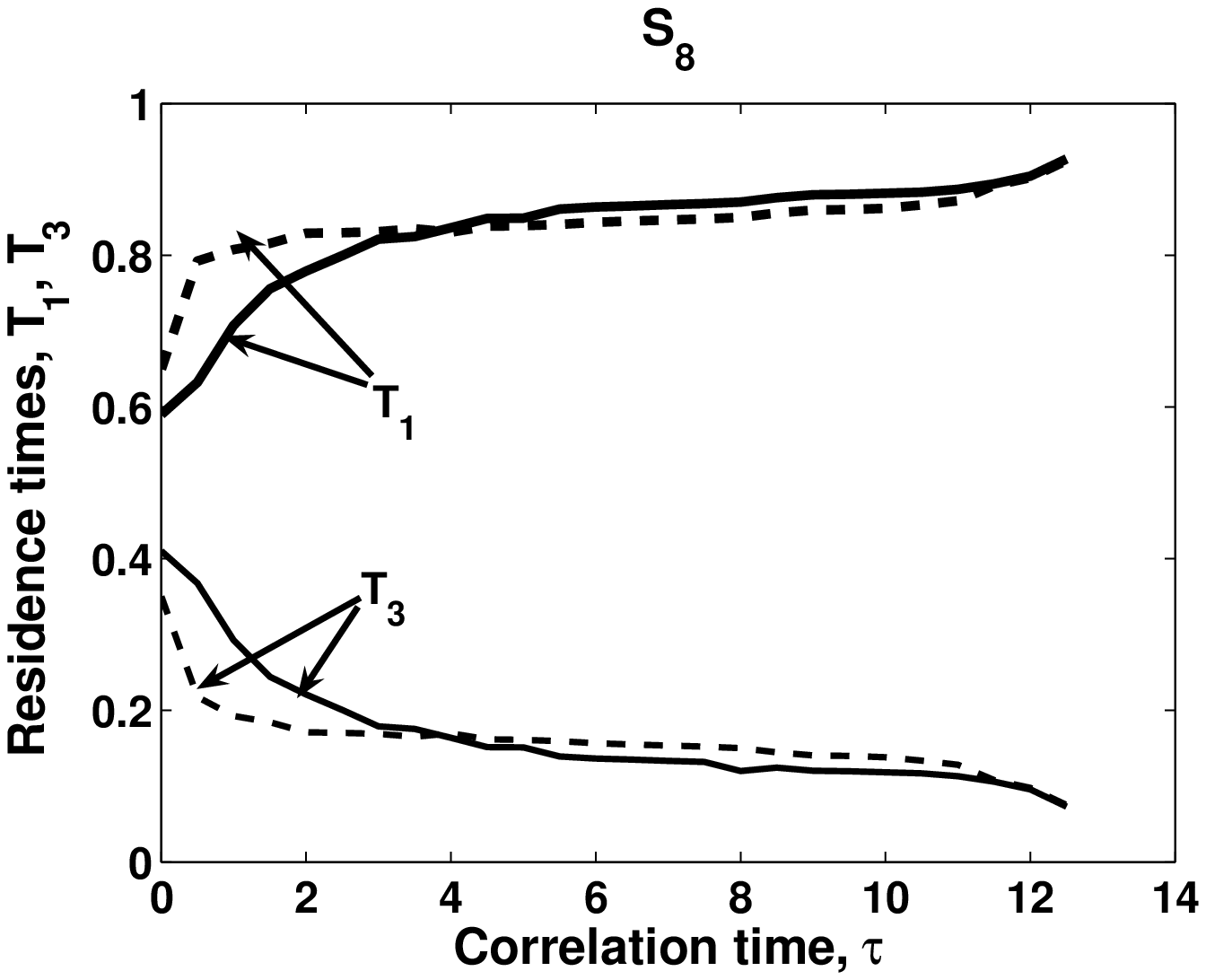}\\
\includegraphics[scale=0.4]{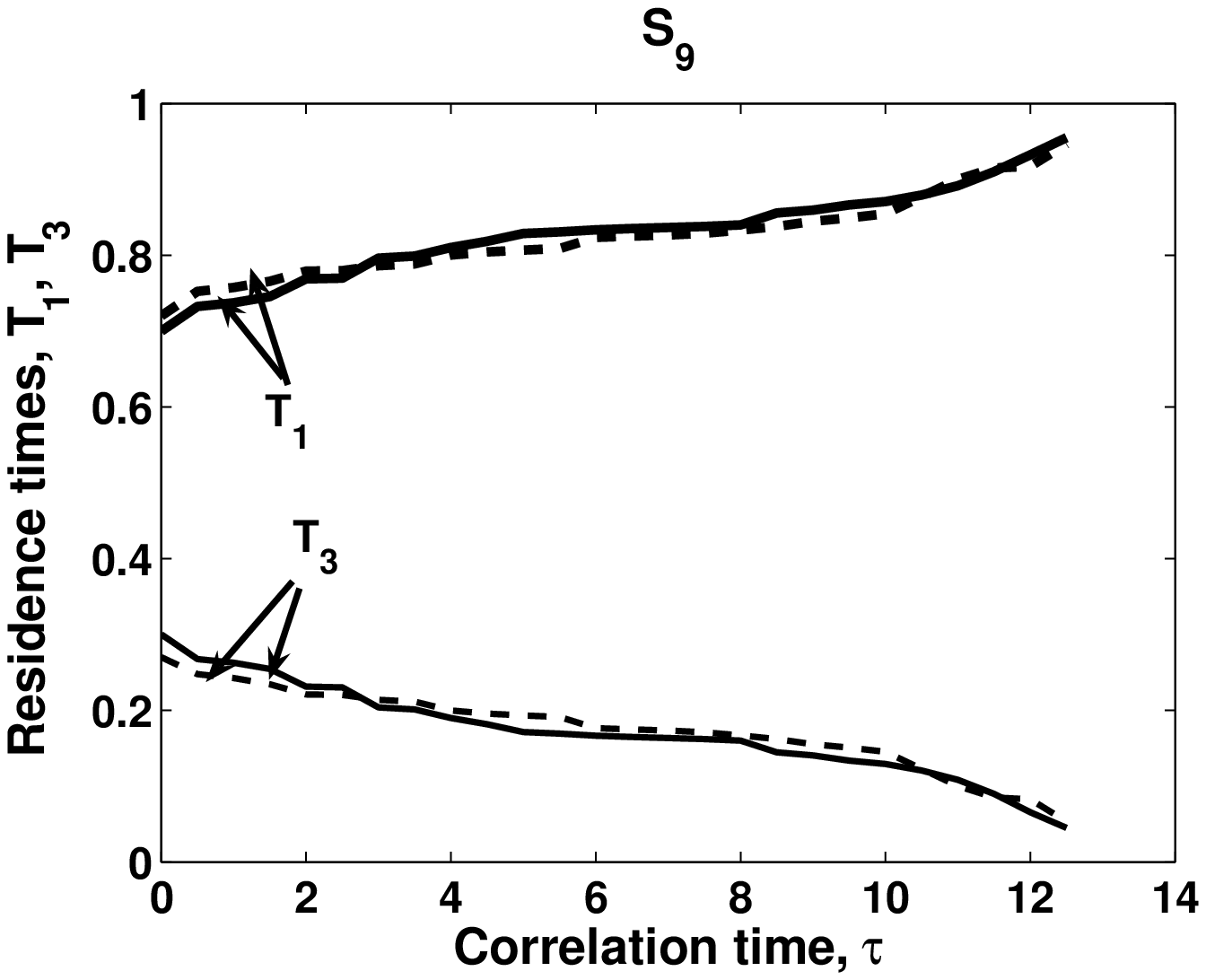}
\includegraphics[scale=0.4]{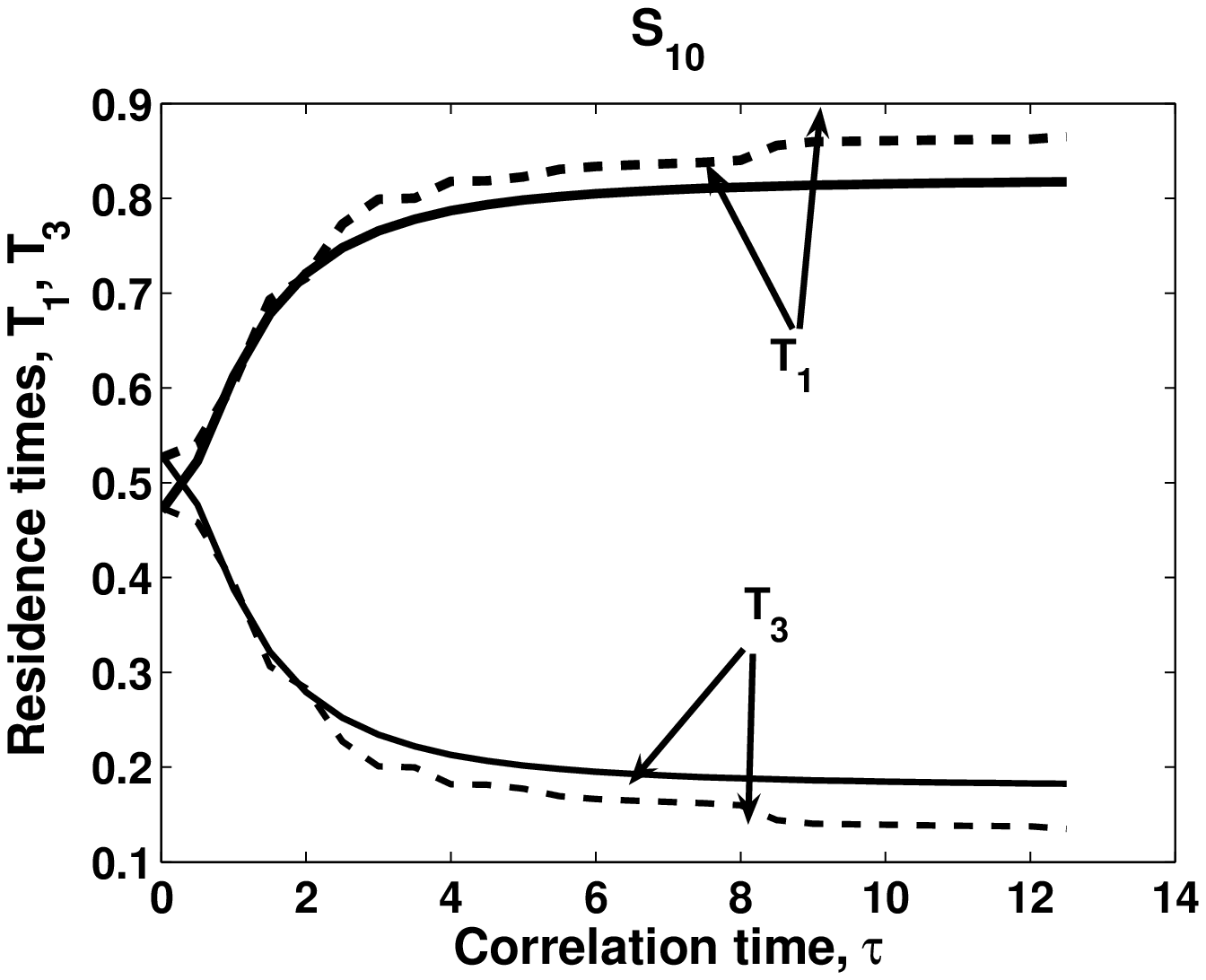}
\caption{\it
Residence times versus the correlation time $\tau$ for different values of the parameter $S_i=(\alpha; \beta)$
when the frequency are different ( $\Omega_{1}\neq \Omega_{3} $).
Solid lines denote the analytical results, while dashed lines denote numerical results.
 Parameters of the system refer to  $ S_{i}$ of Table \ref{different} with $\mu=0.01$, $D=0.2$.}
\label{fig8}
\end{center}
\end{figure}


\begin{figure}[htbp]
\begin{center}
\includegraphics[scale=0.25]{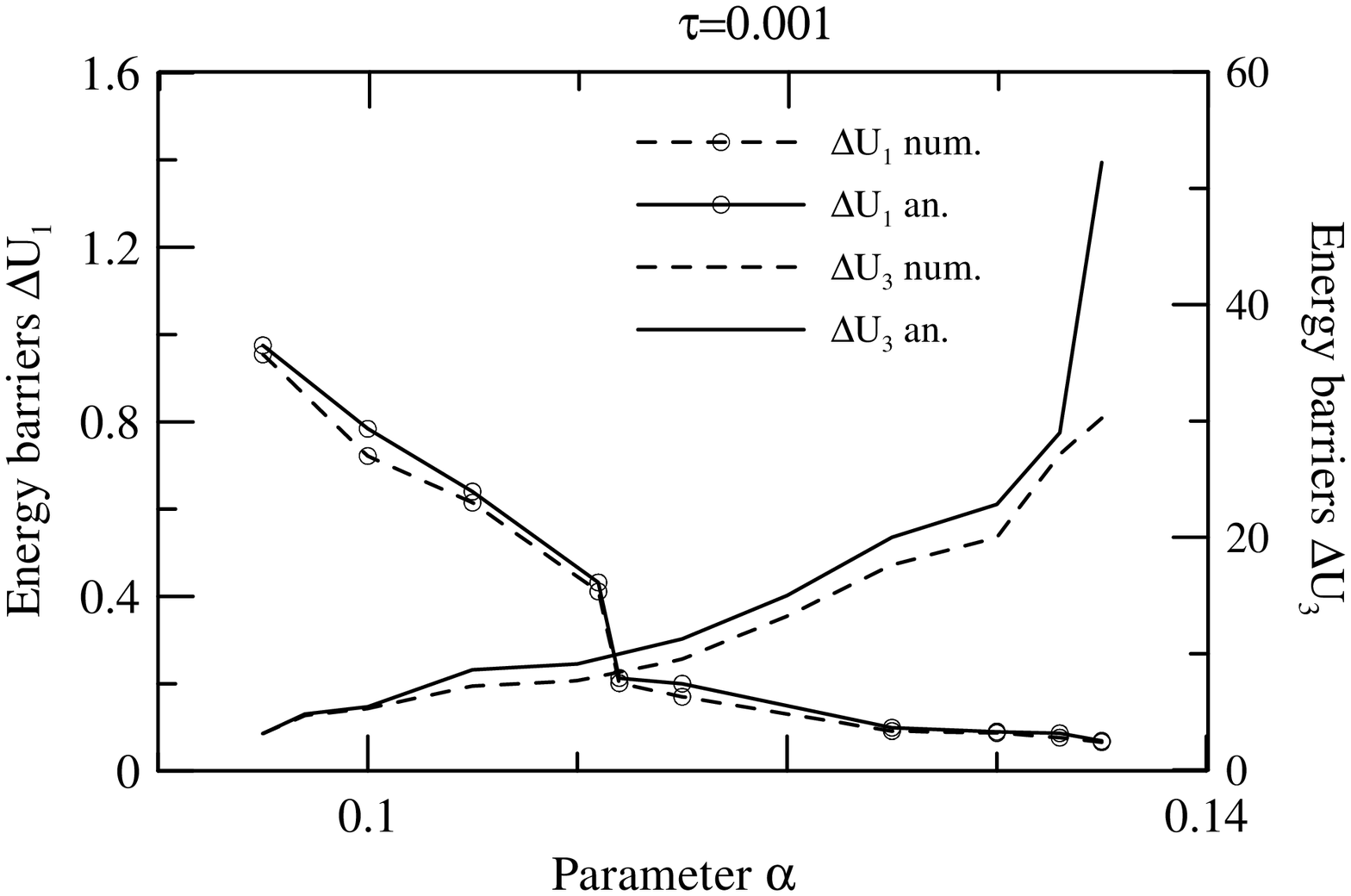} \hspace{0.25cm}
\includegraphics[scale=0.25]{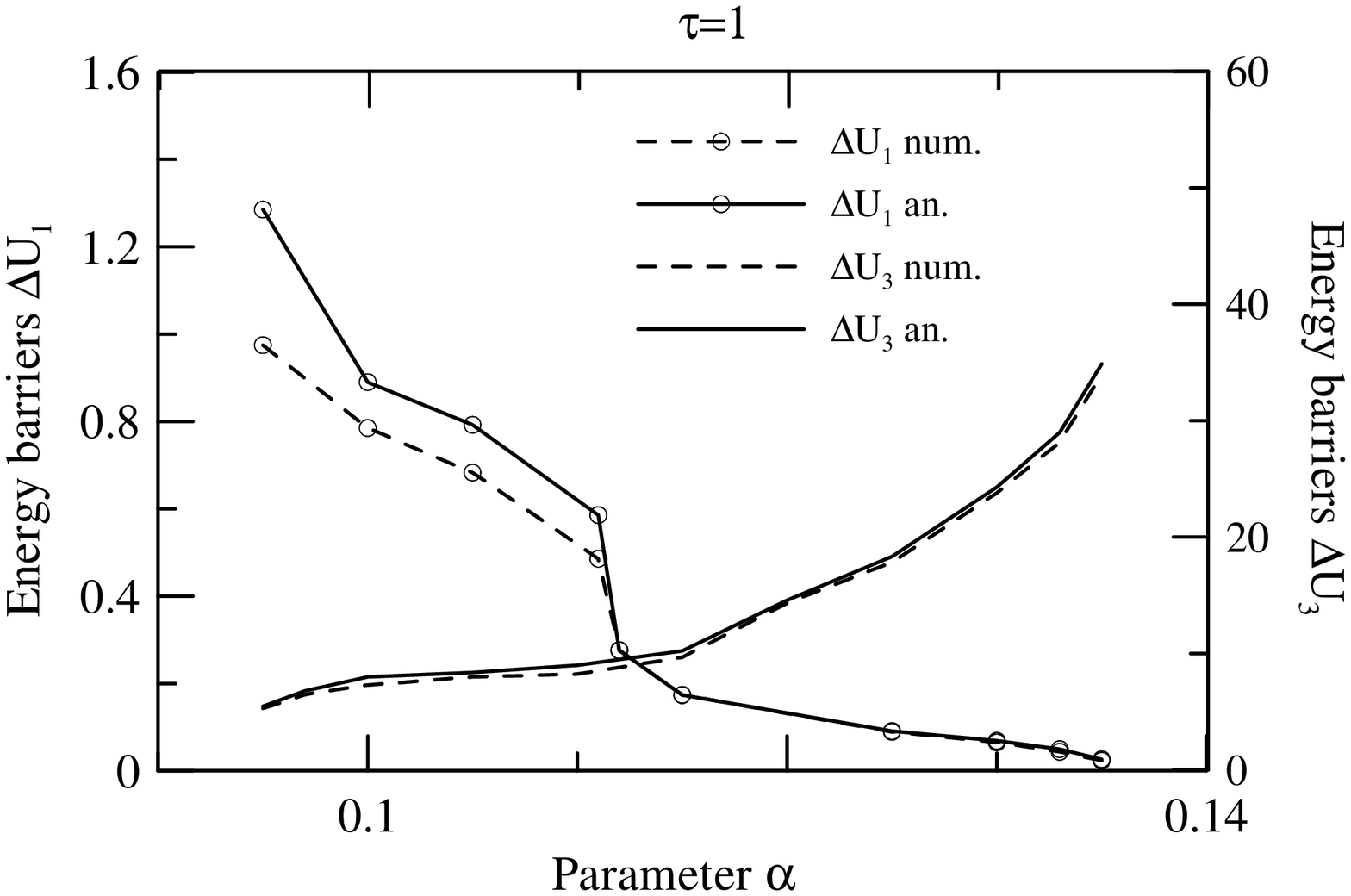}\\
\includegraphics[scale=0.25]{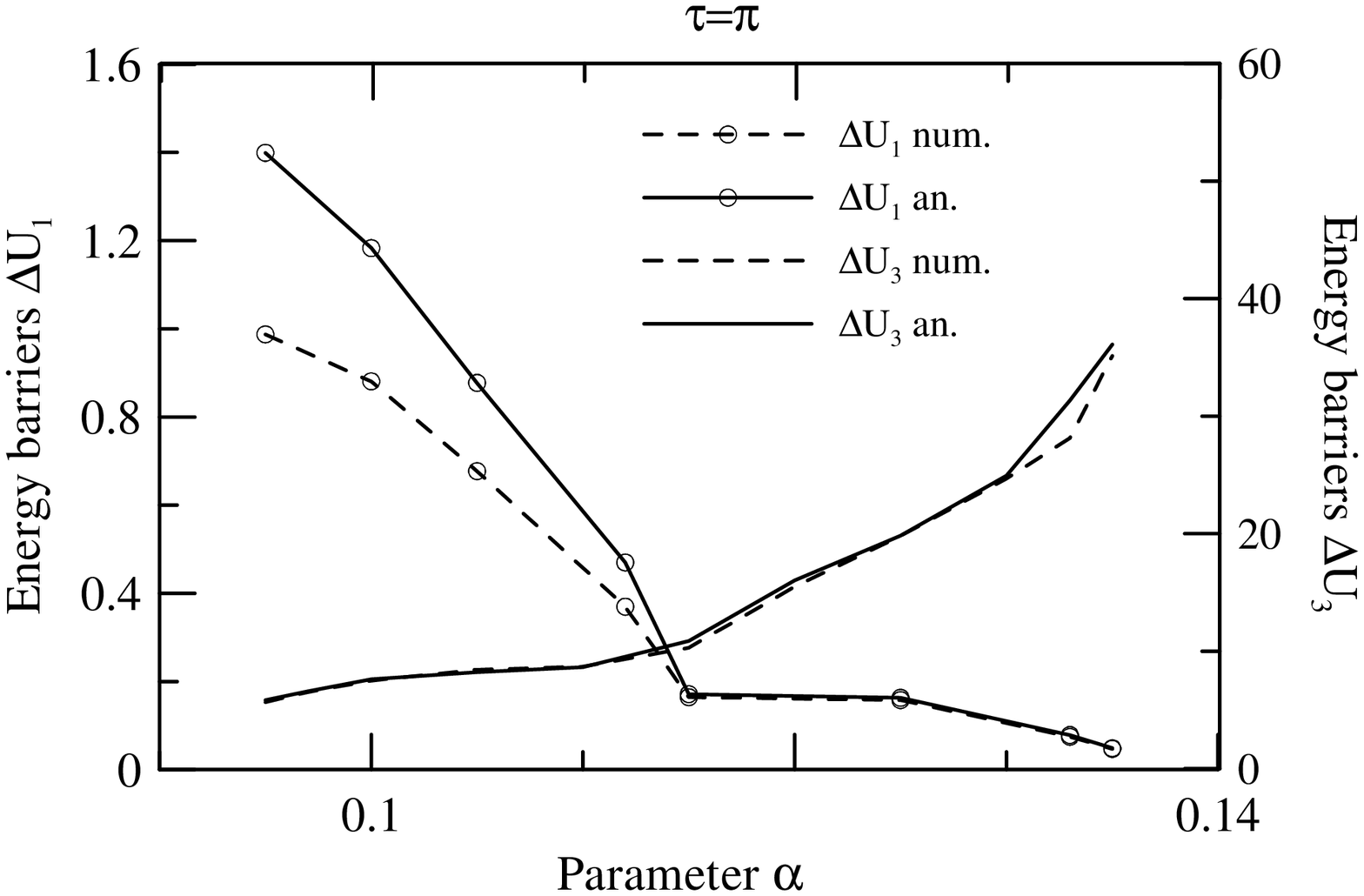} \hspace{0.25cm}
\includegraphics[scale=0.25]{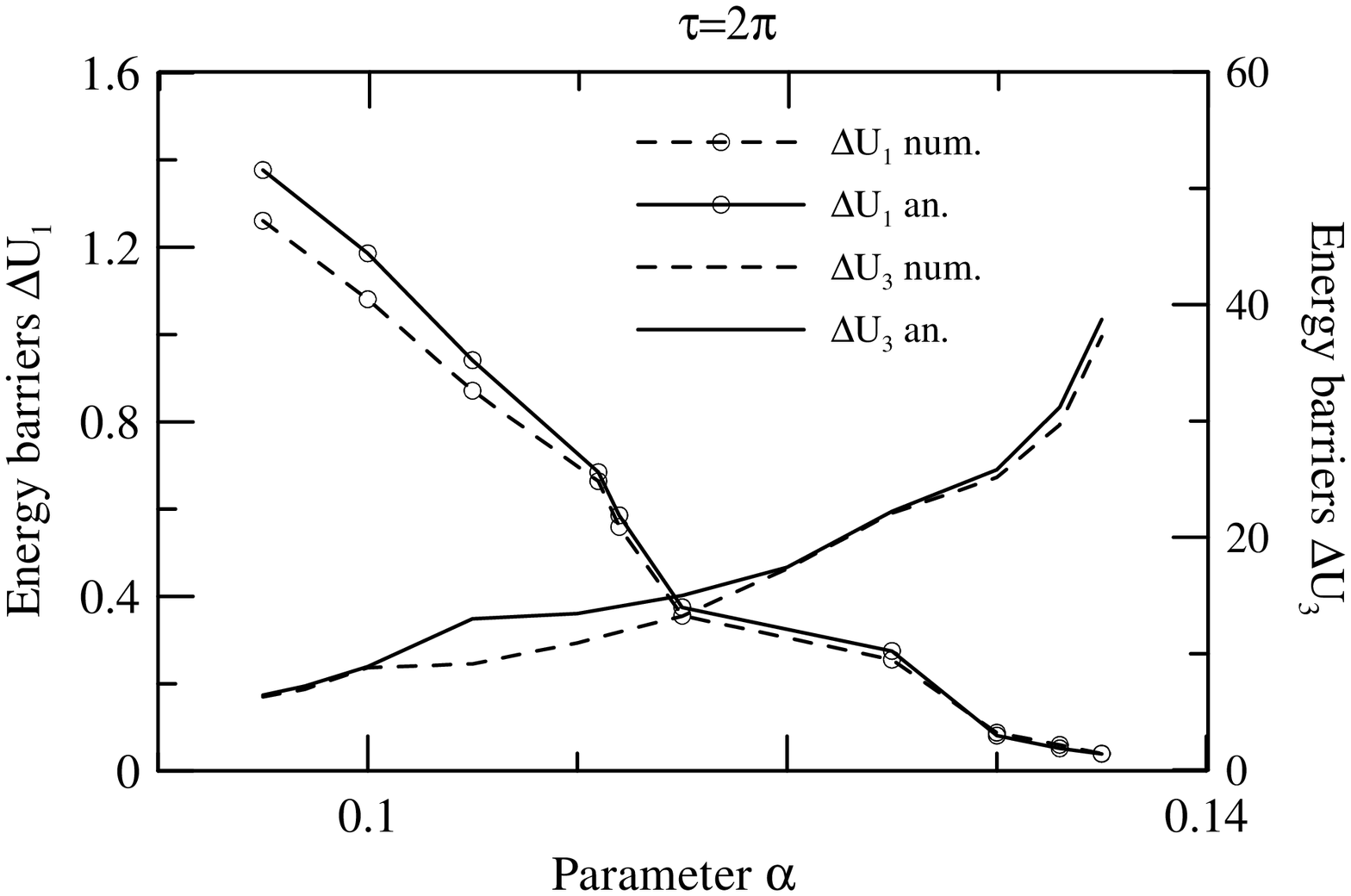}
\caption{\it
Energy barriers $\Delta U_{1,3}$ versus $\alpha$.
The lines with symbols refer to the energy barrier of the inner orbit ($\Delta U_1$, left axis), while the lines without symbols to the outer orbit activation energy /$\Delta U_3$, right axis).
Solid lines denote the estimates as per Eq.(\ref{potential}), while dashed lines denote numerical results.
The other parameters of the system are $\mu=0.01$ and $\beta=0.002$.}
\label{fig12}
\end{center}
\end{figure}

\begin{figure}[ht]
\begin{center}
\includegraphics[scale=0.25]{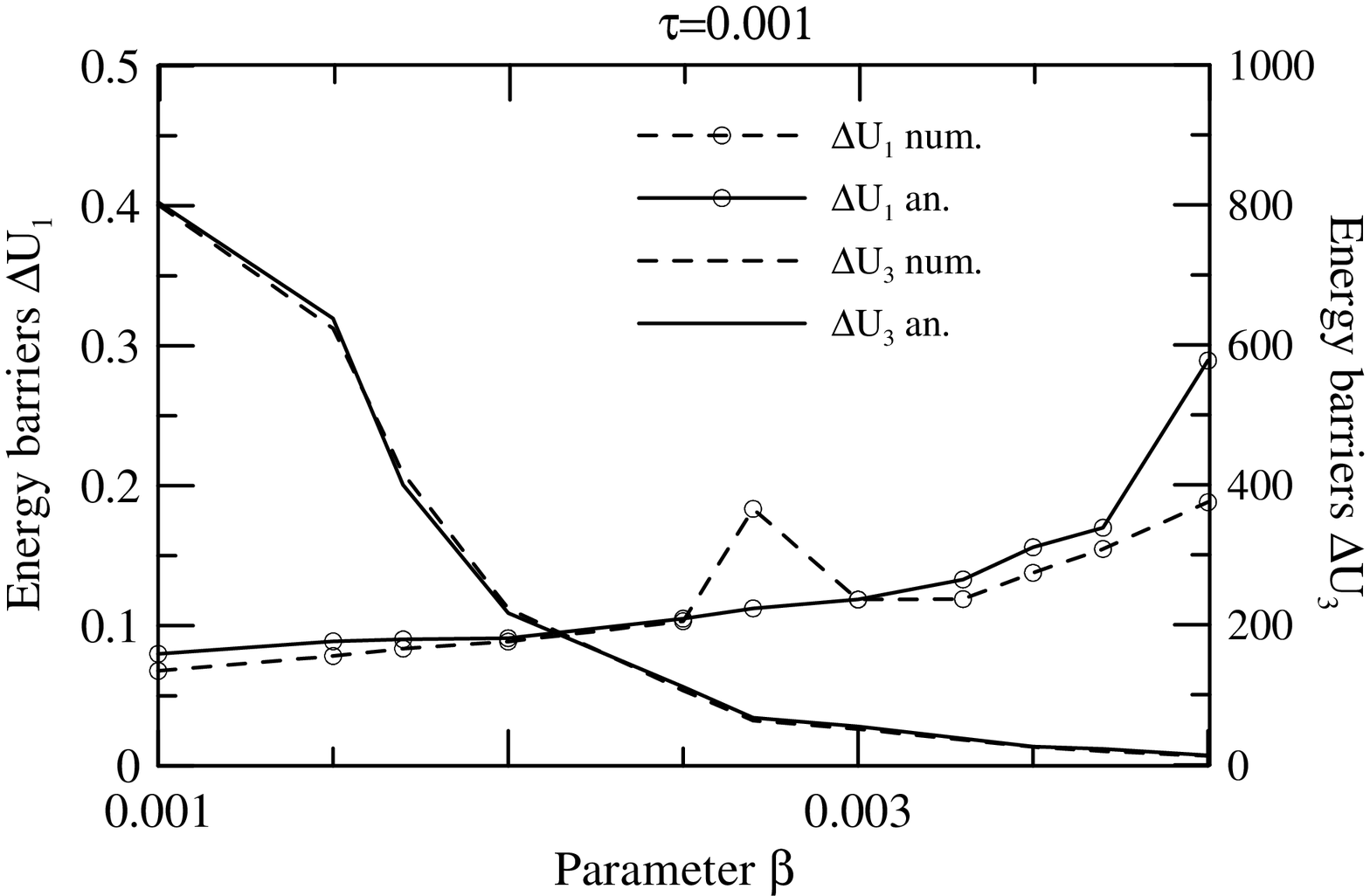}  \hspace{0.25cm}
\includegraphics[scale=0.25]{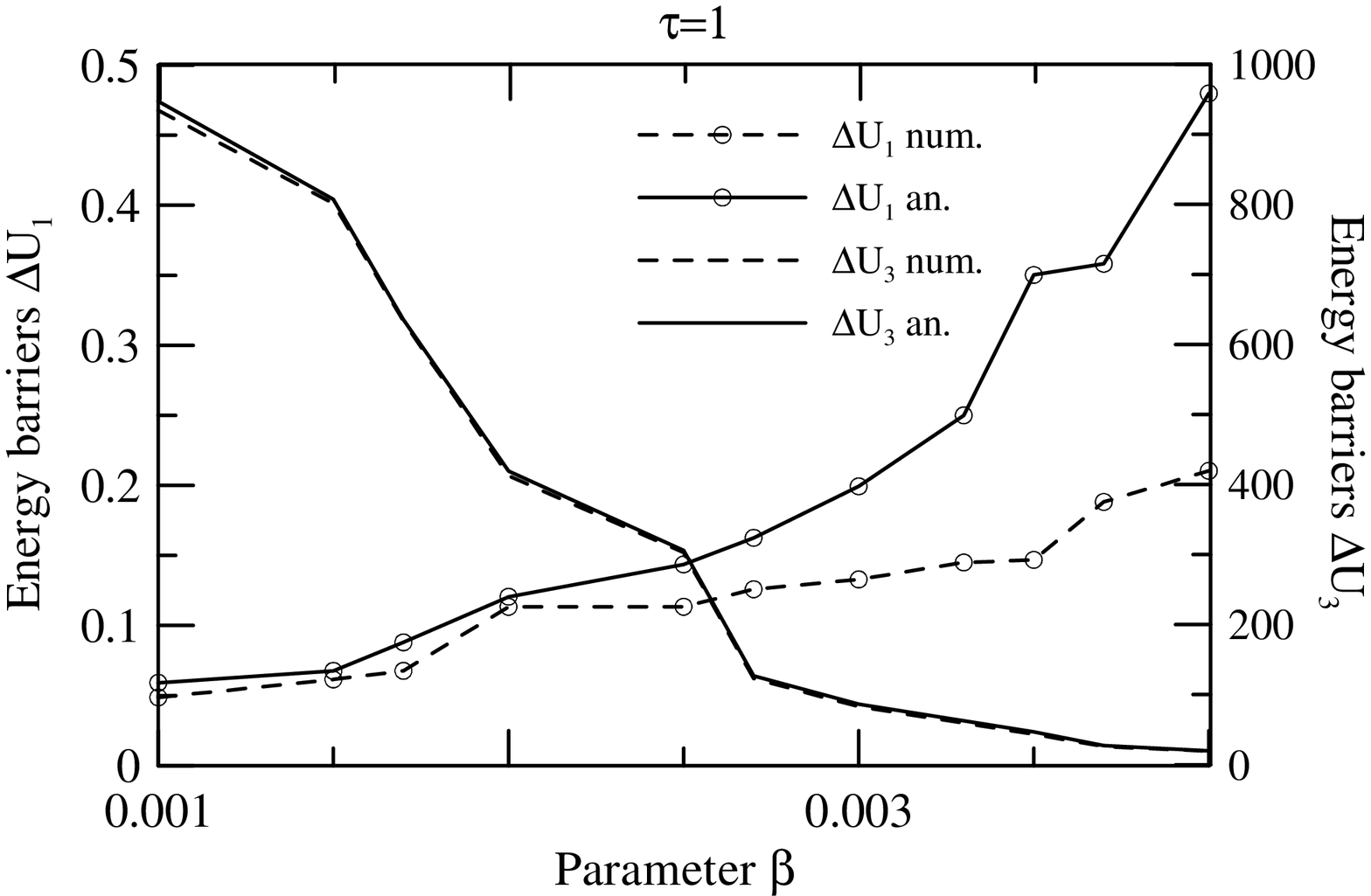}\\
\includegraphics[scale=0.25]{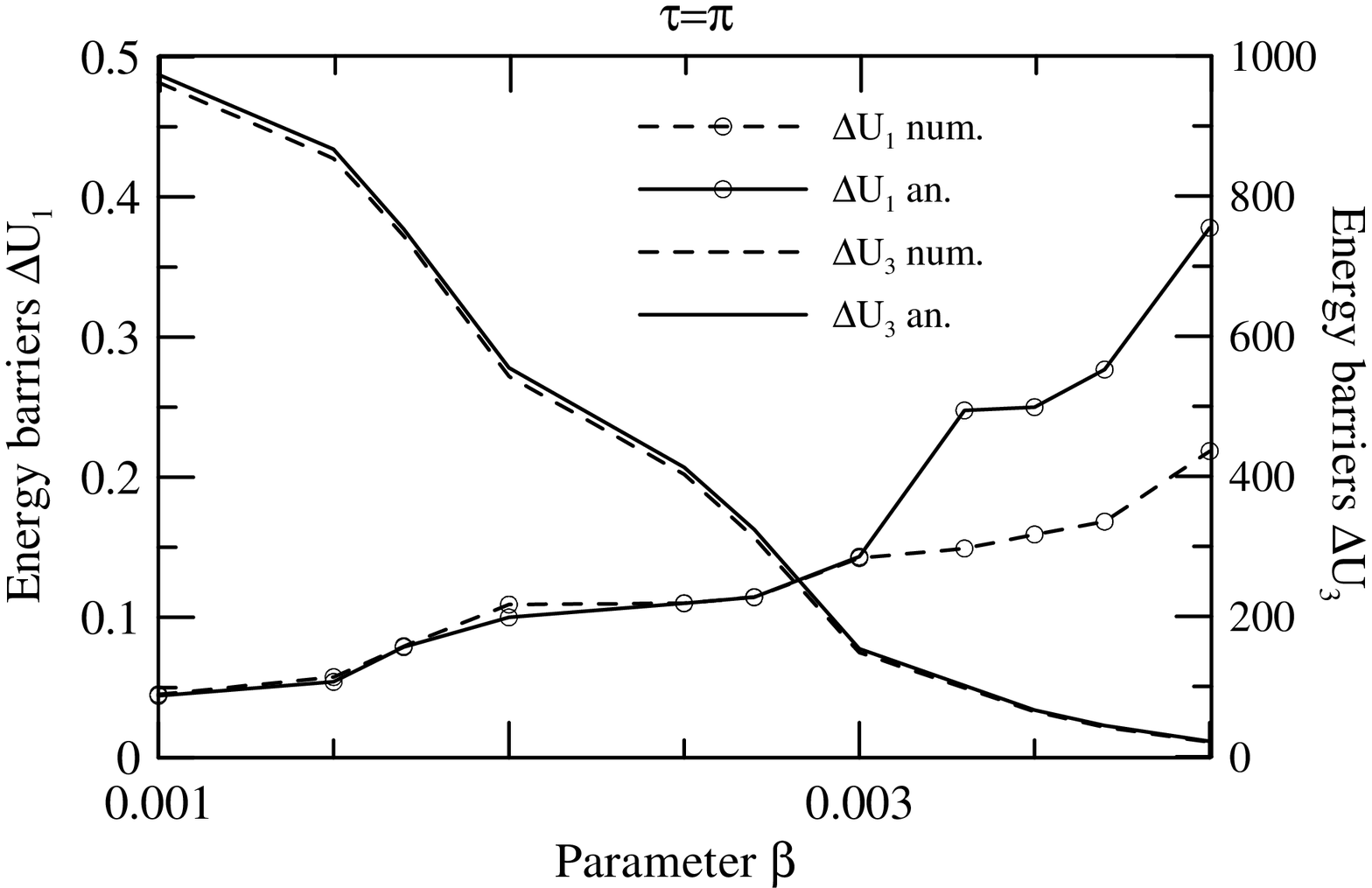} \hspace{0.25cm}
\includegraphics[scale=0.25]{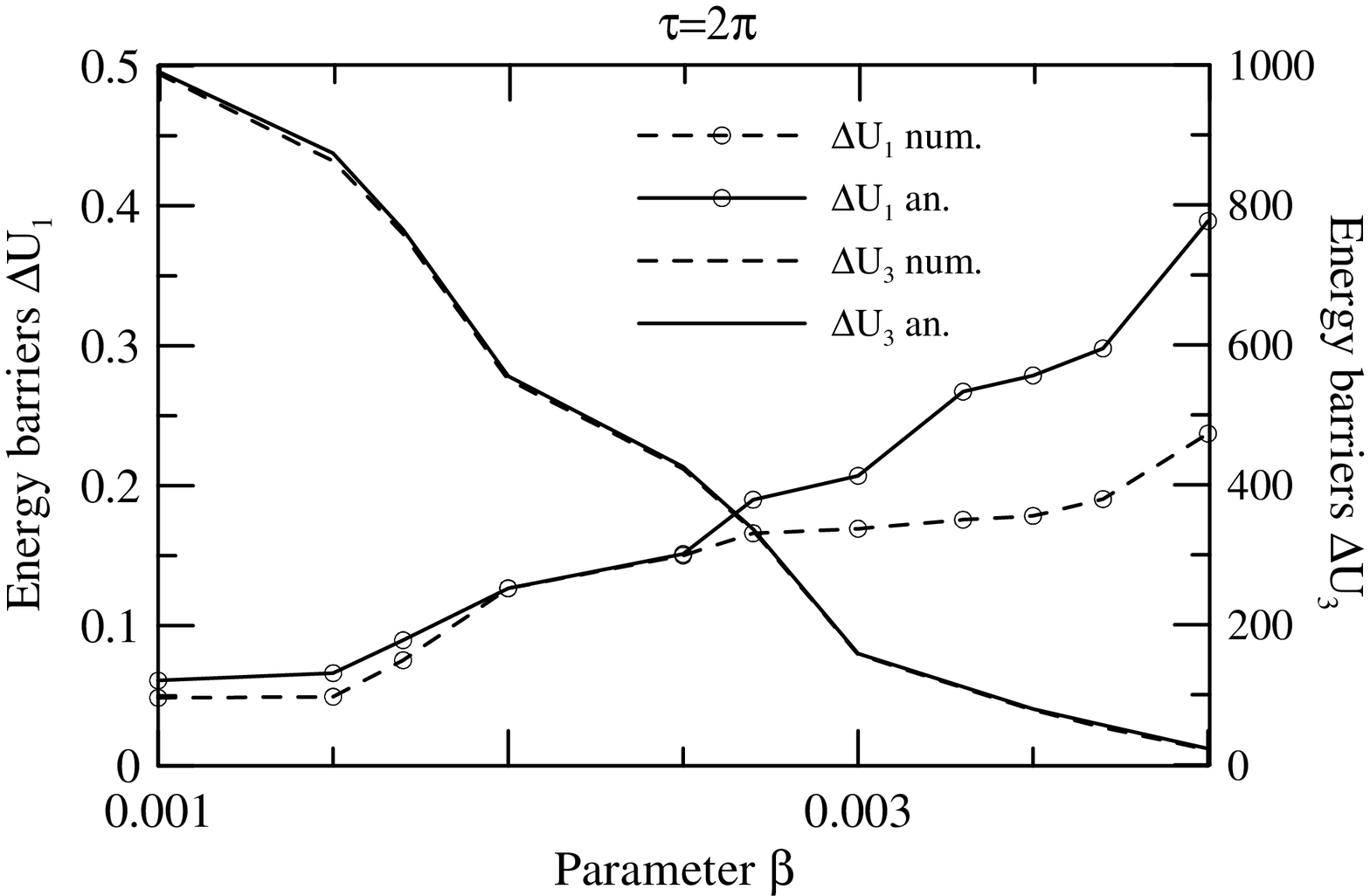}
\caption{\it
Energy barriers $\Delta U_{1,3}$ versus $\beta$.
The lines with symbols refer to the energy barrier of the inner orbit ($\Delta_1$, left axis), while the lines without symbols to the outer orbit activation energy /$\Delta U_3$, right axis).
Solid lines denote the analytical estimates as per Eq.(\ref{potential}), while dashed lines denote numerical results.
The other parameters of the system are $\mu=0.01$ and $\alpha=0.12$.}
\label{fig14}
\end{center}
\end{figure}
\end{widetext}

\begin{itemize}
\item[(i)] We begin with a correlation time that is substantially zero ($\tau=0.001$) and  $T_1$ is larger than $T_3$; in such conditions the attractor of the limit-cycle amplitude $A_1$ is more stable than the limit-cycle amplitude $A_3$.
(This correspond to the set of parameters $S_1$,$S_3$,$ S_5$.)
Increasing the correlation time one observes the symmetric case $T_1 \simeq T_3$: both attractors are equivalent and the system is in a symmetric bistable double well, i.e. it has approximately the same probability to stay in one or the other basins.
As the correlation time is further increased, one finds the reverse situation: $T_1< T_3$, and the first attractor becomes less stable.
Thus, for larger correlation times the system is more likely to stay on the limit-cycle attractor $A_3$.
It is interesting to notice that the agreement between the approximated analysis and the numerical simulations is satisfying for all values of the correlation time, also when the correlation is comparable to the period of the cycle, $2\pi/\Omega\simeq 2\pi$.
\item[(ii)] In the second case when the  correlation time is increased from zero ($\tau>0.001$)
the attractor orbit $A_3$ is first more stable than the attractor orbit $A_1$ (this correspond to the set of parameters $S_2$, $S_4$), and becomes  less stable  with the increases of the correlation time $\tau$.
In other words, this is the reverse transition respect to the first case (i);
\item[(iii)] In the third case (corresponding to the case $S_6$), the attractor orbit $A_3$ is much more stable ($T_3>T_1$), as in (ii).
However, as the correlation time increases, one observes the transition between
asymmetric and symmetric cases, $T_1\simeq T_3$ that seems to be the asymptotic solution for extremely large correlation times.
\end{itemize}
In Figure \ref{fig8}, we instead investigate the case when the two attractors are characterized by different frequencies (Table \ref{different}), we find still another case (for the attractors $S_7$, $S_8$, and $S_9$ ): the inner attractor corresponding to the orbit $A_1$ is more stable when the correlation time is small, and becomes progressively even more stable  when the correlation time is increased.

In conclusion, the effect of the correlation time on the relative stability of the two attractors seems to be very differentiated across the parameter space. However, despite the very different behaviors, the approximated analysis seems to capture the main physical effects.

To outline the behavior as a function of the oscillator parameters, we show in Figs. \ref{fig12}  the
behavior of the activation energies for increasing values of the noise correlation time ($\tau=0.001, 1, \pi, 2\pi$) when  the parameter $\alpha$ increases form $0.095$ to $0.135$ for $\beta = 0.002 $.
In this range of parameters, the solution exhibits (almost) identical frequencies of the two attractors ($\Omega_{1,3} \simeq 1$).
In this change of the parameter $\alpha$ the role of the two activation energies is exchanged: at low
$\alpha$ the lowest energy is the inner energy, while  at the other limit the outer activation energy is the lowest.
However, the approximation (\ref{potential}) is valid all through the parameter change for correlation times
that are as long as the period of the oscillations ($\tau \simeq 2\pi/\Omega$).

Figs. \ref{fig14} describes the same type of analysis keeping a constant value of the parameter $\alpha=0.12$ and varying the parameter $\beta$. Also in this case the effective trapping energy undergoes a drastic change and the roles of the two barrier are exchanged.
Thus, the numerical results of Figs. \ref{fig12},\ref{fig14} confirm that the analytic approach leading to the quasipotential (\ref{potential}) is very accurate in describing the system.

\section{Conclusions}
\label{conclusion}

The main conclusion of this work has been  the existence of a drastic change (a P-bifurcation
\cite{Arnold03}) of the probability distribution function of a birhythmic van der Pol like oscillator
 subject to exponentially correlated noise.
We have particularly investigated the evolution of the modified van der Pol oscillator in the region where birhythmicity occurs.
The characteristics of the birhythmic properties are strongly influenced by the nonlinear coefficients
as well as the noise intensity and correlation.
We have performed a detailed analysis of the bifurcation  diagrams in the parameters plane, and it is evident that  the noise intensity and the correlation time can be treated as bifurcation parameters.
An approximated solution of the Langevin equation based on the Fokker-Planck equation in the quasiharmonic regime \cite{Xu11} gives the stationary probability density $p(A)$ of the instantaneous amplitude.
With this approach, a stochastic bifurcation,  a qualitative change of the stationary solution, occurs when the  intensity or correlation of the noise is changed.
In this system stochastic P-bifurcations \cite{Arnold03} correspond to the appearance and disappearance of one of the maxima of the distribution of the amplitudes $p(A)$.
The boundary of the existence of multi-limit-cycle solutions, in the parametric ($\alpha$ , $\beta$)-plane, is affected in different manners by the noise intensity $D$ and the correlation times $\tau$:
As expected for the standard analysis of correlated noise, to increase the correlation time amounts to decrease the effect of the noise intensity.
The simulations have confirmed that the approximated (obtained through stochastic averaging) Fokker-Planck
equation well describes this type of birhythmic oscillators.
{\hg We therefore expect that real systems (either biological or electronic circuits) governed by birhythmic van der pol type equation are very sensitive to changes in the correlation time of the noise.
In fact, the appearance of correlated noise can drastically change the structure of the attractors, that could possibly be a signature of the presence of significatively correlated noise.
For instance, the noise correlation in Fig.\ref{fig8} or Fig. \ref{fig12} entails that the most stable attractors might become another one, with a different amplitude (Fig. \ref{fig8}) or both different amplitude and frequency (Fig.\ref{fig11}, parameter set $S_{10}$).
}

Summing-up, we have found that the quasiharmonic balance is (surprisingly) effective in predicting the features of the full system, as checked with numerical simulations, even when the correlation time is comparable to the period of the solutions, Fig. \ref{fig10}, (and this is even more surprising) when the two solutions are characterized by different frequencies.
However, we have found (not surprisingly) that when the parameter $\mu$, that tunes the nonlinearity, is increased, the predictions are less accurate, see Fig. \ref{fig11}.
It is therefore tempting to draw the general conclusion that the quasi-harmonic balance is an effective tool for
self-oscillatory systems, and even for birhythmic systems characterized by two different frequencies, if the
nonlinear terms are kept at bay.
Of course, no matter however relevant it might be considered the van der Pol oscillator, one should also bear in mind that the conclusion is based on a specific set of simulations on a single system.
As such, the conclusion should be taken with due care.

\section*{Acknowledgments}

The authors thank E. Tafo Wembe for enriching contributions.

\end{document}